\documentclass[12pt,a4,authoryear]{article}
\usepackage[authoryear]{natbib}
\usepackage{authblk}
\usepackage{amsmath,amssymb}
\usepackage{amsthm}
\usepackage{xcolor}
\usepackage{CJK}
\usepackage{blkarray}
\usepackage{geometry}
\usepackage{mathrsfs}
\usepackage{booktabs}
\usepackage{epstopdf}
\usepackage{diagbox}
\usepackage{amsfonts}
\usepackage[colorlinks,
linkcolor=blue,
anchorcolor=blue,
citecolor=blue
]{hyperref}
\usepackage{multicol}
\usepackage{multirow}
\usepackage{algorithm}
\usepackage{algorithmic}
\usepackage{rotating}
\usepackage{setspace}
\usepackage{graphicx}
\usepackage[toc,page,title,titletoc,header]{appendix}
\usepackage{indentfirst} 
\newtheorem{thm}{Theorem}

\newtheorem{df}{Definition}
\newcommand{\dT}{\mathsf{T}}
\allowdisplaybreaks[4]

\def\squarebox#1{\hbox to #1{\hfill\vbox to #1{\vfill}}}
\def\boxit#1{\vbox{\hrule\hbox{\vrule\kern6pt
          \vbox{\kern6pt#1\kern6pt}\kern6pt\vrule}\hrule}}

\begin{document}
	\title{Bayesian Knockoff Filter}
	\author[1]{Jiaqi Gu}
	\author[2,3]{Guosheng Yin}
	\affil[1]{Department of Neurology and Neurological Sciences, Stanford University
	}
	\affil[2]{Department of Statistics and Actuarial Science, The University of Hong Kong
	}
	\affil[3]{Department of Mathematics, Imperial College London
	}
	\date{}                     
	\setcounter{Maxaffil}{0}
	\renewcommand\Affilfont{\itshape\small}
	\maketitle
	
	\begin{abstract}
		In many scientific fields, researchers are interested in discovering features with substantial effect on the response from a large number of features while controlling the proportion of false discoveries. By incorporating the knockoff procedure in the Bayesian framework, we develop the Bayesian knockoff filter (BKF) for selecting features that have important effect on the response. In contrast to the fixed knockoff variables in a frequentist procedure, we allow the knockoff variables to be continuously updated using the Markov chain Monte Carlo.	Based on the posterior samples and the elaborated greedy selection procedure, our method can distinguish the truly important features
from unimportant ones and the Bayesian false discovery rate
can be controlled at a desirable level. Numerical experiments on both synthetic and real data demonstrate the advantages of our
BKF over existing knockoff methods and Bayesian variable selection approaches, i.e.,
the BKF possesses higher power and yields a lower false discovery rate.
	\end{abstract}
	
	\newpage
	\noindent
	
	\section{Introduction}
	
	Identifying important features that have substantial effect on a
	response variable is one of the most common problems in both
	machine learning and statistics. Traditionally, the importance of a feature
	can be measured by the value of the fitted
	regression coefficient or the $p$-value from
	hypothesis testing of a parameter under a statistical model. Recent decades
	have witnessed the emergence of vast data of high dimensionality, and
	extensive research has been carried out for feature selection.
	Taking the generalized linear model (GLM) as an example, variable selection methods \citep{Efroymson1960,Mitchell1988,Tibshirani1996,Fan2001,Park2008,Carvalho2010},
	and multiple testing procedures \citep{Benjamini1995,Sarkar1997,Yekutieli2001,Storey2002,Leek2008,Whittemore2007,Blanchard2009} represent
	the two main classes of approaches to learning important features
	with respect to the response.
	
	However, the aforementioned methods have several limitations. For variable selection methods, the false discovery rate (FDR) is not controlled and thus their reliability is questionable. Spurious features with no effect on the response might be falsely selected
	because they are correlated with some important features. Although multiple testing procedures \citep{Blanchard2009}
	can be incorporated to control the FDR under an arbitrary dependency structure of
	features, it tends to be conservative for practical use.
	To overcome such limitations,
	a new powerful method called the knockoff filter \citep{Barber2015}
	has been developed recently. The knockoff variables mimic
the dependency structure of original features, while they act as control variables.
By introducing a statistic that overestimates the FDR for
	arbitrary dependency structures among features, the knockoff filter
	is able to conduct feature selection with a well-controlled FDR.
	Inspired by this idea, a series of knockoff methods have been developed \citep{Dai2016,Candes2018,Gimenez2019,Gimenez2019b,Katsevich2019,Barber2019,Sesia2019,Bates2020}, which however
	are all frequentist approaches and their inferences heavily rely upon
	the quality of the sole set of generated knockoff variables. To control the FDR under
	a desired level, the existing methods sacrifice their capability to identify
	the truly important features if a set of poor-quality knockoff variables is
	generated or the number of truly important features is small, and this would further cause power loss.
	Up to now, these two problems have only
	been partially solved by multi-knockoffs \citep{Gimenez2019b} at the cost of power loss in the optimal cases.
	
	In the Bayesian paradigm, we restate the concepts of knockoff variables and FDR
	and develop the Bayesian knockoff filter (BKF).
	With the Markov chain Monte Carlo (MCMC) algorithm, we can estimate the
upper bound of the posterior probability that a particular feature has no effect on the response
and thus compute the Bayesian estimator of the set of important features via an elaborated greedy selection algorithm.
Experiments show that our method outperforms existing single-knockoff methods and is comparable to multi-knockoffs in distinguishing
	important (non-null) features from unimportant ones when
	the FDR is controlled at the same target level. Further, existing Bayesian variable
	selection approaches are unable to control the FDR when the sample size is small. In
comparison with the existing frequentist knockoff methods, the BKF is more robust in terms of power
when the distribution of covariates is misspecified.
		
	The rest of this article is organized as follows. In Section \ref{background},
we formulate the feature selection problem and briefly introduce existing knockoff procedures.
Details of the BKF, including the Bayesian model,
definition of Bayesian FDR, FDR-controlled selection procedure,
MCMC algorithm and its relationships with existing knockoff methods are provided in Section \ref{BKF}.
We conduct experiments on synthetic and real data respectively in Sections \ref{Experiments} and \ref{Real} to investigate performances of
the BKF under different circumstances and compare it with existing methods
in terms of both statistical power and the ability to mitigate false discoveries.
Section \ref{Conclusion} concludes with discussions.
	
\section{Background}\label{background}
	
\subsection{Multiple Testing and Variable Selection}\label{Setup}
	
Consider a dataset $\textbf{D}$ with $n$ independent and
identically distributed (i.i.d.) observations
	$(\textbf{x}_i,y_i), i=1,\ldots,n$, where $\textbf{x}_i$ and $y_i$ are
	copies of feature vector $\textbf{X}=(X_{1},\ldots,X_{p})^{\dT}\in \mathbb{R}^p$
	and response variable $Y\in \mathbb{R}$ respectively.
	It is assumed that response $Y$ only depends on a relatively small subset of features.
	Conditional on this subset of important features,
$Y$ is independent of the remaining features.
Specifically, there are two disjoint subsets of features,
	$\mathcal{H}_{0}$ (null set) and $\mathcal{H}_{1}$ (non-null set), satisfying that
	\begin{itemize}
		\item[(i)] $\mathcal{H}_{0}\cup\mathcal{H}_{1}=\{1,\ldots,p\}$;
		\item[(ii)] $\forall j\in \mathcal{H}_{0}$, $X_j\perp Y|\textbf{X}_{-j}$, where $\perp$
		represents independence between two variables;
		\item[(iii)] $\forall j\in \mathcal{H}_{1}$, $X_j\not\perp Y|\textbf{X}_{-j}$, where $\not\perp$
		represents not independence between two variables;
	\end{itemize}
	where $\textbf{X}_{-j}=(X_{1},\ldots,X_{j-1},X_{j+1},\ldots,X_{p})^{\dT}$. The set $\mathcal{H}_1$ contains all the
	non-null features that have important
	effects on response ${Y}$, while all the
	null features in $\mathcal{H}_0$ are irrelevant to
	the response given the other features. Based on
	the observed data $\textbf{D}=\{(\textbf{x}_i,y_i): i=1,\ldots,n\}$, our goal is to obtain
an estimator $\hat{\mathcal{S}}$ of the non-null set $\mathcal{H}_{1}$ so that:
	\begin{itemize}
		\item[(a)] The false discovery rate (FDR),
		\begin{equation}\label{df:FDR}
			\text{FDR}=\mathbb{E}\Bigg(\frac{|\hat{\mathcal{S}}\cap\mathcal{H}_{0}|}{|\hat{\mathcal{S}}|\vee 1}\Bigg),
		\end{equation}
		is controlled under a desired level $\alpha$.
		\item[(b)] The number of true discoveries, $|\hat{\mathcal{S}}\cap\mathcal{H}_{1}|$, is as large as possible, where
$|\cdot|$ represents the size of a set.
	\end{itemize}
	
	In the framework of hypothesis testing, we are interested in testing the null hypotheses
	\begin{equation}
		\label{hypothese}H_{0j}:X_j\bot Y|\textbf{X}_{-j},\quad j=1,\ldots,p,
	\end{equation}
	by constructing a multiple testing procedure on $H_{01},\ldots,H_{0p}$ with a controlled FDR. The FDR (\ref{df:FDR}) is analogous to the
	type I error rate in traditional single hypothesis testing, while a larger size of
the joint set $\hat{\mathcal{S}}\cap\mathcal{H}_{1}$ implies
	higher power in testing hypotheses (\ref{hypothese}).
By assuming a generalized linear model (GLM) $h(Y|\textbf{X};\boldsymbol{\beta},\boldsymbol{\phi})$
for the conditional distribution $f(Y|\textbf{X})$, we have
	\begin{equation}\label{GLM}
		E\big(Y|\textbf{X}\big)=g^{-1}(\eta),\quad\eta=\sum_{j=1}^{p}X_j\beta_j,
	\end{equation} with a link function $g(\cdot)$.
Thus, testing the null hypothesis $H_{0j}$ is equivalent to testing
	$H^*_{0j}:\beta_j=0$ for $j=1,\ldots,p$ under the GLM,
which is equivalent to
conducting variable selection on features $X_{1},\ldots,X_{p}$ \citep{Candes2018}.
	
Various methods have been proposed
	in the literature for multiple testing on hypotheses $H_{01},\ldots,$ $H_{0p}$
	under the GLM. Existing methods
	can be classified into two popular paradigms.
	One is variable selection in regression analysis under both frequentist and Bayesian frameworks. Frequentist methods in this class include step-wise regression \citep{Efroymson1960} and penalized regression methods with different penalty functions, such as ridge regression, Lasso \citep{Tibshirani1996}, SCAD \citep{Fan2001} and their variants. By fitting a GLM to $(\textbf{X},Y)$ with
	penalties or step-wise selection procedures to enhance sparsity on $\beta_1,\ldots,\beta_p$, we can obtain the estimator $\hat{\mathcal{S}}=\{j:\hat{\beta}_j\neq 0\}$ for the set of non-null features. Although the asymptotic guarantees of these methods,
such as model selection consistency, have been established even under high-dimensional settings, they still suffer
an uncontrollable FDR with finite sample size. Bayesian variable selection approaches,
such as the spike-and-slab prior \citep{Mitchell1988},
Bayesian Lasso \citep{Park2008}, the horseshoe estimator \citep{Carvalho2010} and the recent
	iterative Bayesian stepwise selection (IBSS) procedure \citep{Wang2020}, compute the
	posterior distribution of $\mathcal{H}_1$ and yield
	the Bayesian estimator $\hat{\mathcal{S}}$ by minimizing a particular
	posterior expected loss. However, commonly used Bayesian estimators $\hat{\mathcal{S}}$,
	including the highest probability model and the median probability model,
do not take the FDR into consideration and hence their control of false discoveries are questionable.
	
	Following the multiple testing procedure
	\citep{Benjamini1995}, another class of approaches \citep{Sarkar1997,Yekutieli2001,Leek2008}
	calculate $p$-value for each hypothesis $H_{0j}$ ($j=1,\ldots,p$)
	or equivalently $H^*_{0j}$, which is only feasible under low-dimensional GLMs. Moreover, these
	methods can only control the FDR theoretically when these $p$-values possess
	an independent property or positive regression dependency on a subset \citep{Yekutieli2001},
	which is difficult to verify under GLMs. To overcome such shortcomings of frequentist multiple testing procedures,
	Bayesian methods have been developed to control the Bayesian FDR \citep{Storey2002,Whittemore2007}
	for multiple comparisons \citep{Scott2006,Miranda-Moreno2007,Efron2008}.
	We incorporate the recently developed knockoff methods \citep{Barber2015,Candes2018} to Bayesian multiple
	testing procedures and develop a fully Bayesian approach with
	Gibbs sampled knockoffs for feature selection.
	
	\subsection{Knockoffs}\label{KF}
	
	To control the finite-sample FDR in feature selection,
	\citet{Barber2015} propose a (fixed-$\textbf{X}$) knockoff filter to estimate $\mathcal{H}_1$ without imposing any assumptions
	on dependency structures among features $X_{1},\ldots,X_{p}$ under a linear model,
	\begin{equation}\label{OLM}
		Y=\sum_{j=1}^{p}X_j\beta_j+\epsilon,\quad\epsilon\sim N(0,\sigma^2),
	\end{equation}
	where features are assumed to be fixed.
	In contrast, the
	model-$\textbf{X}$ knockoff filter \citep{Candes2018} makes an extension by
	assuming the distribution $f(\textbf{X})$ to be known
	but $f(Y|\textbf{X})$ unknown. With the distribution of features $f(\textbf{X})$,
a joint model $f(\textbf{X},\tilde{\textbf{X}})$ satisfying the following Definition \ref{DF:Knockoff} is constructed and the model-$\textbf{X}$ knockoff variables $\tilde{\mathbb{X}}=(\tilde{\textbf{x}}_1,\ldots,\tilde{\textbf{x}}_n)^{\dT}$ are generated
	conditional on the observed features $\mathbb{X}=({\textbf{x}}_1,\ldots,{\textbf{x}}_n)^{\dT}$.

	\begin{df}
		\label{DF:Knockoff}
		\textbf{Model-X knockoff \rm{\citep{Candes2018}}: }For random variables ${\bf X}=(X_{1},\ldots,X_{p})^{\dT}$ of any families and response $Y$, the random variables ${\tilde{\bf X}}=(\tilde{X}_{1},\ldots,\tilde{X}_{p})^{\dT}$ are the model-X knockoff variables for ${\bf X}$ if
		\begin{enumerate}
			\item[(1)]\label{cond1} for any subset $\mathcal{S}\subset\{1,\ldots,p\}$,
			\begin{equation}\label{Exchange}
				({\bf X},{\tilde{\bf X}})_{{\rm Swap}(\mathcal{S})}\ \ {\displaystyle\mathop{=\joinrel=}^{\cal D}}\ \ ({\bf X},{\tilde{\bf X}}),
			\end{equation}
			where $({\bf X},{\tilde{\bf X}})_{{\rm Swap}(\mathcal{S})}$ is obtained by swapping elements $X_j$ and $\tilde{X}_j$ of $({\bf X},{\tilde{\bf X}})$ for all $j\in\mathcal{S}$ and ${\displaystyle\mathop{=\joinrel=}^{\cal D}}$ denotes equality in distribution;
			\item[(2)]\label{cond2} conditional on ${\bf X}$, ${\tilde{\bf X}}$ and $Y$ are independent, i.e., ${\tilde{\bf X}}\bot Y|{\bf X}$.
		\end{enumerate}
	\end{df}
	
	If $f({\bf X})$ is unknown, \citet{Candes2018} provide an approximate construction on the basis of the first two moments of features with a graphical Lasso estimator of the covariance matrix.
	That is, a Gaussian graphical model is fitted to
features $\mathbb{X}=({\textbf{x}}_1,\ldots,{\textbf{x}}_n)^{\dT}$ and the joint model $f({\bf X},\tilde{ {\bf X}})$ is
	constructed based on the fitted graphical model.
	\citet{Barber2019} further incorporate a
	screening procedure prior to implementation of
	the knockoff filter to make it feasible in high-dimensional
	settings. \citet{Dai2016,Katsevich2019} extend the knockoff method
	in a way that prior knowledge of group structures among features can be utilized
	in feature selection and the FDR can be controlled at
	both the feature level and group level.
	
	With the knockoff sample $\tilde{\mathbb{X}}$,
	the model-\textbf{X} knockoff filter computes feature statistics $W_1,\ldots,W_p$ to evaluate the evidence in $\textbf{D}$ against hypotheses $H_{01},\ldots,H_{0p}$.  As suggested by \citet{Candes2018}, $W_j$ should satisfy the
	flip-sign property under swapping of the $j$-th feature with its knockoff. Typical ways to construct feature statistics $W_1,\ldots,W_p$ include the Lasso signed max \citep{Barber2015} and Lasso coefficient-difference \citep{Candes2018} statistic.
	A large value of $W_j$ usually implies strong evidence in $\textbf{D}$ against $H_{0j}$.
	The estimator $\hat{\mathcal{S}}$ is then obtained as
	$\{j:W_j\geq \tau\}$ where the threshold $\tau$ is chosen as
	\begin{equation}
		\label{threshold}
		\tau=\min\{t>0:\widehat{\text{FDP}}(t)\leq \alpha\},
	\end{equation}
	where the estimator of the false discovery proportion
	\begin{equation*}
		\widehat{\text{FDP}}(t)=\frac{1+|\{j:W_j\leq -t\}|}{|\{j:W_j\geq t\}|\vee1}
	\end{equation*}
	is shown to overestimate the overall FDR for all $t\in (0,\infty)$ \citep{Candes2018}.
	
	However, all the aforementioned knockoff filters are frequentist methods and their inferences are based on one knockoff sample $\tilde{\mathbb{X}}$ only. Because $\widehat{\text{FDP}}(t)$ overestimates FDR, it is
	a conservative way to control the overall FDR by adopting the estimator of FDP, and
	the inference is highly dependent on the quality of the generated knockoff sample. With a low-quality $\tilde{\mathbb{X}}$, such procedures may incur power loss. In addition, all the methods rely on the assumption that the dimension of the feature vector ($p$) is large, so that the selection procedures force the size of $\hat{\mathcal{S}}$ to be zero or not smaller than $\lfloor1/\alpha\rfloor$ if nonzero. This property results in power loss when the size of the true $\mathcal{H}_1$ is small. Up to now, the two problems
	have only been partially resolved by multi-knockoffs \citep{Gimenez2019b} at the cost of power loss for large $\mathcal{H}_1$.
	
	
	\section{Bayesian Knockoff Filter}
	\label{BKF}
	
	
	\subsection{Bayesian Model}\label{BHM}
	To overcome the weaknesses of existing knockoff filters, we develop
	the Bayesian knockoff filter (BKF).
	Similar to the model-$\textbf{X}$ knockoff filter, we assume the distribution $f(\textbf{X})$ is known and establish the joint distribution $f(\textbf{X},\tilde{\textbf{X}})$ which is invariant to swaps for all subsets $\mathcal{S}\subset\{1,\ldots,p\}$. For example, if the original $\textbf{X}$ follows a multivariate Gaussian distribution $ \text{MVN}(\mathbf{0},\boldsymbol{\Sigma})$, the joint distribution $f(\textbf{X},\tilde{\textbf{X}})$ satisfying {Definition \ref{DF:Knockoff}} is
	\begin{equation}\label{Joint}
		\begin{aligned}
			\begin{pmatrix}
				\textbf{X}\\\tilde{\textbf{X}}
			\end{pmatrix}\sim \text{MVN}\Bigg[\begin{pmatrix}
				\mathbf{0}\\\mathbf{0}
			\end{pmatrix},\textbf{G}\Bigg],\quad
			\quad\text{with }\textbf{G}=\begin{bmatrix}
				\boldsymbol{\Sigma}&\boldsymbol{\Sigma}-\text{diag}\{\textbf{s}\}\\
				\boldsymbol{\Sigma}-\text{diag}\{\textbf{s}\}&\boldsymbol{\Sigma}\\
			\end{bmatrix},
		\end{aligned}
	\end{equation}
	where the diagonal matrix $\text{diag}\{\textbf{s}\}$ satisfies the condition that $2\text{diag}\{\textbf{s}\}-\text{diag}\{\textbf{s}\}\boldsymbol{\Sigma}\text{diag}\{\textbf{s}\}$
is a positive semi-definite matrix \citep{Barber2015}.
	
	In the case of Gaussian covariates, (\ref{Joint}) can be used
	to construct knockoff variables. However, if covariates are non-Gaussian or even not continuous, it is challenging to deduce the joint distribution $f(\textbf{X},\tilde{\textbf{X}})$. Although there are several existing methods to generate $\tilde{\textbf{X}}$ without an explicit expression of  $f(\textbf{X},\tilde{\textbf{X}})$ \citep{Sesia2019,Bates2020}, they rely on either a specific structure of $f(\textbf{X})$
(e.g., a hidden Markov model or a graphical model) or a
computational strategy (multiple-try Metropolis).
If the distribution of the original $\textbf{X}$ is unknown, we
	require $(\textbf{X},\tilde{\textbf{X}})_{{\rm Swap}(\mathcal{S})}$ and $(\textbf{X},\tilde{\textbf{X}})$ to
	have the same first two moments rather than the same distribution for any subset $\mathcal{S}$.
	We adopt the second-order approximation construction in \citet{Candes2018} to approximate
	$f(\textbf{X})$ with a Gaussian model $\text{MVN}(\hat{\boldsymbol{\mu}},\hat{\boldsymbol{\Sigma}})$ with
	the estimated mean $\hat{\boldsymbol{\mu}}$ and covariance matrix $\hat{\boldsymbol{\Sigma}}$ to
	construct the joint distribution $f(\textbf{X},\tilde{\textbf{X}})$.
	As a result, the conditional generative model $f(\tilde{\textbf{X}}|\textbf{X})$
	can be deduced.
	
Under the GLM assumption (\ref{GLM}) for $f(Y|\textbf{X})$,
we parameterize the extended conditional distribution $f(Y|\textbf{X},\tilde{\textbf{X}})$ as
an extended GLM $h(Y|\textbf{X},\tilde{\textbf{X}};\boldsymbol{\beta},\tilde{\boldsymbol{\beta}},\boldsymbol{\phi})$
with
	\begin{equation}\label{EGLM}
		E(Y|\textbf{X},\tilde{\textbf{X}};\boldsymbol{\beta},\tilde{\boldsymbol{\beta}},\boldsymbol{\phi})
		=g^{-1}(\eta),\quad\eta=\sum_{j=1}^{p}(X_j\beta_j+\tilde{X}_j\tilde{\beta}_j),
	\end{equation}
where $\boldsymbol{\phi}$ represents nuisance parameters,
	such as variance $\sigma^2$ in a normal distribution or the dispersion parameter
	in an over-dispersed Poisson distribution \citep{Nelder1972}.
In the Bayesian paradigm, if the prior distribution is denoted by
	$f(\boldsymbol{\beta},\tilde{\boldsymbol{\beta}},\boldsymbol{\phi})$, the joint posterior density of
	knockoff variables and parameters
is given by
	\begin{equation}\label{Jointposterior}
		\begin{aligned}
			&f(\tilde{\textbf{x}}_1,\ldots,\tilde{\textbf{x}}_n,
			\boldsymbol{\beta},\tilde{\boldsymbol{\beta}},\boldsymbol{\phi}|\textbf{D}
			)\propto f(\boldsymbol{\beta},\tilde{\boldsymbol{\beta}},\boldsymbol{\phi})\prod_{i=1}^{n}h(y_i|\textbf{x}_i,\tilde{\textbf{x}}_i;\boldsymbol{\beta},\tilde{\boldsymbol{\beta}},\boldsymbol{\phi})f(\tilde{\textbf{x}}_i|\textbf{x}_i).
		\end{aligned}
	\end{equation}
	
	\subsection{Feature Selection with Bayesian FDR}\label{bfdr}
	
	Analogous to the existing frequentist work, our
	goal is to obtain a Bayesian estimator $\hat{\mathcal{S}}$
	for the set of non-null features $\mathcal{H}_1$ with the Bayesian FDR controlled and the posterior expected number of true discoveries as large as possible. However, under the Bayesian paradigm, the sets $\mathcal{H}_{1}$ and $\mathcal{H}_0$ are both assumed to
	be random and thus the Bayesian FDR of any subset $\mathcal{S}$ is defined as follows.
	\begin{df}
		\label{BFDR}
		\textbf{Bayesian FDR \rm{\citep{Storey2002,Whittemore2007}}:} For all possible subsets ${\mathcal{S}}\subset\{1,\ldots,p\}$,
the Bayesian false discovery rate {\rm (BFDR)} is
		\begin{equation}
			\label{BayesianFDR1}
			\text{\rm BFDR}({\mathcal{S}})=\mathbb{E}\Bigg(\frac{|{\mathcal{S}}\cap\mathcal{H}_{0}|}{|{\mathcal{S}}|\vee 1}\Bigg|{\bf D}\Bigg).
		\end{equation}
		
	\end{df}

	If the posterior probabilities ${\mathbb{P}}(H_{01}|\textbf{D}),\ldots,{\mathbb{P}}(H_{0p}|\textbf{D})$ are known, it is clear that
	$$
	\mathbb{E}\big(|{\mathcal{S}}\cap\mathcal{H}_{0}|\big|\textbf{D}\big)
	=\mathbb{E}\Bigg(\sum_{j\in{\mathcal{S}}}I[j\in\mathcal{H}_{0}]\Bigg|\textbf{D}\Bigg)\\
	=\sum_{j\in{\mathcal{S}}}\mathbb{P}(H_{0j}|\textbf{D}).\\
	$$
	As a result, we can obtain an equivalent definition of the Bayesian FDR for all possible ${\mathcal{S}}\subset\{1,\ldots,p\}$ as
	\begin{equation}
		\label{BayesianFDR2}
		\text{BFDR}({\mathcal{S}})=\frac{1}{|{\mathcal{S}}|\vee 1}\sum_{j\in{\mathcal{S}}}\mathbb{P}(H_{0j}|\textbf{D}),
	\end{equation}
	and estimating $\mathcal{H}_1$ can be translated into a decision problem to select a subset,
	\begin{equation}
		\label{Loss}
		\begin{aligned}
			\hat{\mathcal{S}}&=\mathop{\arg\min}_{\mathcal{S}\in \{1,\ldots,p\}}E[L(\mathcal{H}_1,{\mathcal{S}})|\textbf{D}],\quad s.t. \quad\text{BFDR}({\mathcal{S}})\leq \alpha,
		\end{aligned}
	\end{equation}
	where the loss function $L(\mathcal{H}_1,{\mathcal{S}})=-|{\mathcal{S}}\cap \mathcal{H}_1|$.
	Under the Bayesian paradigm, the importance of feature $X_j$ is
characterized by the posterior probability $\mathbb{P}(H_{1j}|\textbf{D})$,
	and minimizing the posterior expected loss $E[L(\mathcal{H}_1,{\mathcal{S}})|\textbf{D}]=-\sum_{j\in{\mathcal{S}}}\mathbb{P}(H_{1j}|\textbf{D})$ is equivalent to maximizing the overall importance of features in the subset $\mathcal{S}$. Following the equivalent definition (\ref{BayesianFDR2}) and $\mathbb{P}(H_{1j}|\textbf{D})=1-\mathbb{P}(H_{0j}|\textbf{D})$, when posterior probabilities $\mathbb{P}(H_{0j}|\textbf{D})$ ($j=1,\ldots,p$) are known, the constrained optimization problem can be viewed as a knapsack problem to include as many features in $\hat{\mathcal{S}}$ as possible while keeping ${{\text{BFDR}}}({\mathcal{S}})\leq \alpha$, whose solution can be easily
obtained via a greedy selection algorithm described in Algorithm \ref{alg_greedy}.
	
	\begin{algorithm}[h]
		\caption{Greedy selection algorithm to obtain the Bayesian estimator $\hat{\mathcal{S}}$.}\label{alg_greedy}
		\begin{algorithmic}[1]
			\STATE {\bfseries Input:} Posterior probabilities ${\mathbb{P}}(H_{01}|\textbf{D}),\ldots,{\mathbb{P}}(H_{0p}|\textbf{D})$ and the target level $\alpha$.
			\STATE Initialize $\mathcal{S}_0=\varnothing$ and ${{\text{BFDR}}}(\mathcal{S}_0)=0$.
			\STATE Sort ${\mathbb{P}}(H_{01}|\textbf{D}),\ldots,{\mathbb{P}}(H_{0p}|\textbf{D})$ in an increasing order: ${\mathbb{P}}(H_{0(1)}|\textbf{D})\leq\cdots\leq {\mathbb{P}}(H_{0(p)}|\textbf{D}).$
			\FOR{$i=1,\ldots,n$}
			\STATE {Calculate ${{\text{BFDR}}}(\mathcal{S}_j)$ by (\ref{BayesianFDR2}) where
				$\mathcal{S}_j=\{j':{\mathbb{P}}(H_{0j'}|\textbf{D})\leq {\mathbb{P}}(H_{0(j)}|\textbf{D})\}$.}
			\ENDFOR
			\STATE {\bfseries Output:} Bayesian estimator $\hat{\mathcal{S}}=\mathcal{S}_k$ where $k=\max\{j:{{\text{BFDR}}}(\mathcal{S}_j)\leq \alpha\}$.
		\end{algorithmic}
	\end{algorithm}

	As shown by \citet{Mueller2004}, Algorithm \ref{alg_greedy} is optimal in the sense that it maximizes the statistical power while controlling the Bayesian FDR under $\alpha$.
	
	However, when the prior does not possess a point mass at $\beta_j=0$, posterior probabilities $\mathbb{P}(\beta_j=0|\textbf{D})$ ($j=1,\ldots,p$) equal $0$ for all possible observed data $\textbf{D}$ and thus cannot be used for meaningful inference. Alternatively, we use a set of random variables with the flip-sign property to compute the approximate upper bounds of $\mathbb{P}(H_{0j}|\textbf{D})$ ($j=1,\ldots,p$) under the posterior density (\ref{Jointposterior}) by restating the flip-sign property in the Bayesian framework as follows.

	
	\begin{df}
		\label{anti-symmetry}
		\textbf{The flip-sign property:} A family of random variables $\{W_j:j=1,\ldots,p\}$ are
		said to obey the flip-sign property if for all possible $\mathcal{H}_0\subset\{1,\ldots,p\}$, the posterior distribution of ${\rm \bf W}=(W_1,\ldots,W_p)^{\dT}$ conditional on $\mathcal{H}_0$ satisfies
		\begin{equation*}
			f({\bf W}|{\bf D},\mathcal{H}_0)=f({\bf W}_{\mathcal{S}}|{\bf D},\mathcal{H}_0)
		\end{equation*}
		for all subsets $\mathcal{S}\subset\mathcal{H}_0$ where ${\rm \bf W}_{\mathcal{S}}=(W_{1,\mathcal{S}},\ldots,W_{p,\mathcal{S}})^{\dT}$ and $$W_{j,\mathcal{S}}=\begin{cases}
			W_j,&j\notin\mathcal{S},\\
			-W_j,&j\in\mathcal{S},\\
		\end{cases}$$
		for any $\mathcal{S}\subset\mathcal{H}_0$.
	\end{df}
	
	Given the definition of model-$\textbf{X}$ knockoff, Theorem \ref{thm1} offers us a way to construct a family of random variables which obey the flip-sign property.
	
	\begin{thm}\label{thm1}
		Suppose that
		\begin{itemize}
			\item[(i)] observed features {\rm $\textbf{x}_1,\ldots,\textbf{x}_n$} and random knockoffs {\rm $\tilde{\textbf{x}}_1,\ldots,\tilde{\textbf{x}}_n$} follow a joint distribution {\rm $f(\textbf{X},\tilde{\textbf{X}})$} satisfying Definition \ref{DF:Knockoff} and;
			\item[(ii)] the extended conditional distribution {\rm$f(Y|\textbf{X},\tilde{\textbf{X}})$} is 	
			a GLM {\rm $h(Y|\textbf{X},\tilde{\textbf{X}};\boldsymbol{\beta},\tilde{\boldsymbol{\beta}},\boldsymbol{\phi})$} satisfying (\ref{EGLM}).
		\end{itemize}
		If the marginal prior $f(\boldsymbol{\beta},\tilde{\boldsymbol{\beta}})$
is invariant to swaps for any subset $\mathcal{S}\subset\{1,\ldots,p\}$,
the feature statistics ${{\rm \bf W}}=(W_1,\ldots,W_p)^{\dT}$ obey the flip-sign property as long as $W_j$
is antisymmetric with respect to $\beta_j$ and $\tilde{\beta}_j$,	
$$W_j(\beta_j,\tilde{\beta}_j)=-W_j(\tilde{\beta}_j,\beta_j),\quad j=1,\ldots,p.$$
	\end{thm}
	
	The proof of Theorem \ref{thm1} is provided in Appendix \ref{proof1}. Based on Theorem \ref{thm1} and antisymmetric feature statistics $W_1,\ldots,W_p$, the Bayesian estimator $\hat{\mathcal{S}}$ can be obtained via the greedy selection algorithm (Algorithm \ref{alg_greedy}) with $\hat{\mathbb{P}}(H_{01}|\textbf{D}),\ldots,\hat{\mathbb{P}}(H_{0p}|\textbf{D})$, which are
	the upper bounds of ${\mathbb{P}}(H_{01}|\textbf{D}),\ldots,{\mathbb{P}}(H_{0p}|\textbf{D})$ estimated as follows.
	
With the marginal prior $f(\boldsymbol{\beta},\tilde{\boldsymbol{\beta}})$ invariant
to swaps, the posterior distribution of
$W_j$ is symmetric with respect to $0$ conditional on $H_{0j}$,
	\begin{equation}\label{symm}
		\mathbb{P}(W_j<0|\textbf{D},H_{0j})=\mathbb{P}(W_j>0|\textbf{D},H_{0j}),
	\end{equation} for $j=1,\ldots,p$. By the law of total probability, we have
\begin{subequations}
		\begin{align}
			\mathbb{P}(W_j<0|\textbf{D})&=\mathbb{P}(W_j<0|\textbf{D},H_{0j})
			\mathbb{P}(H_{0j}|\textbf{D}) +\mathbb{P}(W_j<0|\textbf{D},H_{1j})\mathbb{P}(H_{1j}|\textbf{D}),\label{LTP1}\\
			\mathbb{P}(W_j>0|\textbf{D})&=\mathbb{P}(W_j>0|\textbf{D},H_{0j})
			\mathbb{P}(H_{0j}|\textbf{D}) +\mathbb{P}(W_j>0|\textbf{D},H_{1j})\mathbb{P}(H_{1j}|\textbf{D}).\label{LTP2}
		\end{align}
	\end{subequations}
If antisymmetric $W_j$ is well defined so that $\mathbb{P}(W_j>0|\textbf{D},H_{1j})>\mathbb{P}(W_j<0|\textbf{D},H_{1j})$ and (\ref{symm}) holds, we can subtract (\ref{LTP1}) from (\ref{LTP2}), which leads to
\begin{equation}		
\mathbb{P}(W_j>0|\textbf{D})-\mathbb{P}(W_j<0|\textbf{D})=\mathbb{P}(H_{1j}|\textbf{D})
\bigg\{\mathbb{P}(W_j>0|\textbf{D},H_{1j})-\mathbb{P}(W_j<0|\textbf{D},H_{1j})\bigg\}.
\end{equation}
	Due to the fact that $0<\mathbb{P}(W_j>0|\textbf{D},H_{1j})-\mathbb{P}(W_j<0|\textbf{D},H_{1j})\leq 1$, an upper bound of $\mathbb{P}(H_{0j}|\textbf{D})$ can be obtained as
	\begin{equation}\label{Inequality}
		\begin{aligned}
			\mathbb{P}(H_{0j}|\textbf{D})&=1-\mathbb{P}(H_{1j}|\textbf{D})\leq1-\mathbb{P}(W_j>0|\textbf{D})+\mathbb{P}(W_j<0|\textbf{D}).
		\end{aligned}
	\end{equation}
	
We list some examples of marginal priors that are invariant
to swaps and antisymmetric $W_j$ that satisfies
$\mathbb{P}(W_j>0|\textbf{D},H_{1j})>\mathbb{P}(W_j<0|\textbf{D},H_{1j})$ and
thus leads to inequality (\ref{Inequality}).
\begin{itemize}
		\item Invariant prior:\begin{enumerate}
			\item Flat prior: $f(\boldsymbol{\beta},\tilde{\boldsymbol{\beta}})\propto1$;
			\item Normal prior: $f(\boldsymbol{\beta},\tilde{\boldsymbol{\beta}})\propto\prod_{j=1}^p\exp\{-(\beta_j^2+\tilde{\beta}_j^2)/(2\sigma_j^2)\}$;
			\item Laplace prior: $f(\boldsymbol{\beta},\tilde{\boldsymbol{\beta}})\propto\prod_{j=1}^p\exp\{-\lambda_j|\beta_j|-\lambda_j|\tilde{\beta}_j|\}$;
			\item Modified spike-and-slab prior \citep{Candes2018}: \begin{equation}\label{MSS}
				\begin{aligned}
					&f(\boldsymbol{\beta},\tilde{\boldsymbol{\beta}})=\prod_{j=1}^pf(\beta_j,\tilde{\beta}_j),\ \ \text{ where }f(\beta_j,\tilde{\beta}_j)=\begin{cases}
						I[\beta_j=\tilde{\beta}_j=0]&\text{w.p. }(1-\xi),\\
						\phi(\beta_j;0,\tau^2)I[\tilde{\beta}_j=0]&\text{w.p. }\xi/2,\\
						\phi(\tilde{\beta}_j;0,\tau^2)I[{\beta}_j=0]&\text{w.p. }\xi/2.\\
					\end{cases}
				\end{aligned}
			\end{equation}
		\end{enumerate}
		\item Antisymmetric $W_j$:\begin{enumerate}
			\item $W_j=|\beta_j|-|\tilde{\beta}_j|$ or $W_j=|\beta_j|^2-|\tilde{\beta}_j|^2$;
			\item $W_j=|\beta_j+\tilde{\beta}_j|\cdot sign(|\beta_j|-|\tilde{\beta}_j|)$.
		\end{enumerate}
	\end{itemize}
	
	In practice, with posterior samples $(\tilde{\mathbb{X}}^{(t)},\boldsymbol{\beta}^{(t)},
	\tilde{\boldsymbol{\beta}}^{(t)},\boldsymbol{\phi}^{(t)})$ drawn from MCMC algorithms (detailed in Section \ref{mcmc}),
	$\textbf{W}^{(t)}=(W_1^{(t)},\ldots,W_p^{(t)})^{\dT}$ can be computed for $t=1,\ldots,T$ and thus
the upper bounds (\ref{Inequality}) can be estimated by
	$$\begin{aligned}
		\widehat{{\mathbb{P}}}(H_{0j}|\textbf{D})=1-\frac{1}{T}\sum_{t=1}^{T}I(W_j^{(t)}>0)+\frac{1}{T}\sum_{t=1}^{T}I(W_j^{(t)}<0),\quad j=1,\ldots,p.
	\end{aligned}$$
	
	\subsection{Markov Chain Monte Carlo}\label{mcmc}
	
	Following (\ref{Jointposterior}), Algorithm \ref{alg1} provides a Gibbs sampling framework
	to draw posterior samples\\
	$\{(\tilde{\mathbb{X}}^{(t)},\boldsymbol{\beta}^{(t)},
	\tilde{\boldsymbol{\beta}}^{(t)},\boldsymbol{\phi}^{(t)}):t=1,\ldots,T\}$,
	from the full conditional distributions.
	
	\textcolor{black}{\begin{algorithm}[h]
		\caption{Gibbs sampler}\label{alg1}
		\begin{algorithmic}[1]
			\STATE {\bfseries Input:} Observed data $\textbf{D}=\{(\textbf{x}_i,y_i);i=1,\ldots,n\}$.
			\STATE Initialize $\tilde{\textbf{x}}_i\sim f(\tilde{\textbf{x}}_i|\textbf{x}_i)$ for $i=1,\ldots,n$.
			\REPEAT
			\STATE {Sample $(\boldsymbol{\beta},\tilde{\boldsymbol{\beta}})\sim f(\boldsymbol{\beta},\tilde{\boldsymbol{\beta}}|\tilde{\textbf{x}}_1,\ldots,\tilde{\textbf{x}}_n,\boldsymbol{\phi},\textbf{D})$;}
			\STATE Sample $\boldsymbol{\phi}\sim f(\boldsymbol{\phi}|\tilde{\textbf{x}}_1,\ldots,\tilde{\textbf{x}}_n,\boldsymbol{\beta},\tilde{\boldsymbol{\beta}},\textbf{D})$;
			\FOR{$i=1,\ldots,n$} \STATE {Sample $\tilde{\textbf{x}}_i\sim f(\tilde{\textbf{x}}_i|\boldsymbol{\beta},\tilde{\boldsymbol{\beta}},\boldsymbol{\phi},\textbf{x}_i,y_i)$.} \ENDFOR
			\UNTIL{convergence}
		\end{algorithmic}
	\end{algorithm}}
	
	\begin{figure}[b]
		\centering
		\includegraphics[width=0.49\linewidth]{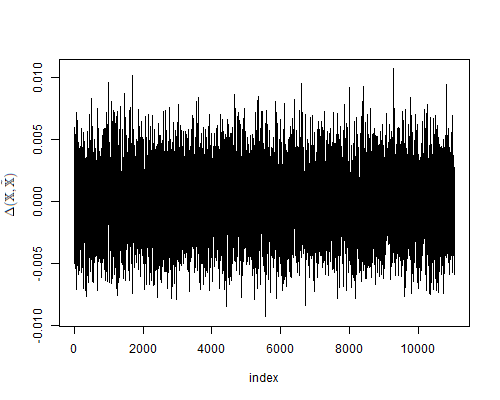}
		\includegraphics[width=0.49\linewidth]{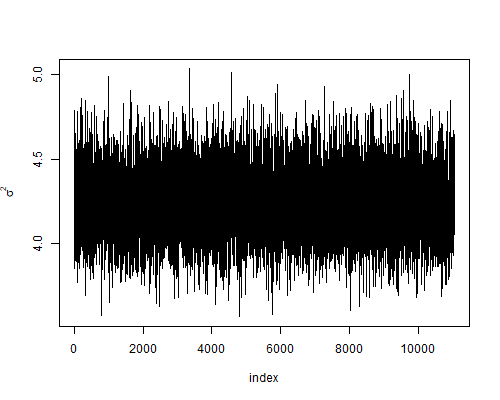}
		\includegraphics[width=0.49\linewidth]{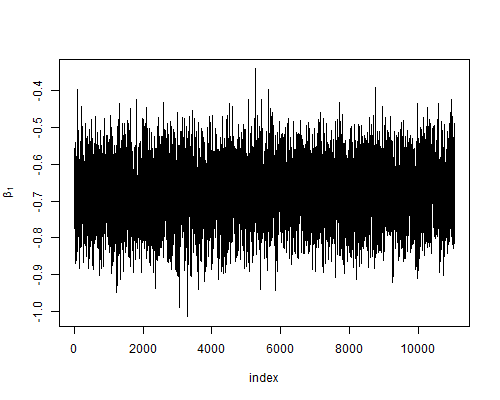}
		\includegraphics[width=0.49\linewidth]{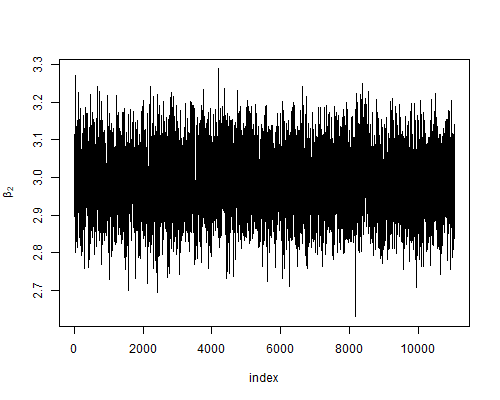}
		\caption{Trace plots of posterior samples of $\Delta(\mathbb{X},\tilde{\mathbb{X}})$, $\sigma^2$, 
$\beta_1$, and $\beta_2$ for the simulated dataset under settings
			in Section \ref{NandA} with $n=1000$ and $a=2$.}
		\label{fig:knockoffdetection}
	\end{figure}
	
In practice, the sampling of $\boldsymbol{\beta}$, $\tilde{\boldsymbol{\beta}}$, $\boldsymbol{\phi}$ and $\tilde{\textbf{x}}_i$
	corresponding to steps 4--7 can be implemented by existing computational
	methods, such as Metropolis--Hasting or rejection sampling algorithms \citep{Sesia2019,Bates2020}. For example, if the joint distribution $f(\textbf{X},\tilde{\textbf{X}})$ is (\ref{Joint}) and the extended conditional distribution {\rm$f(Y|\textbf{X},\tilde{\textbf{X}})$} is $$h(Y|\textbf{X},\tilde{\textbf{X}};\boldsymbol{\beta},\tilde{\boldsymbol{\beta}},\sigma^2)=\frac{1}{\sqrt{2\pi \sigma^2}}\exp\Big\{-\frac{1}{2\sigma^2}(Y-\textbf{X}^{\dT}\boldsymbol{\beta}
	-\tilde{\textbf{X}}^{\dT}\tilde{\boldsymbol{\beta}})^2\Big\},$$ where the nuisance parameter ($\boldsymbol{\phi}$) is the variance parameter $\sigma^2$, the joint posterior distribution (\ref{Jointposterior}) becomes \begin{equation}\label{B1}
		\begin{aligned}
			&f(\tilde{\textbf{x}}_1,\ldots,\tilde{\textbf{x}}_n,
			\boldsymbol{\beta},\tilde{\boldsymbol{\beta}},\sigma^2|\textbf{D})\\
			&\propto f(\boldsymbol{\beta},\tilde{\boldsymbol{\beta}},\sigma^2)
			\frac{1}{\sigma^n}
			\exp\Bigg\{-\frac{1}{2\sigma^2}\sum_{i=1}^{n}(y_i-\textbf{x}_i^{\dT}\boldsymbol{\beta}
			-\tilde{\textbf{x}}_i^{\dT}\tilde{\boldsymbol{\beta}})^2-\frac{1}{2}\sum_{i=1}^{n}(\textbf{x}_i^{\dT},\tilde{\textbf{x}}_i^{\dT})
			\textbf{G}^{-1}\begin{pmatrix}
				\textbf{x}_i\\\tilde{\textbf{x}}_i
			\end{pmatrix}\Bigg\}.
		\end{aligned}
	\end{equation}
where $\textbf{x}_{i}=(x_{i1},\ldots,x_{ip})^\dT$ and $\tilde{\textbf{x}}_{i}=(\tilde{x}_{i1},\ldots,\tilde{x}_{ip})^\dT$
for $i=1,\ldots,n$. Let  $\boldsymbol{\beta}_{-j}=(\beta_{1},\ldots,\beta_{j-1},\beta_{j+1},\ldots,\beta_{p})^\dT$, $\tilde{\boldsymbol{\beta}}_{-j}=(\tilde{\beta}_{1},\ldots,\tilde{\beta}_{j-1},\tilde{\beta}_{j+1},\ldots,\tilde{\beta}_{p})^\dT$
for $j=1,\ldots,p$. When the modified spike-and-slab prior (\ref{MSS}) is used,
the posterior samples of $\boldsymbol{\beta}$, $\tilde{\boldsymbol{\beta}}$, $\sigma^2$ and $\tilde{\textbf{x}}_i$ can be drawn from
their full conditionals detailed in Algorithm \ref{alg2}.
	
	\begin{algorithm}
		\caption{Gibbs sampler under the modified spike-and-slab prior (\ref{MSS}).}\label{alg2}
		\begin{algorithmic}[1]
			\STATE {\bfseries Input:} Observed data $\textbf{D}=\{(\textbf{x}_i,y_i);i=1,\ldots,n\}$.
			\STATE Initialize $\tilde{\textbf{x}}_i\sim\text{MVN}({\textbf{x}}_i-\boldsymbol{\Sigma}^{-1}\text{diag}\{\textbf{s}\}{\textbf{x}}_i,
2\text{diag}\{\textbf{s}\}-\text{diag}\{\textbf{s}\}\boldsymbol{\Sigma}^{-1}\text{diag}\{\textbf{s}\})$ for $i=1,\ldots,n$.
			\REPEAT
			\FOR{$j=1,\ldots,p$}
			\STATE Let $z_{ij}=y_i-\sum_{j'\neq j}x_{ij'}\beta_{j'}-\sum_{j'\neq j}\tilde{x}_{ij'}\tilde{\beta}_{j'}$, \ $i=1,\ldots,n$.
			\STATE Compute 	{\small$$\begin{aligned}
					\tau_j^2&=\Big(1/\tau^2+\sum_{i=1}^nx_{ij}^2/\sigma^2\Big)^{-1},\quad
					\tilde{\tau}_j^2=\Big(1/\tau^2+\sum_{i=1}^n\tilde{x}_{ij}^2/\sigma^2\Big)^{-1},\\
					\mu_j^2&=\tau_j^2\sum_{i=1}^nx_{ij}z_{ij}/\sigma^2,\quad
					\tilde{\mu}_j^2=\tilde{\tau}_j^2\sum_{i=1}^n\tilde{x}_{ij}z_{ij}/\sigma^2,\\
					 \xi_j&=\frac{\xi\sqrt{2\pi\tau_j^2}\exp\big\{\mu_j^2/(2\tau_j^2)\big\}}{2(1-\xi)+\xi\sqrt{2\pi\tau_j^2}\exp\big\{\mu_j^2/(2\tau_j^2)\big\}+\xi\sqrt{2\pi\tilde{\tau}_j^2}\exp\big\{\tilde{\mu}_j^2/(2\tilde{\tau}_j^2)\big\}},\\
					 \tilde{\xi}_j&=\frac{\xi\sqrt{2\pi\tilde{\tau}_j^2}\exp\big\{\tilde{\mu}_j^2/(2\tilde{\tau}_j^2)\big\}}{2(1-\xi)+\xi\sqrt{2\pi\tau_j^2}\exp\big\{\mu_j^2/(2\tau_j^2)\big\}+\xi\sqrt{2\pi\tilde{\tau}_j^2}\exp\big\{\tilde{\mu}_j^2/(2\tilde{\tau}_j^2)\big\}}.\\
				\end{aligned}$$}
			\STATE Generate a random number $r\sim \text{Unif}(0,1)$ and then
			\begin{itemize}
				\item[] sample $\beta_j\sim \text{N}(\mu_j,\tau_j^2)$ and let $\tilde{\beta}_j=0$ if $r\leq \xi_j$;
				\item[] sample $\tilde{\beta}_j\sim \text{N}(\tilde{\mu}_j,\tilde{\tau}_j^2)$ and let ${\beta}_j=0$ if $\xi_j<r\leq \xi_j+\tilde{\xi}_j$;
				\item[] let $\beta_j=\tilde{\beta}_j=0$ otherwise.
			\end{itemize}
			\ENDFOR
			\STATE Sample $\sigma^2\sim \text{IG}\big({n}/{2},\sum_{i=1}^n(y_i-\textbf{x}_i^{\dT}\boldsymbol{\beta}-\tilde{\textbf{x}}_i^{\dT}\tilde{\boldsymbol{\beta}})^2/2\big)$;
			\FOR{$i=1,\ldots,n$} \STATE {Sample $\tilde{\textbf{x}}_i\sim \text{MVN}(\tilde{\boldsymbol{\mu}}_i,\tilde{\boldsymbol{\Sigma}})$ where $$
				\begin{aligned}
					\tilde{\boldsymbol{\Sigma}}&=\Big(
					\textbf{A}+\frac{1}{\sigma^2}\tilde{\boldsymbol{\beta}}\tilde{\boldsymbol{\beta}}^{\dT}\Big)^{-1},\\
					 \tilde{\boldsymbol{\mu}}_i&=\tilde{\boldsymbol{\Sigma}}\Big[\big(\text{diag}\{\textbf{s}\}^{-1}-\textbf{A}-\frac{1}{\sigma^2}\tilde{\boldsymbol{\beta}}\boldsymbol{\beta}^\dT\big)\textbf{x}_i+\frac{1}{\sigma^2}\tilde{\boldsymbol{\beta}}y_i\Big],\\
					\textbf{A}&=\big(2\text{diag}\{\textbf{s}\}-\text{diag}\{\textbf{s}\}\boldsymbol{\Sigma}^{-1}\text{diag}\{\textbf{s}\}\big)^{-1}.
				\end{aligned}
				$$} \ENDFOR
			\UNTIL{convergence}
		\end{algorithmic}
	\end{algorithm}
	
	To investigate whether the posterior samples
	$\tilde{\mathbb{X}}^{(1)},\ldots,\tilde{\mathbb{X}}^{(T)}$ generated by {Algorithm \ref{alg1}}
	satisfy {Definition \ref{DF:Knockoff}}, we apply {Algorithm \ref{alg1}} to the data generated under settings
	in Section \ref{NandA} (with sample size $n=1000$ and signal strength $a=2$) and compute the statistic
	\begin{equation}\label{Quality}
		\begin{aligned}
			\Delta(\mathbb{X},\tilde{\mathbb{X}})=&\frac{1}{n}\sum_{i=1}^{n}\sum_{1\leq j<k\leq p}(\tilde{x}_{ij}\tilde{x}_{ik}
			+2{x}_{ij}\tilde{x}_{ik}-3{x}_{ij}{x}_{ik}).
		\end{aligned}
	\end{equation}
Because it is required $E[\Delta(\tilde{\mathbb{X}})]=0$ for valid knockoffs,
	$\Delta(\mathbb{X},\tilde{\mathbb{X}})$ would
	fluctuate around $0$ if knockoff variables $\tilde{\mathbb{X}}$ generated by Algorithm \ref{alg1} are valid.
	In our experiments,
	we keep $T=10,000$ posterior samples after 1,000 burn-in iterations.
	As shown in Figure \ref{fig:knockoffdetection}, the Markov chains
	are stable, well mixed and $\Delta(\mathbb{X},\tilde{\mathbb{X}})$ fluctuates
	around $0$, indicating that $\tilde{\mathbb{X}}^{(1)},\ldots,\tilde{\mathbb{X}}^{(T)}$
	are valid knockoff variables.

	\subsection{Relationships with Existing Methods}
	
	Although the BKF is developed along the inspiring idea of knockoffs \citep{Barber2015},
	we elaborate on its differences from existing methods as follows.
	
	\begin{itemize}
\item[-] \textbf{Model-$\textbf{X}$ knockoff \citep{Candes2018}}:
In contrast to the fixed-\textbf{X} knockoff \citep{Barber2015} which assumes that $f(Y|\textbf{X})$ is fully known as a linear model while the distribution of covariates $f(\textbf{X})$ is unknown,
our method follows a similar setup of the model-\textbf{X} knockoff with the assumption that
$f(\textbf{X})$ is known and a GLM is suitable for the conditional distribution $f(Y|\textbf{X})$.
However, there are fundamental differences between the model-$\textbf{X}$ knockoff and our BKF.
The model-\textbf{X} knockoff is a frequentist method, which relies upon only one set of knockoff variables.
		It uses the Bayesian variable selection (BVS) method as one of possible ways to compute
		feature statistics and still controls the frequentist FDR in feature selection.
Due to the usage of BVS feature statistics, it can leverage prior knowledge via the probability point mass
		at $\beta_j=0$ in the prior specification. In contrast, our BKF is a fully Bayesian data augmentation approach where
		the observed data $\textbf{D}$ are treated as fixed while the set of non-null features $\mathcal{H}_1$,
		knockoffs $\tilde{\textbf{x}}_1,\ldots,\tilde{\textbf{x}}_n$ and parameters
		$\boldsymbol{\beta},\tilde{\boldsymbol{\beta}},\boldsymbol{\phi}$ are all treated as random.
		Incorporating the knockoffs as missing data and FDR into a Bayesian framework,
		our BKF repeatedly samples knockoff variables in MCMC for more stable inference and
		conducts feature selection based on the Bayesian FDR control, no matter whether
the priors have a point mass at $\beta_j=0$.
		
		\item[-] \textbf{Metropolized knockoff sampling \citep{Bates2020}}:
Although \cite{Bates2020} use MCMC as a mechanism to generate knockoff variables,
		the Metropolized knockoff sampling is still a frequentist method, relying upon only one set of knockoff variables
		for inference. In contrast, our BKF treats knockoff variables as missing data and continuously samples
		them via data augmentation in MCMC for more stable Bayesian inference.
		
		\item[-] \textbf{Multiple knockoffs \citep{Gimenez2019b}}:
The multiple knockoffs generate more than one set of knockoff variables.
Taking the 2 multi-knockoffs under a Gaussian case as an example, the joint model of original features $\textbf{X}$ and knockoff features $(\tilde{\textbf{X}}_1,\tilde{\textbf{X}}_2)$ in \citet{Gimenez2019b} is exchangeable for any generalized swaps among
the three sets of features. Because exchangeability is assumed simultaneously
across $\textbf{X}$ and multiple knockoffs, this would make $\textbf{X}$ and all knockoffs
similar and thus lower the power. In contrast, we only define the joint model of
the original features $\textbf{X}$ and one set of
posterior sampled knockoff features at a time for each posterior iteration in the MCMC.
It requires exchangeability for any swaps between the original features and that set of
knockoff features. As a result, the difference between original features and knockoff features would be greater in BKF,
which leads to higher power, especially when strong correlations exist among original features.
	\end{itemize}
	
	\section{Numerical Experiments}\label{Experiments}
	
	We conduct numerical experiments on synthetic data: (i) to evaluate the performance of BKF in controlling FDR and detecting
	true discoveries (power) under various
	data generation settings;
	(ii) to compare BKF with existing frequentist knockoff methods on power and FDR; and (iii) to illustrate the advantage of BKF in controlling false discoveries over existing Bayesian feature selection approaches.
	We also apply our method to real data to demonstrate the practical performance of BKF.
	
	\subsection{Comparisons with Existing Knockoffs}\label{Freq}
	
	We first compare the performance of BKF with three existing knockoff methods, including the fixed-$\textbf{X}$ knockoff
	\citep{Barber2015}, the model-$\textbf{X}$ knockoff \citep{Candes2018} and the multi-knockoffs \citep{Gimenez2019b}.
	For multi-knockoffs, we use the 2 multi-knockoffs in the experiments. For all the existing methods, we use their
	SDP (semidefinite program)
	constructions of knockoff and the Lasso coefficient difference as the importance statistic for inference.
	To evaluate the performances of different methods, two criteria are considered: (a) the overall FDR defined by (\ref{df:FDR});
	and (b) statistical power $|\hat{\mathcal{S}}\cap\mathcal{H}_1|\big/|\mathcal{H}_1|$,
	where $\hat{\mathcal{S}}$ corresponds to the estimator obtained by each of the
	knockoff methods in each replication. We consider different settings to examine the effects of
	sample size, signal strength, variance structures among features, dimensionality and the size of $\mathcal{H}_1$. In addition, we also evaluate the robustness of BKF and existing knockoff methods to misspecification of $f(\textbf{X})$.
	
	For fair comparison between our BKF with existing frequentist methods,
	we use the flat prior in the implementation of BKF. We
	discard the first 500 iterations as burn-ins and keep $T=2,000$ posterior samples for inference.
	
	\subsubsection{Sample Size and Signal Strength}\label{NandA}
	
	\begin{figure}[b]
		\centering
		\includegraphics[width=0.4\linewidth]{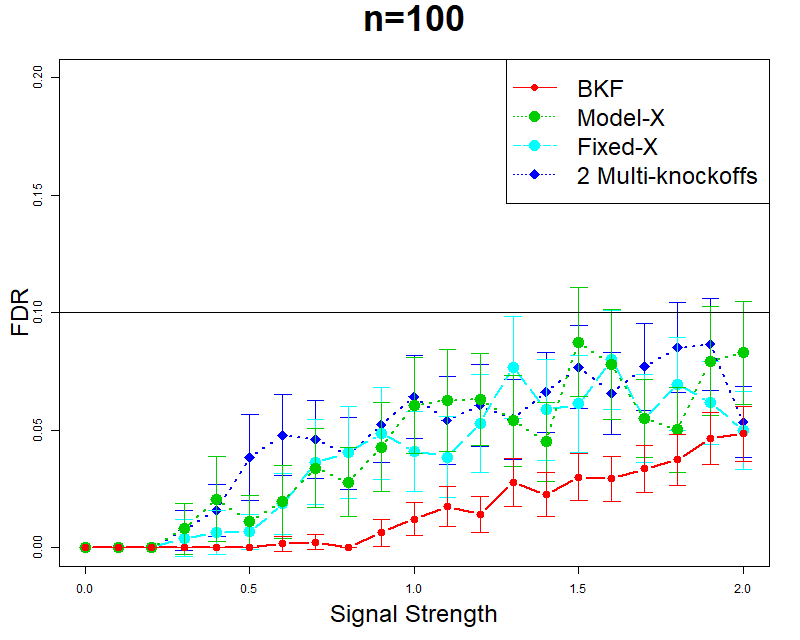}
		\includegraphics[width=0.4\linewidth]{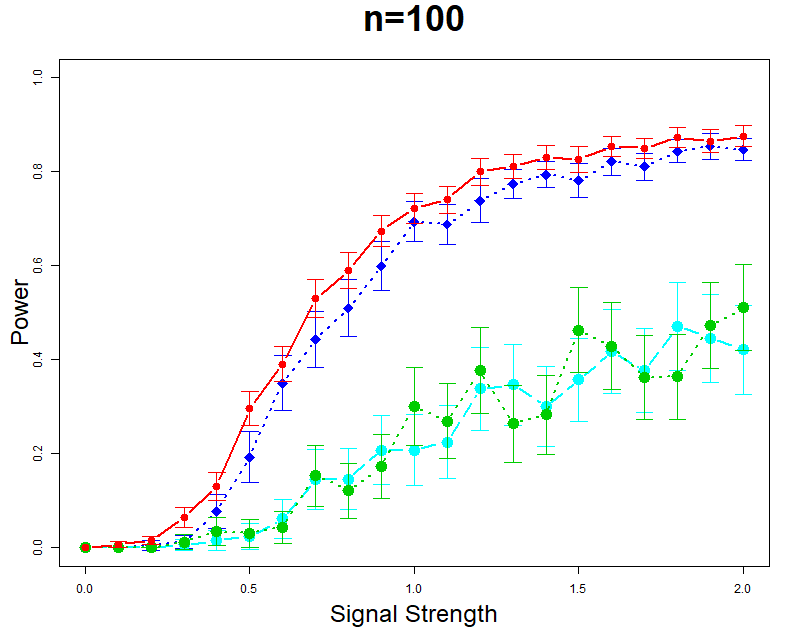}
		\includegraphics[width=0.4\linewidth]{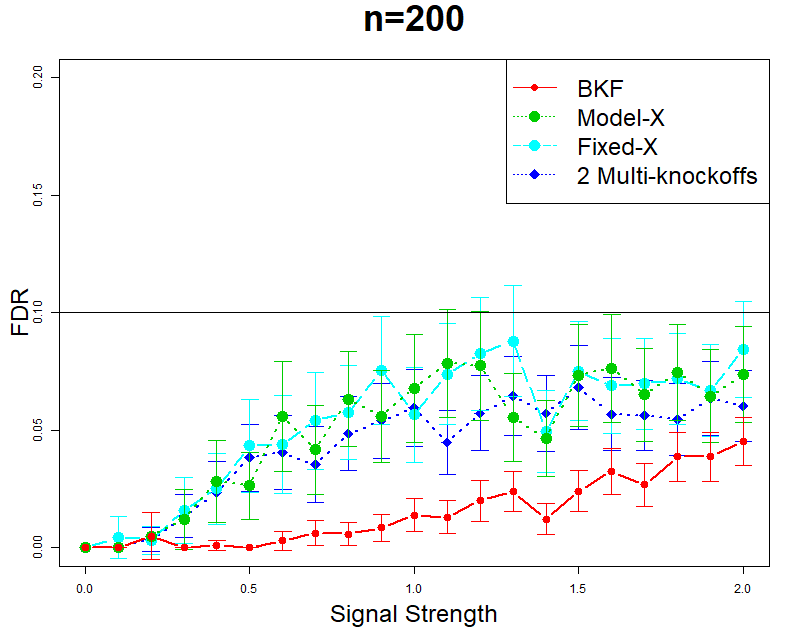}
		\includegraphics[width=0.4\linewidth]{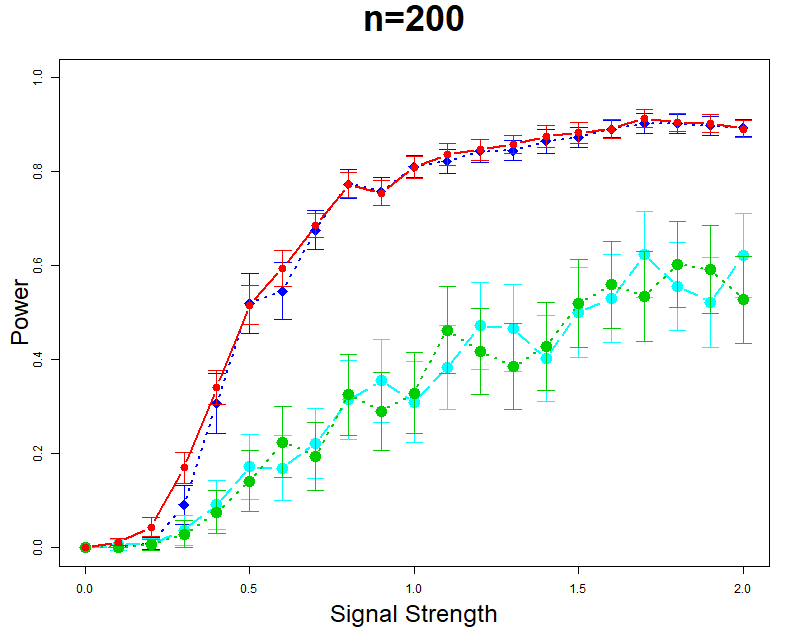}
		\includegraphics[width=0.4\linewidth]{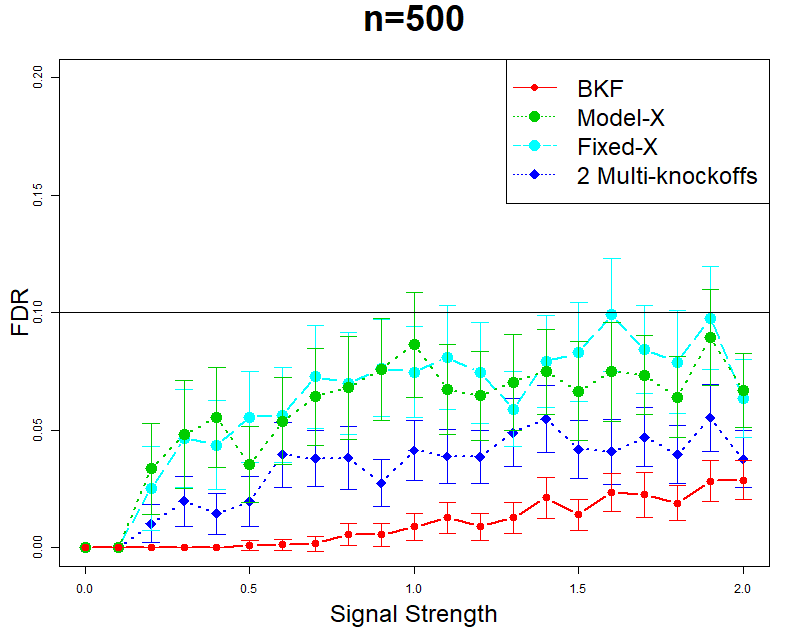}
		\includegraphics[width=0.4\linewidth]{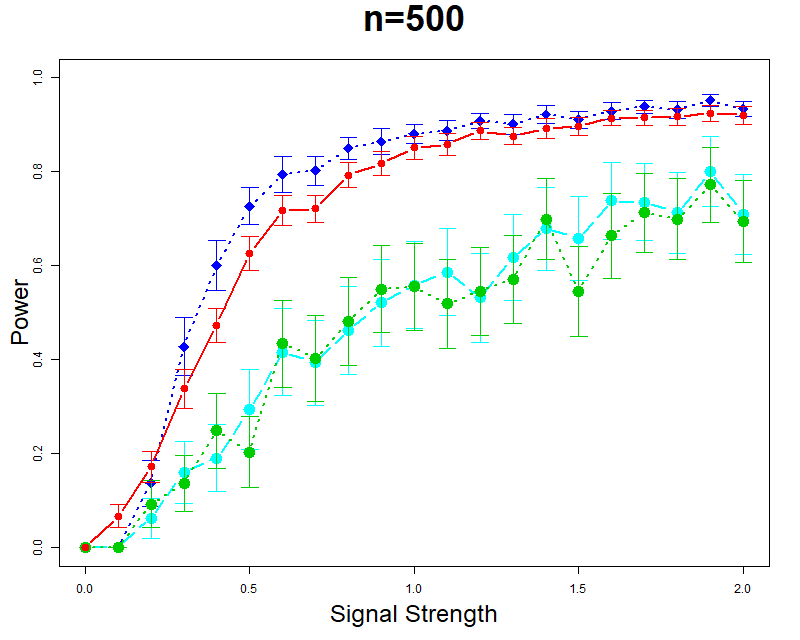}
		\includegraphics[width=0.4\linewidth]{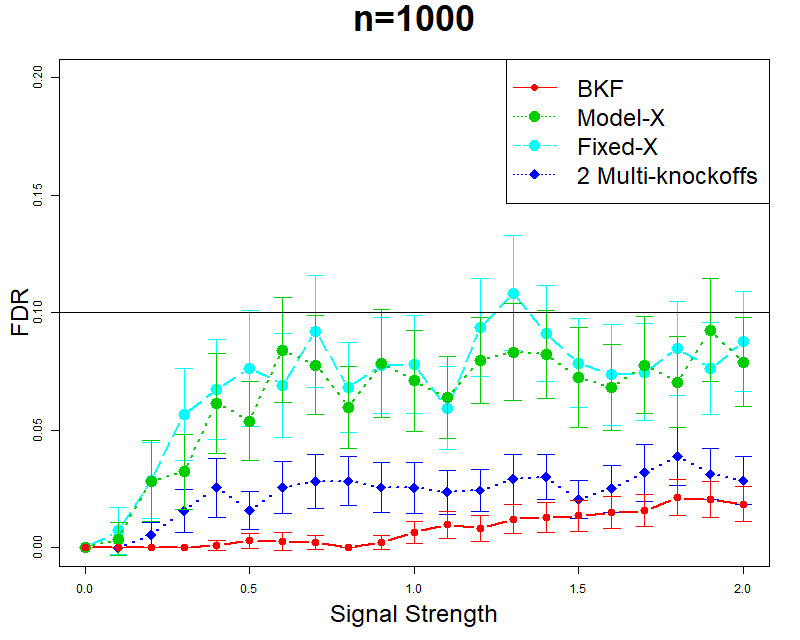}
		\includegraphics[width=0.4\linewidth]{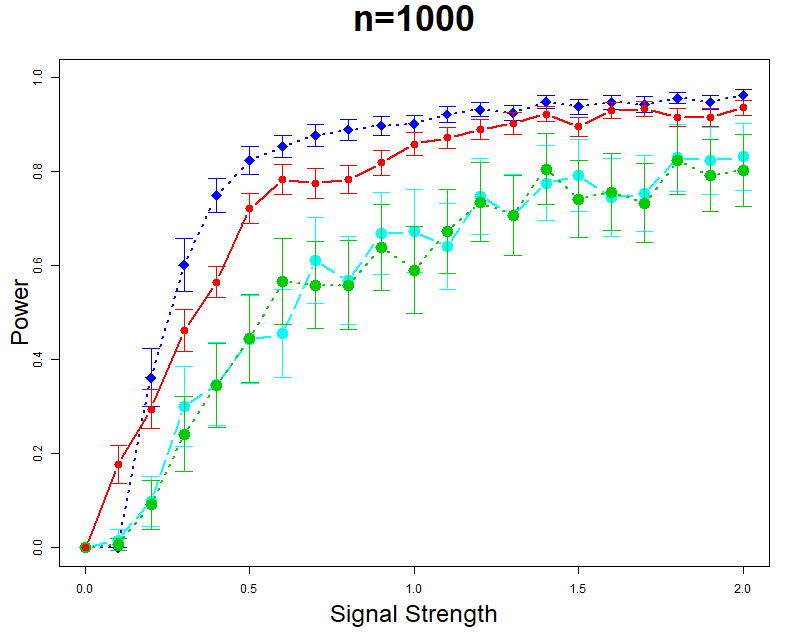}
		\caption{Comparisons of the FDR and power of the BKF and other knockoff methods under different
			sample sizes $n$ and signal strengths $a$. Each point is averaged over 100 replications.}
		\label{fig:NandA}
	\end{figure}
	
	To examine how BKF performs under different sample sizes and signal strengths,
	we simulate 100 datasets $\textbf{D}=\{(\textbf{x}_i,y_i): i=1,\ldots,n\}$ with
	a fixed number of features $p=30$ and random $\mathcal{H}_{1}$ whose size is fixed as $|\mathcal{H}_{1}|=10$. For each dataset, $\mathcal{H}_{1}$ is randomly drawn from all $\binom{p}{10}$ possible subsets containing $10$ elements of $\{1,\ldots,p\}$, features $\textbf{x}_1,\ldots,\textbf{x}_n$ are generated as i.i.d. samples from $\text{MVN}(\mathbf{0},\textbf{I})$ and responses  $y_1,\ldots,y_n$ are
	generated from a Gaussian linear model (\ref{OLM}) with coefficients $\beta_1,\ldots,\beta_p$.
	The true non-null features in subset $\mathcal{H}_1$ are randomly chosen where
$$
\mbox{if} \ j\in \mathcal{H}_1, \ \beta_j=\begin{cases}
		a,&\text{w.p. }1/2,\\
		-a,&\text{w.p. }1/2,
	\end{cases}
$$
otherwise $\beta_j=0$. By varying the sample size $n\in\{100,200,500,1000\}$ and
	signal strength $a\in\{0.1,0.2,\ldots,2\}$ with a fixed noise level $\sigma^2=4$, we
	can evaluate the effects of sample size and signal strength on the performances of BKF and
	other existing methods. In our experiments, we use model (\ref{Joint}) to generate knockoffs and the desired level of FDR is $\alpha=0.1$.
	
	Figure \ref{fig:NandA} shows that similar to other knockoff methods,
	the power of BKF grows as the signal strength $a$ or sample size $n$ increases.
	At the same time, the overall FDR of BKF is controlled under the desired level,
	indicating BKF is valid for multiple testing on hypotheses (\ref{hypothese}).
	Compared with existing knockoff methods, the overall FDR of BKF is significantly lower,
	implying that our method is more reliable in avoiding discovery
	of false signals. Similar to the 2 multi-knockoffs, the capability of BKF to
identify the true $H_{1j}$ grows much faster and becomes more stable as
	the signal strength or sample size increases than the
	single-knockoff methods (i.e., fixed-$\textbf{X}$ knockoff and
	model-$\textbf{X}$ knockoff). For example, when $a=1$, the power of BKF
	reaches $0.8$ with sample size $n=500$, while the single-knockoff methods
	can only identify about $60\%$ of
	the true signals even when the sample size is doubled. Compared with the 2 multi-knockoffs,
our BKF performs better when sample size is small ($n=100$) and only suffers slight power loss when sample size is large ($n=1000$).
In the aspect of robustness in power, BKF is comparable to the 2 multi-knockoffs
	and better than the other two methods. As a summary,
	BKF performs well and stable at detecting important signals yet with smaller chance of making false discoveries.
		
	\subsubsection{Variance Structure}\label{Var}
	
	\begin{figure}[b]
		\centering
		\includegraphics[width=0.4\linewidth]{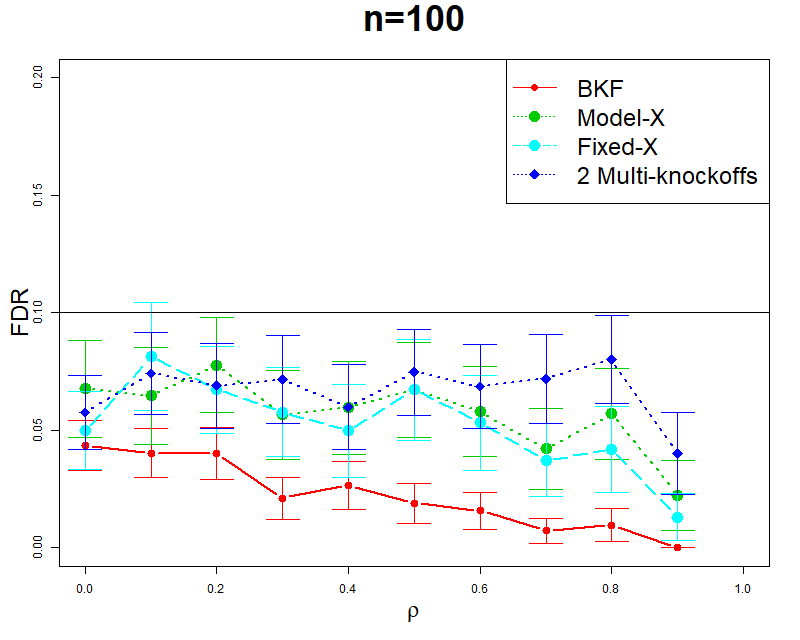}
		\includegraphics[width=0.4\linewidth]{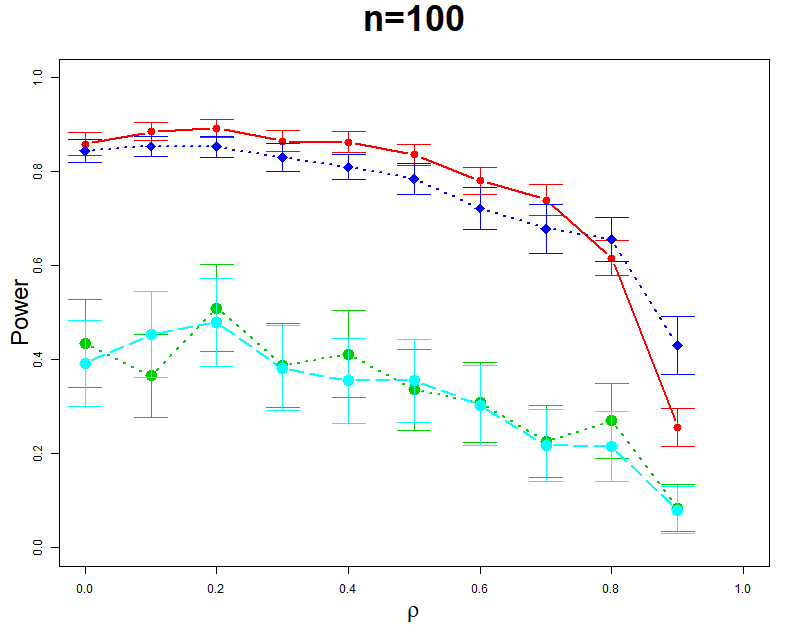}
		\includegraphics[width=0.4\linewidth]{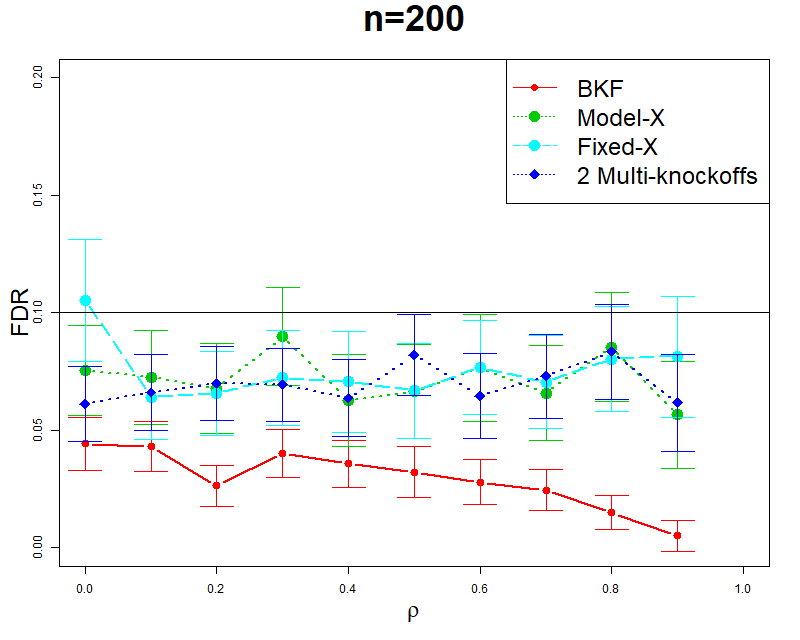}
		\includegraphics[width=0.4\linewidth]{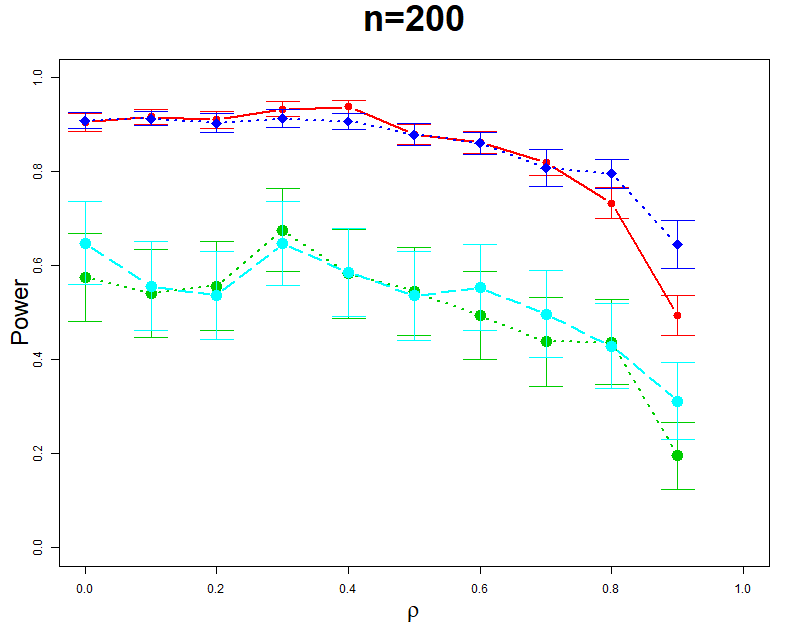}
		\includegraphics[width=0.4\linewidth]{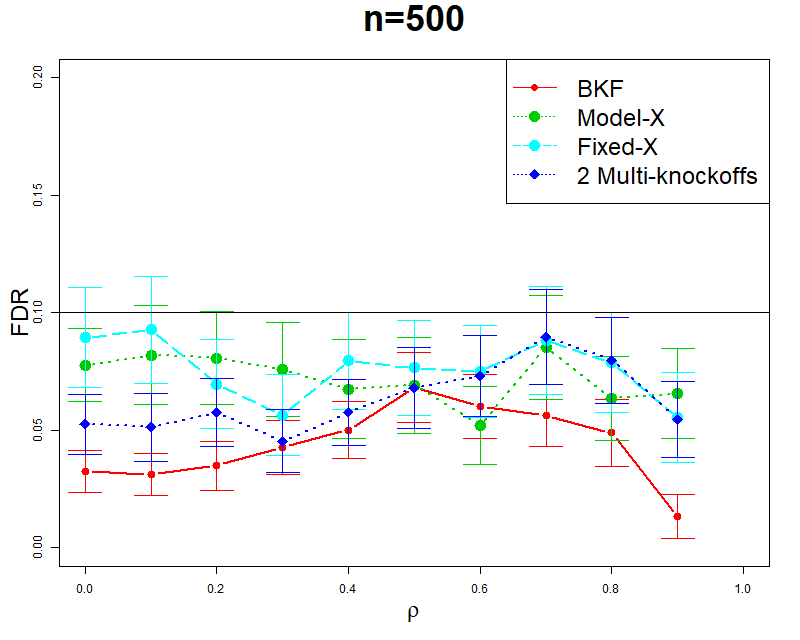}
		\includegraphics[width=0.4\linewidth]{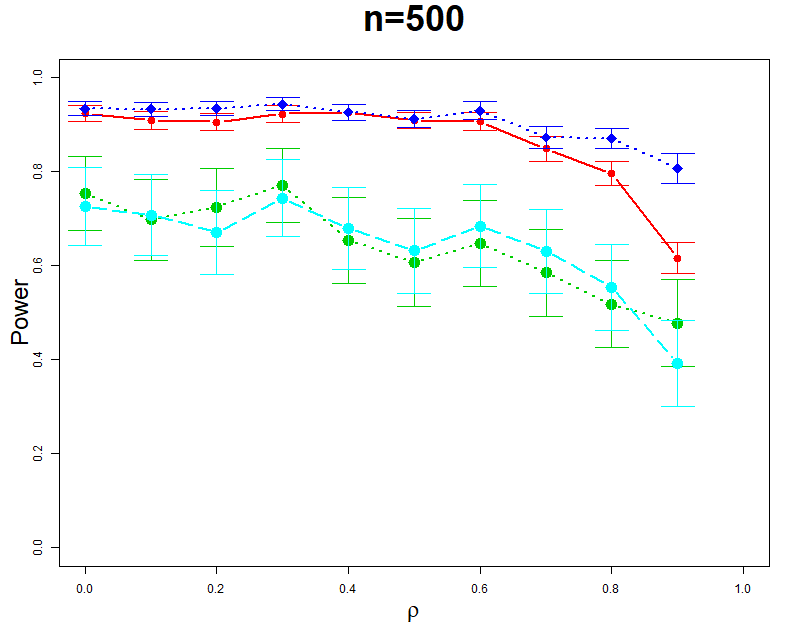}
		\includegraphics[width=0.4\linewidth]{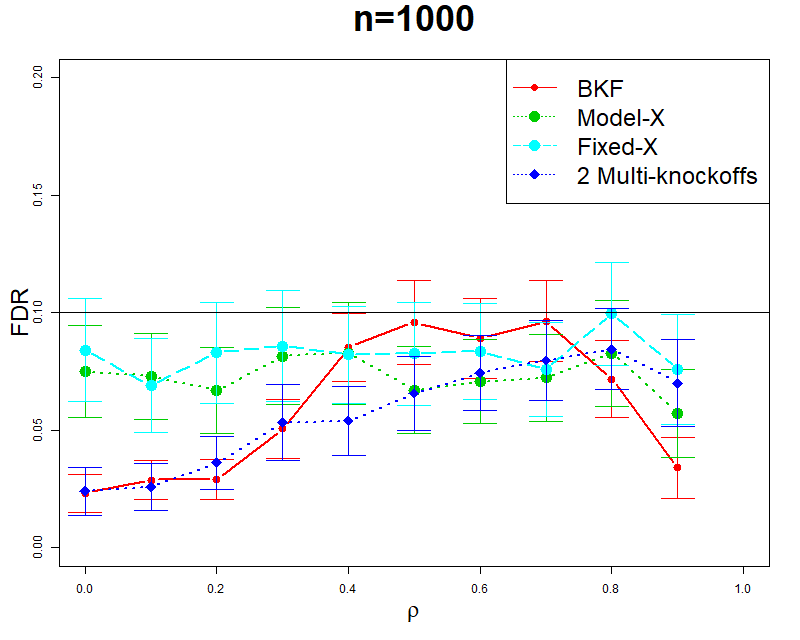}
		\includegraphics[width=0.4\linewidth]{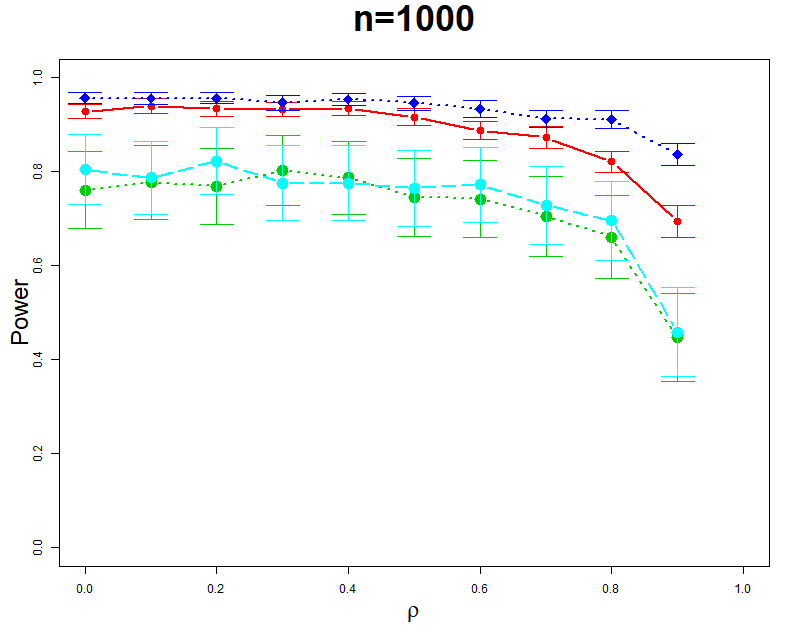}
		\caption{Comparisons of the FDR and power of the BKF and other knockoff methods
under different sample sizes $n$ and correlations $\rho$ in Case (i). Each point is averaged over 100 replications.}
		\label{fig:casei}
	\end{figure}
	
	\begin{figure}[b]
		\centering
		\includegraphics[width=0.4\linewidth]{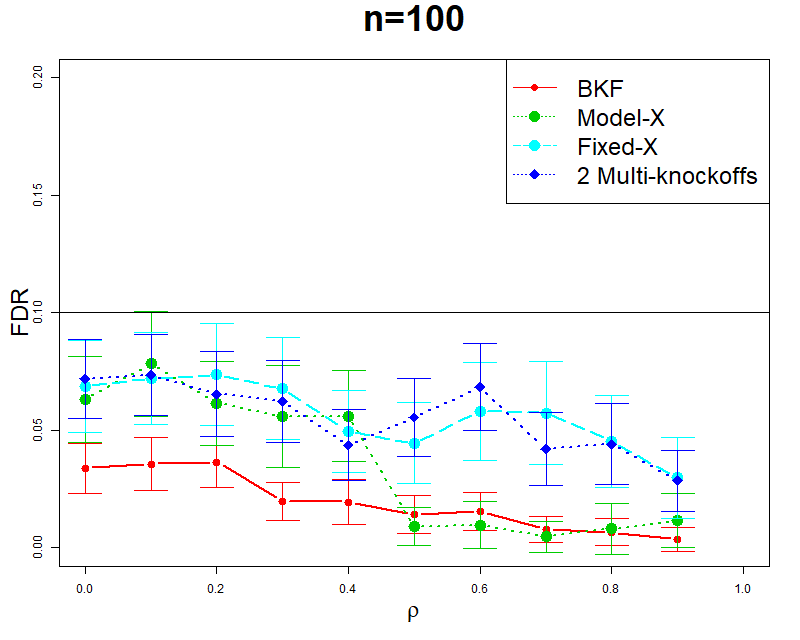}
		\includegraphics[width=0.4\linewidth]{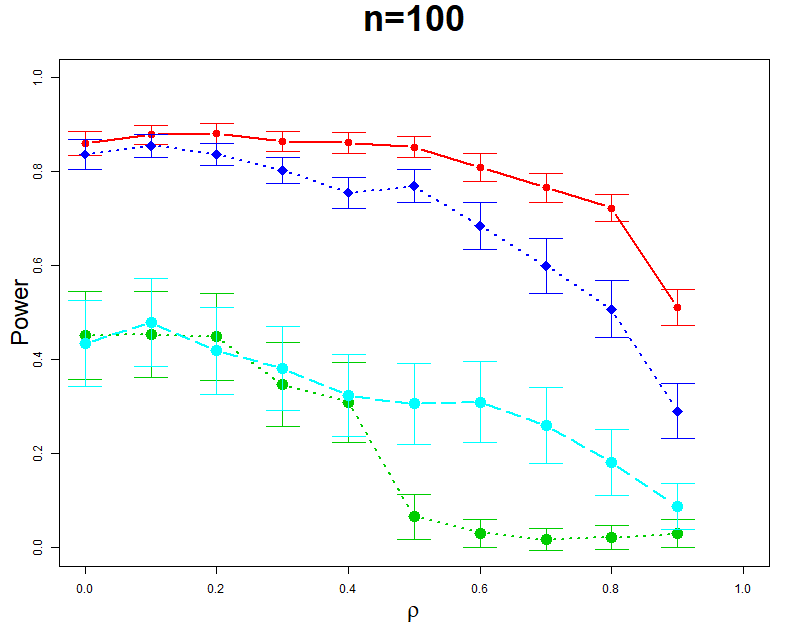}
		\includegraphics[width=0.4\linewidth]{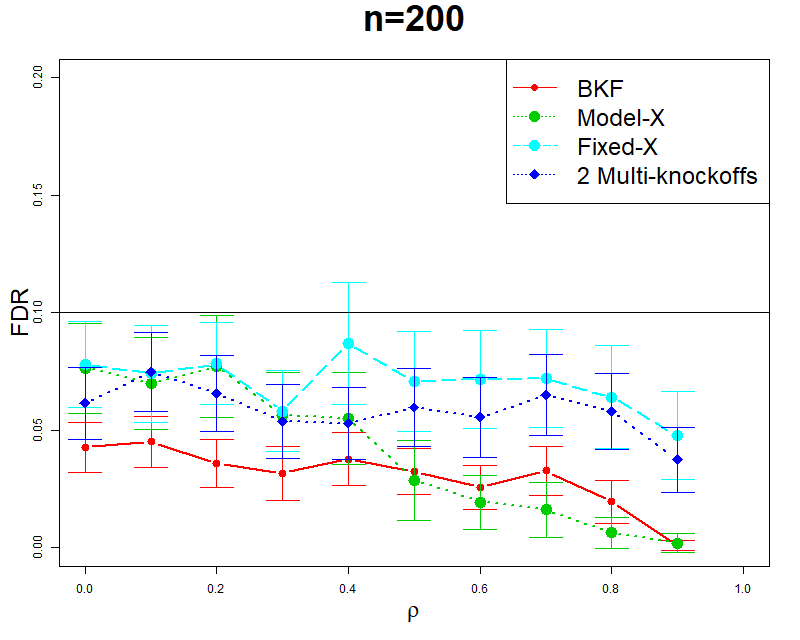}
		\includegraphics[width=0.4\linewidth]{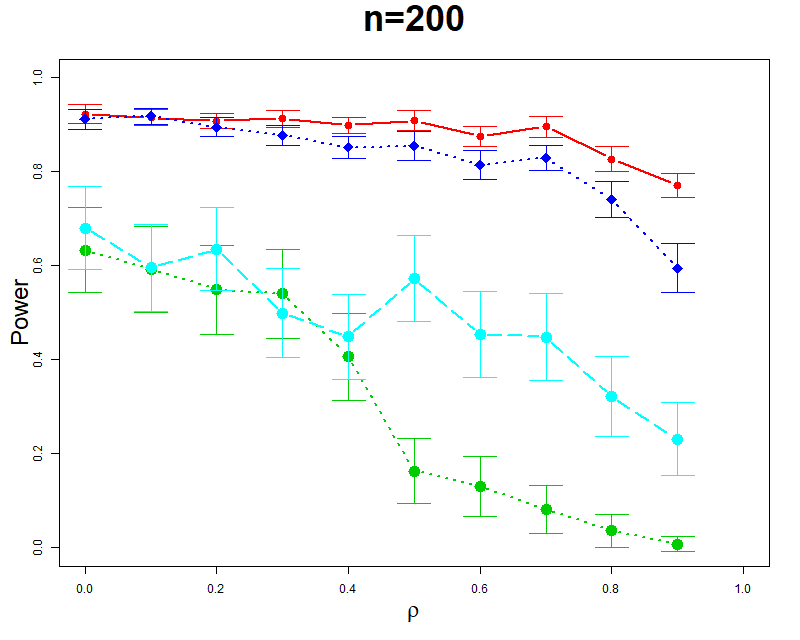}
		\includegraphics[width=0.4\linewidth]{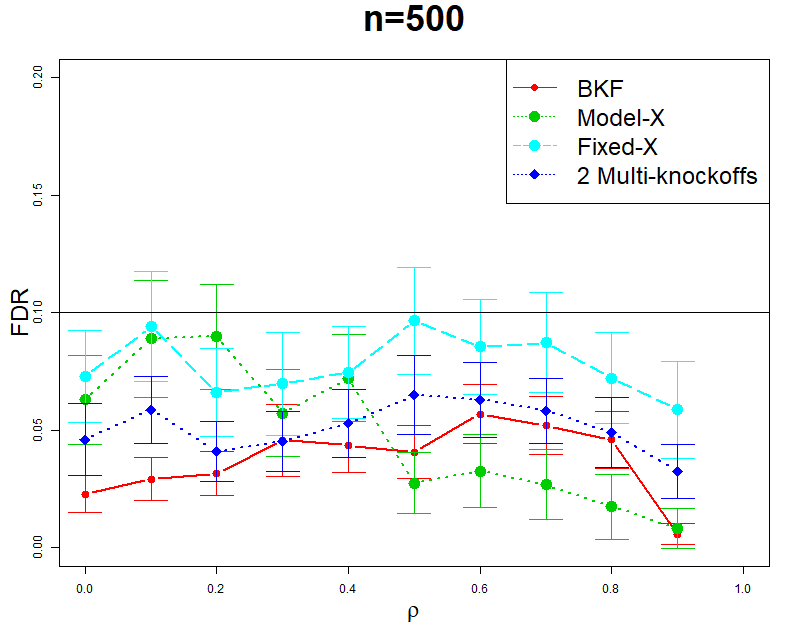}
		\includegraphics[width=0.4\linewidth]{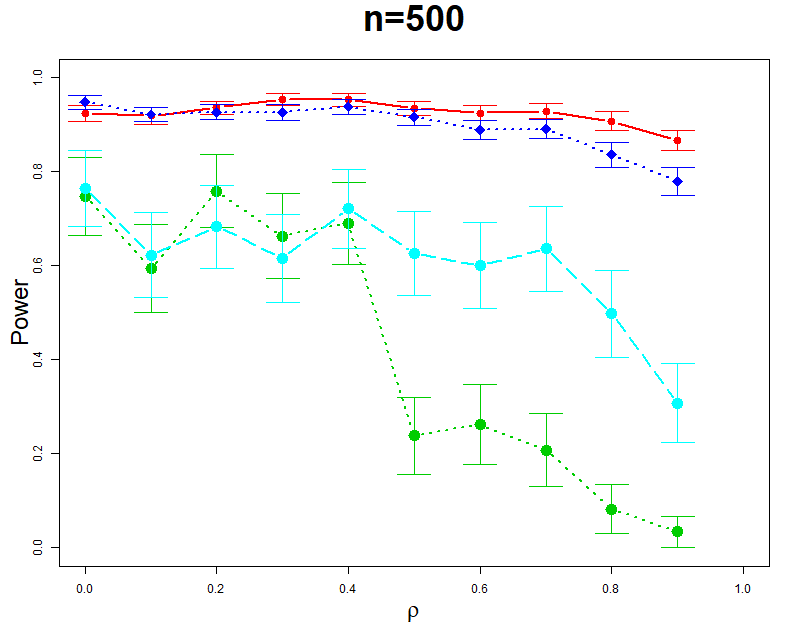}
		\includegraphics[width=0.4\linewidth]{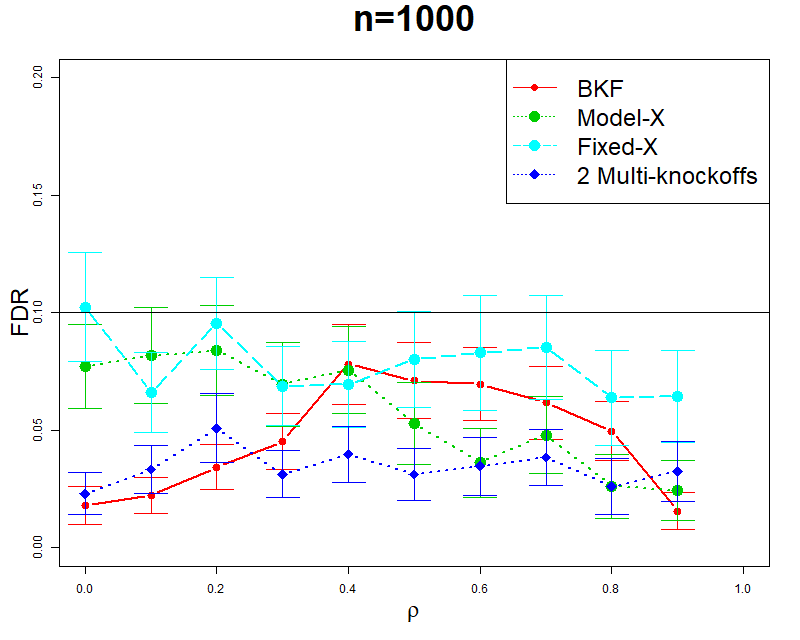}
		\includegraphics[width=0.4\linewidth]{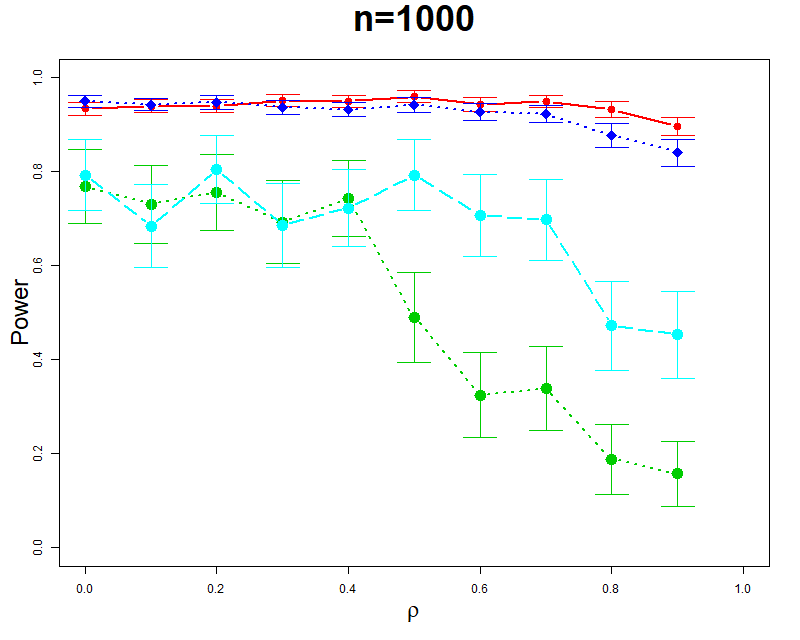}
		\caption{
Comparisons of the FDR and power of the BKF and other knockoff methods under
different sample sizes $n$ and correlations $\rho$ in Case (ii). Each point is averaged over 100 replications.}
		\label{fig:caseii}
	\end{figure}
	
	We also investigate the impact of the variance structure among features $X_1,\ldots,X_p$
	on the performance of BKF. Toward this goal, we apply BKF to obtain $\hat{\mathcal{S}}$
	for datasets generated from model (\ref{OLM}) with dimension $p=30$ and random $\mathcal{H}_{1}$ whose size is fixed as $|\mathcal{H}_{1}|=10$.
	We consider two cases for the variance-covariance matrix.
	Features $\textbf{x}_1,\ldots,\textbf{x}_n$ are generated from $\text{MVN}(\mathbf{0},\boldsymbol{\Sigma}_\rho)$ where
	$$\boldsymbol{\Sigma}_\rho=[\sigma_{\rho,ij}]_{p\times p}\quad\text{and}\quad\sigma_{\rho,ij}=\begin{cases}
		\rho^{|i-j|},&\text{Case (i)},\\
		\rho^{I[i\neq j]},&\text{Case (ii)}.
	\end{cases}$$
	Case (i) corresponds to the situation where features are sampled from an auto-correlated
	time series and Case (ii) indicates an equal correlation among all features. For each dataset, $\mathcal{H}_{1}$ is randomly drawn from all $\binom{p}{10}$ possible subsets containing $10$ elements of $\{1,\ldots,p\}$ and the true non-null features in subset $\mathcal{H}_1$ of size $10$ are randomly chosen
	where
$$\mbox{if} \ j\in \mathcal{H}_1, \ \beta_j=\begin{cases}
		2,&\text{w.p. }1/2,\\
		-2,&\text{w.p. }1/2,\\
	\end{cases}$$
otherwise $\beta_j=0$. Given features $\textbf{x}_1,\ldots,\textbf{x}_n$, responses $y_1,\ldots,y_n$ are generated from model (\ref{OLM}).
	For each combination of sample size $n\in\{100,200,500,1000\}$ and Case (i) and (ii) covariance structures,
	100 datasets are simulated for each value of the correlation coefficient $\rho\in\{0,\ldots,0.9\}$.
	
	The power and overall FDR of our BKF as well as
	those of existing methods are presented in Figures \ref{fig:casei}  and  \ref{fig:caseii}.
	It can be observed that correlations among features would lower
	the power of all methods. With the overall FDR under control, the power of
	all methods decreases as the correlation $\rho$ increases for
	both covariance structures. In all scenarios, both BKF and the 2 multi-knockoffs possess
	higher power than the fixed-$\textbf{X}$ knockoff and the model-$\textbf{X}$ knockoff with the overall FDR controlled under $\alpha=0.1$. However, there are still differences between BKF and the 2 multi-knockoffs. When features are generated under Case (i), the power of BKF is between the 2 multi-knockoffs and other single-knockoff methods. As $\rho$ increases, the power of BKF is closer to
that of the single-knockoff methods. However, when correlations among features are the same, i.e., Case (ii), BKF is more powerful than
	the 2 multi-knockoffs, especially when $\rho$ is large. Therefore, BKF is more powerful to
	identify false $H_{0j}$ in the equal correlation Case (ii) while
	the 2 multi-knockoffs method has higher power in the auto-correlated Case (i). The reason is that in the case of equal correlations, the difference between the 2 multi-knockoffs and original variables diminishes much faster than
that in BKF as $\rho$ increases, leading to higher power of BKF for large $\rho$. In contrast, the quality of the 2 multi-knockoffs does not decrease vastly in auto-correlated cases and thus the 2 multi-knockoffs method performs slightly better.
	
	\subsubsection{Dimensionality and Size of $\mathcal{H}_1$}
	
Under single-knockoff methods, including the fixed-$\textbf{X}$ knockoff
	and the model-$\textbf{X}$ knockoff, their inferences rely upon only one knockoff sample $\tilde{\mathbb{X}}$.
	As a result, these methods may incur power loss and instability when the size of $\mathcal{H}_1$ is small.
	The multi-knockoffs method \citep{Gimenez2019b}, on the other hand, makes inference based on
	more than one knockoff samples. However, as the number of knockoff samples grows,
	the difference between the original features and the knockoffs would diminish, leading to
	power loss when the size of $\mathcal{H}_1$ is large.
	To demonstrate the robustness of BKF with respect to both the size of $\mathcal{H}_1$
	(denoted as $|\mathcal{H}_1|=v$) and the number of features ($p$), we compare the performances of BKF
	with the model-$\textbf{X}$ knockoff and 2 multi-knockoffs on datasets with different numbers
	of features and different sizes of $\mathcal{H}_1$. In all datasets, the sample size is fixed as
	$n=200$ and vectors $\textbf{x}_1,\ldots,\textbf{x}_n$ are generated under the setup of Section
	\ref{NandA} and Case (i) of Section \ref{Var} with $\rho=0.6$. With $p\in\{100,200,500,1000\}$
	and $v\in\{1,\ldots,30\}$, $\mathcal{H}_{1}$ of each dataset is randomly drawn from all $\binom{p}{v}$ 
possible subsets containing $v$ elements of $\{1,\ldots,p\}$. 
Responses $y_1,\ldots,y_n$ of each dataset are generated from model
	(\ref{OLM}) with randomly chosen true non-null features in subset $\mathcal{H}_1$ where
$$\mbox{if} \ j\in \mathcal{H}_1, \ \beta_j=\begin{cases}
		2,&\text{w.p. }1/2,\\
		-2,&\text{w.p. }1/2,\\
\end{cases}$$
otherwise $\beta_j=0$.
	Since the flat prior would make the joint posterior density (\ref{Jointposterior}) degenerated, 
the modified spike-and-slab prior with $\xi=0.1$
and $\tau=1$ \citep{Candes2018} is used for the BKF. 
	
	\begin{figure}[b]
		\centering
		\includegraphics[width=0.4\linewidth]{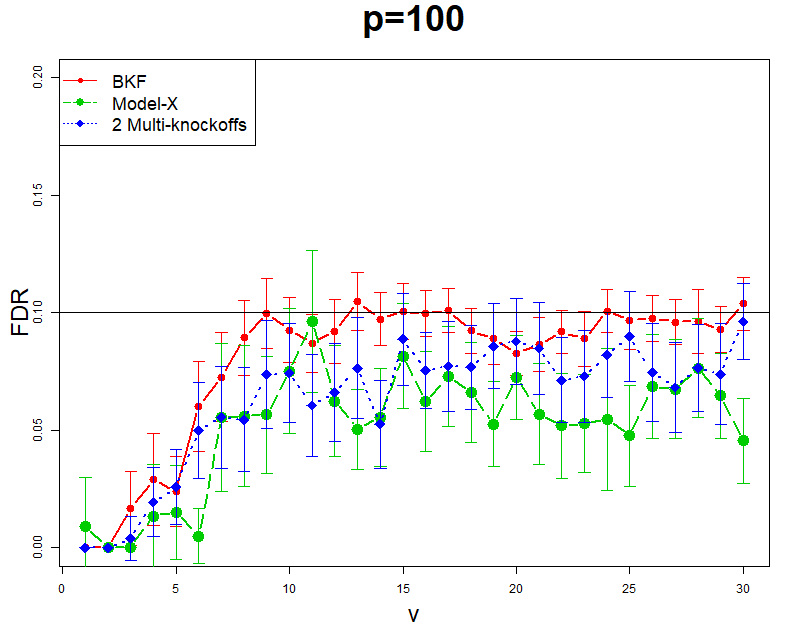}
		\includegraphics[width=0.4\linewidth]{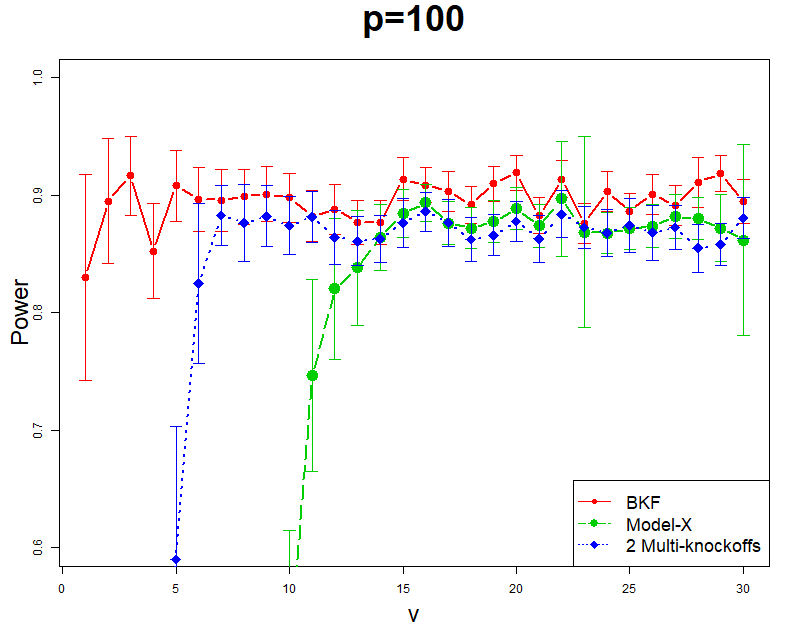}
		\includegraphics[width=0.4\linewidth]{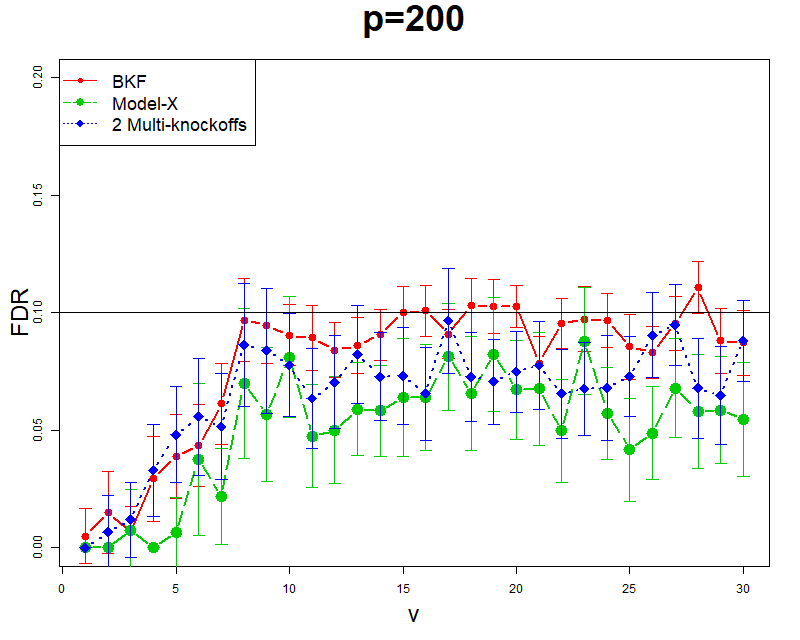}
		\includegraphics[width=0.4\linewidth]{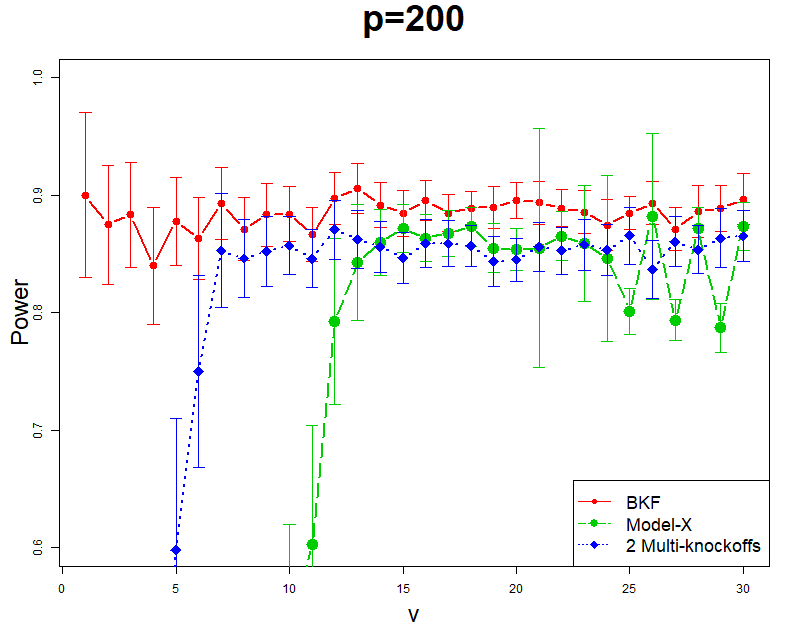}
		\includegraphics[width=0.4\linewidth]{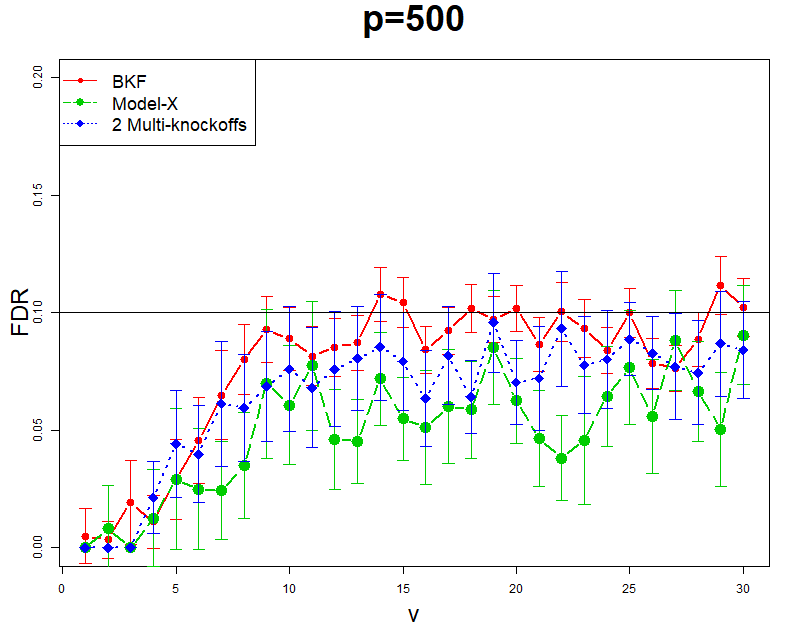}
		\includegraphics[width=0.4\linewidth]{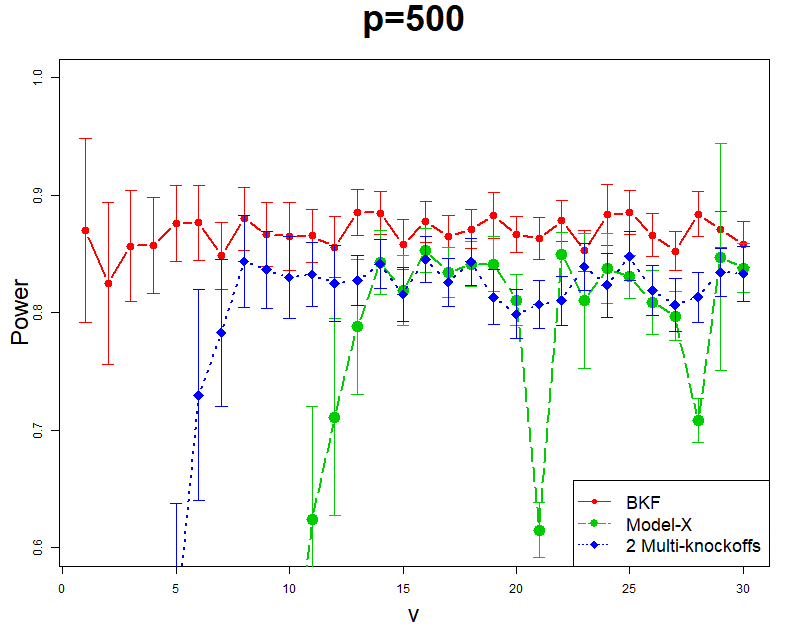}
		\includegraphics[width=0.4\linewidth]{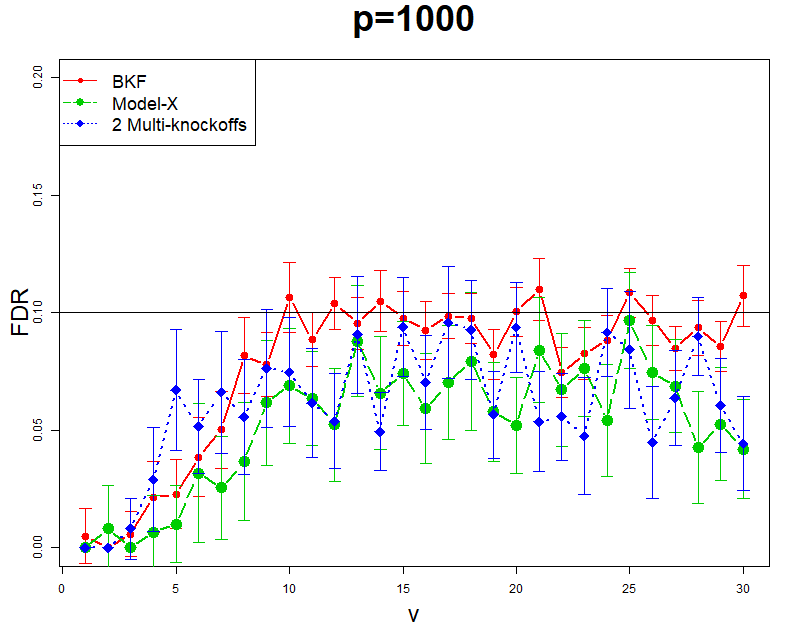}
		\includegraphics[width=0.4\linewidth]{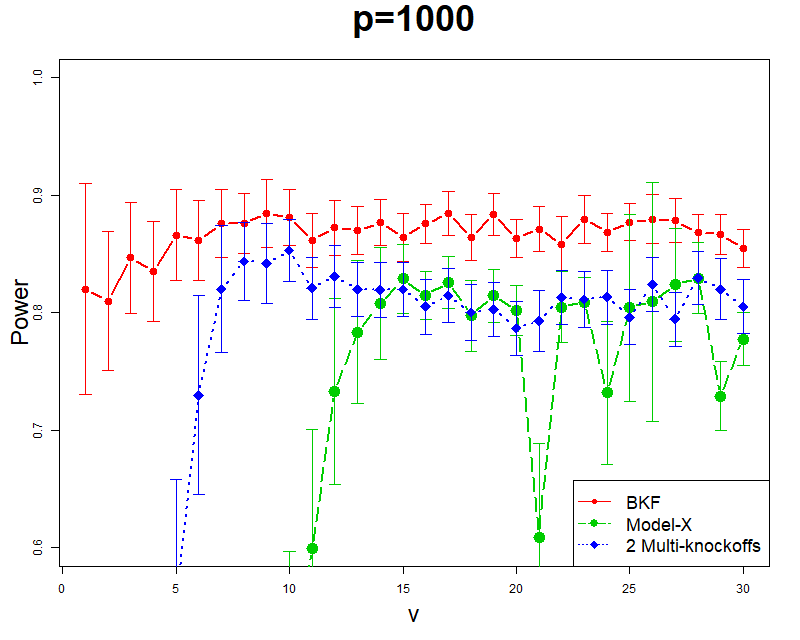}
		\caption{Comparisons of the FDR and power of the BKF and other knockoff methods under different numbers of features ($p$),
			different sizes of $\mathcal{H}_1$ ($v$) and the case of independent features. Each point is averaged over 100 replications.}
		\label{fig:p100fdr}
	\end{figure}
	
	\begin{figure}[b]
		\centering
		\includegraphics[width=0.4\linewidth]{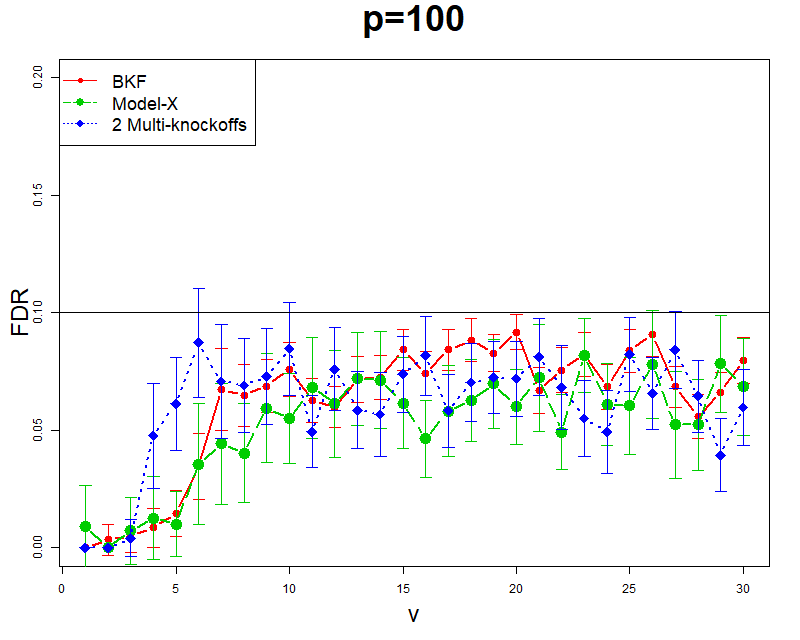}
		\includegraphics[width=0.4\linewidth]{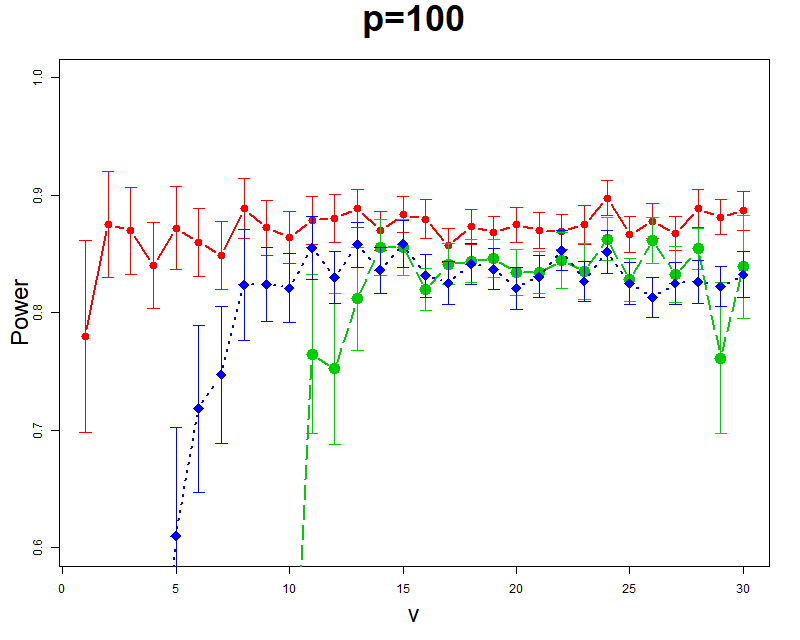}
		\includegraphics[width=0.4\linewidth]{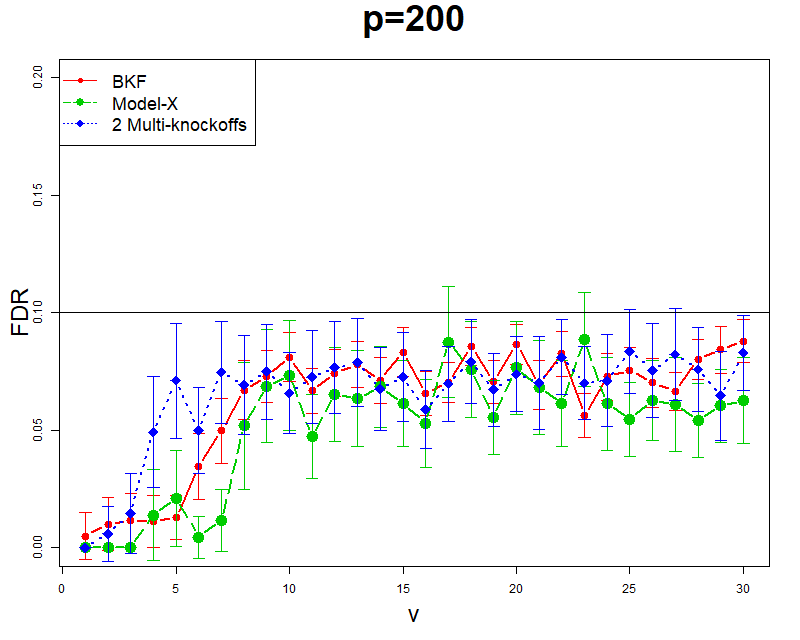}
		\includegraphics[width=0.4\linewidth]{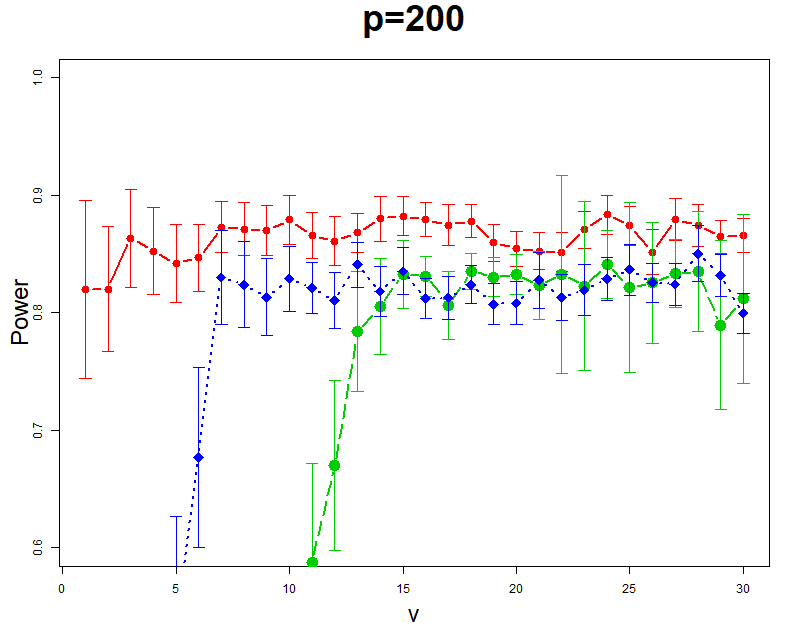}
		\includegraphics[width=0.4\linewidth]{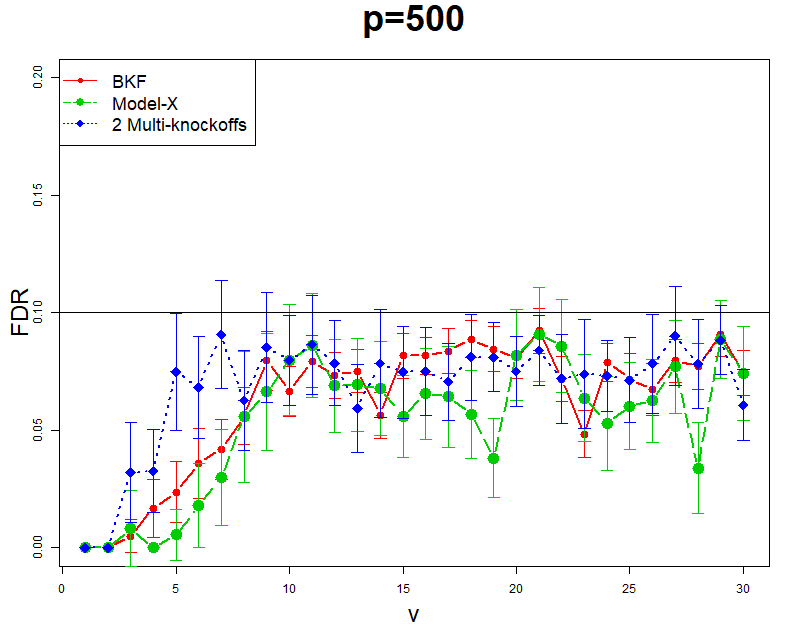}
		\includegraphics[width=0.4\linewidth]{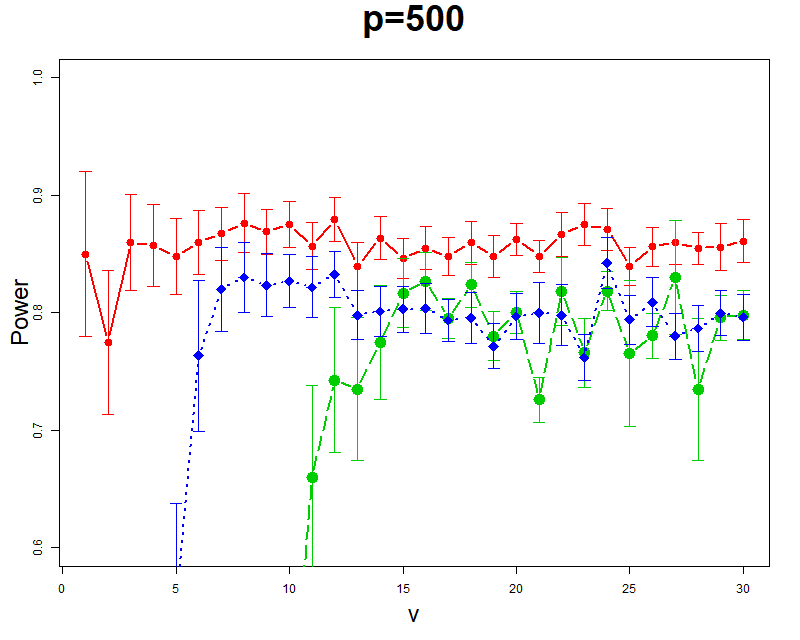}
		\includegraphics[width=0.4\linewidth]{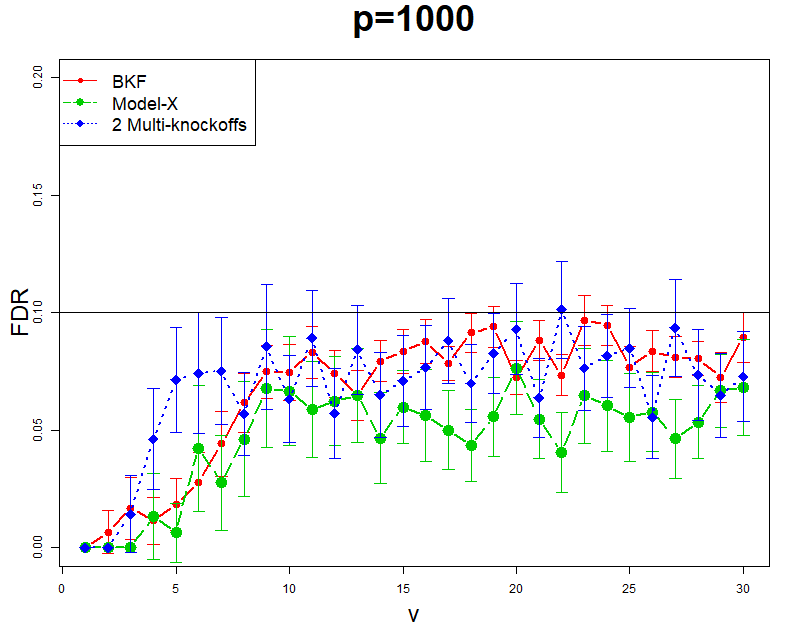}
		\includegraphics[width=0.4\linewidth]{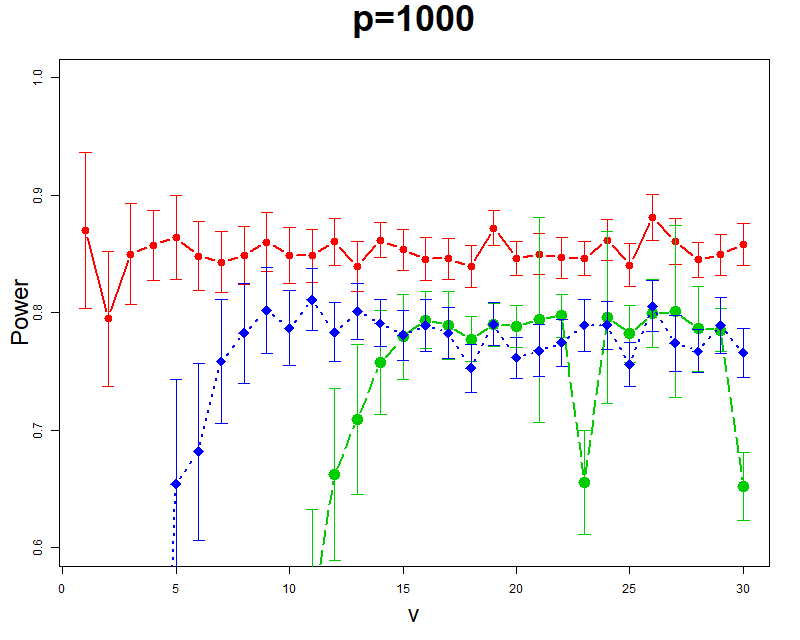}
		\caption{Comparisons of the FDR and power of the BKF and other knockoff methods under different numbers of features ($p$),
			different sizes of $\mathcal{H}_1$ ($v$) and the case of auto-correlated features. Each point is averaged over 100 replications.}
		\label{fig:decay_p100fdr}
	\end{figure}
	
	As $p$ and $v$ vary, the performances of the BKF, model-$\textbf{X}$ knockoff and 2 multi-knockoffs are exhibited
	in Figures \ref{fig:p100fdr} and \ref{fig:decay_p100fdr}. Generally speaking, all
	methods can control the FDR under $\alpha=0.1$.
	For all three methods, there is
	slight power loss when the number of features ($p$) is large. However,
	there exist substantial differences in their ability to detect false $H_{0j}$
	when $v$ decreases. As discussed in Section \ref{KF}, the selection procedure of 
the model-$\textbf{X}$ knockoff forces $|\hat{\mathcal{S}}|$ to be zero or not smaller than $\lfloor1/\alpha\rfloor$ if nonzero. This can be reflected by the phenomenon that it possesses comparable power only when $v$ is larger than $\lfloor1/\alpha\rfloor=10$. Although the 2 multi-knockoffs
	method improves the power
	when $v$ is between $5$ and $10$, it still suffers from power loss for extremely small $v$.
	The improvement in power is achieved at the
	sacrifice of the ability to detect false $H_{0j}$ when the true $\mathcal{H}_1$ is large. 
In addition, the performance deteriorates when features are correlated. In contrast, BKF maintains its power above 80\% for all
	values of $v$ and both covariance structures of features, indicating that it is
	not susceptible to power loss caused by the small
	size of $\mathcal{H}_1$ as well as dependency among features.
	
	\subsubsection{Robustness to Misspecified $f(\textbf{X})$}\label{misspecification}
	
	Similar to the
	model-$\textbf{X}$ knockoff filter \citep{Candes2018}, our BKF assumes that the distribution $f(\textbf{X})$ is known and thus establishes the joint distribution $f(\textbf{X},\tilde{\textbf{X}})$. If $f(\textbf{X})$ is unknown, the BKF can approximate
	$f(\textbf{X})$ by a Gaussian model $\text{MVN}(\hat{\boldsymbol{\mu}},\hat{\boldsymbol{\Sigma}})$ with
	the estimated mean $\hat{\boldsymbol{\mu}}$ and covariance matrix $\hat{\boldsymbol{\Sigma}}$. To investigate how misspecification of $f(\textbf{X})$ influences the performance of BKF, we simulate 100 datasets under the same settings as in Section \ref{NandA}, except that features ${x}_{ij} \ (i=1, \ldots, n; j=1, \ldots, p)$ are generated as i.i.d. samples from $t$-distribution with degree of freedom 3 and then divided by $\sqrt{3}$ to maintain the same signal-to-noise ratio. We still use the Gaussian approximation $\text{MVN}(\hat{\boldsymbol{\mu}},\hat{\boldsymbol{\Sigma}})$ of $f(\textbf{X})$ to establish the joint distribution $f(\textbf{X},\tilde{\textbf{X}})$ and generate knockoffs for inference.
	
	\begin{figure}[b]
		\centering
		\includegraphics[width=0.4\linewidth]{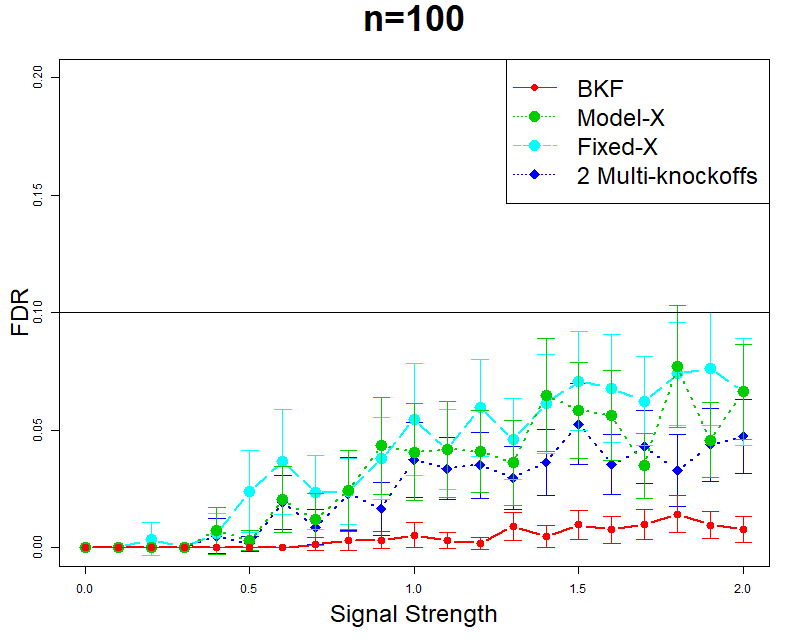}
		\includegraphics[width=0.4\linewidth]{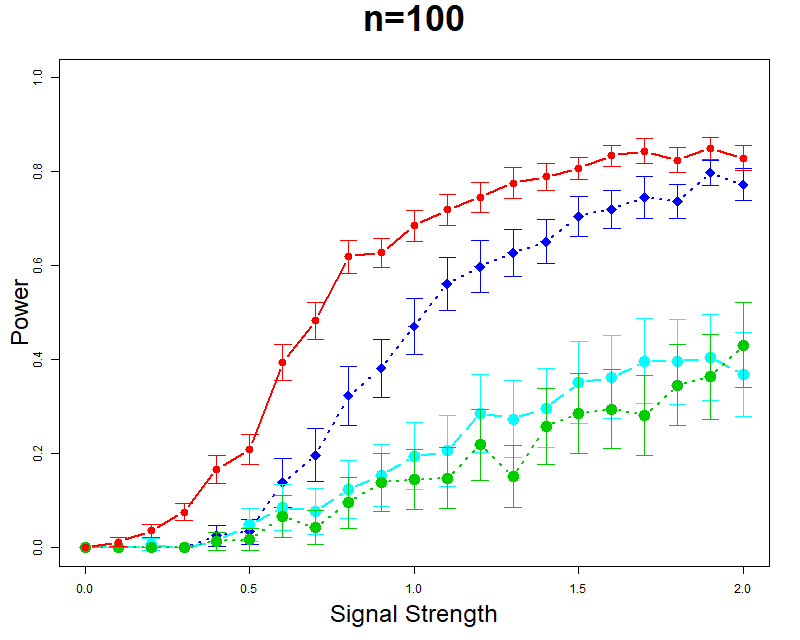}
		\includegraphics[width=0.4\linewidth]{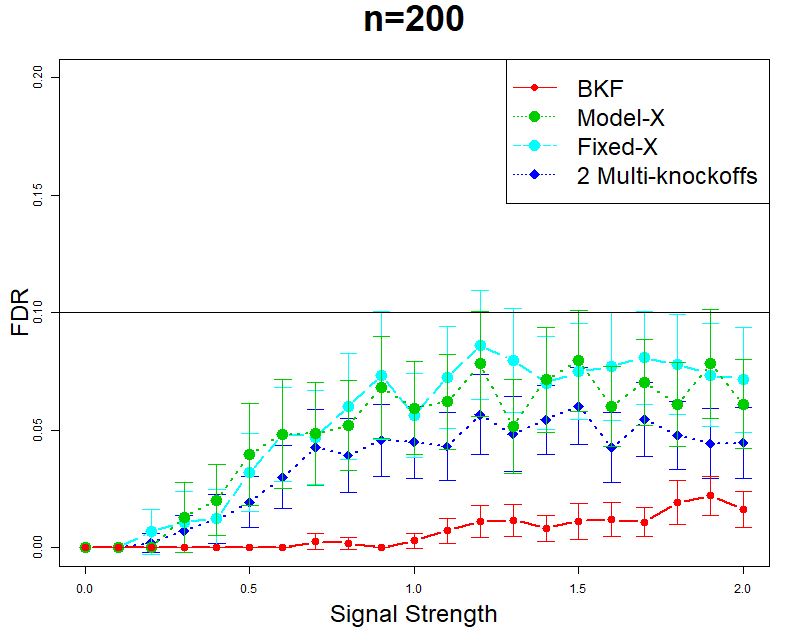}
		\includegraphics[width=0.4\linewidth]{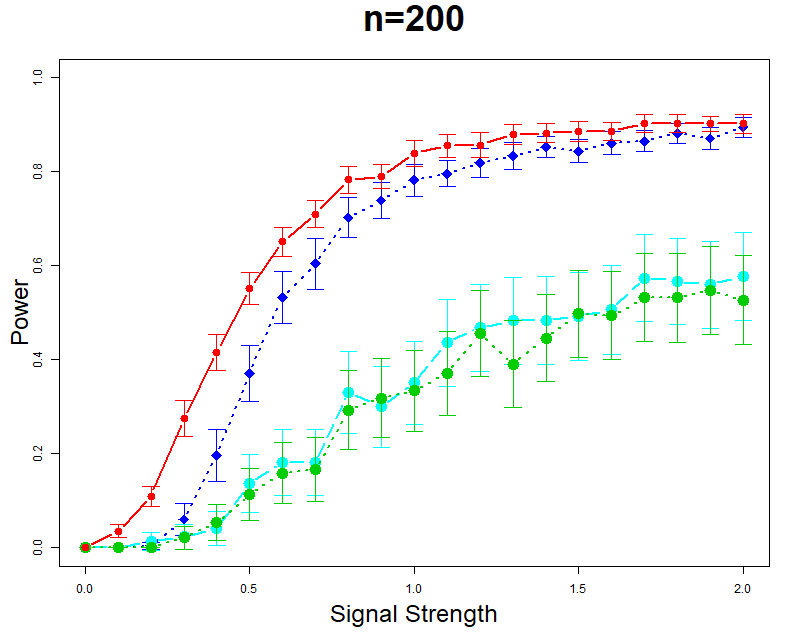}
		\includegraphics[width=0.4\linewidth]{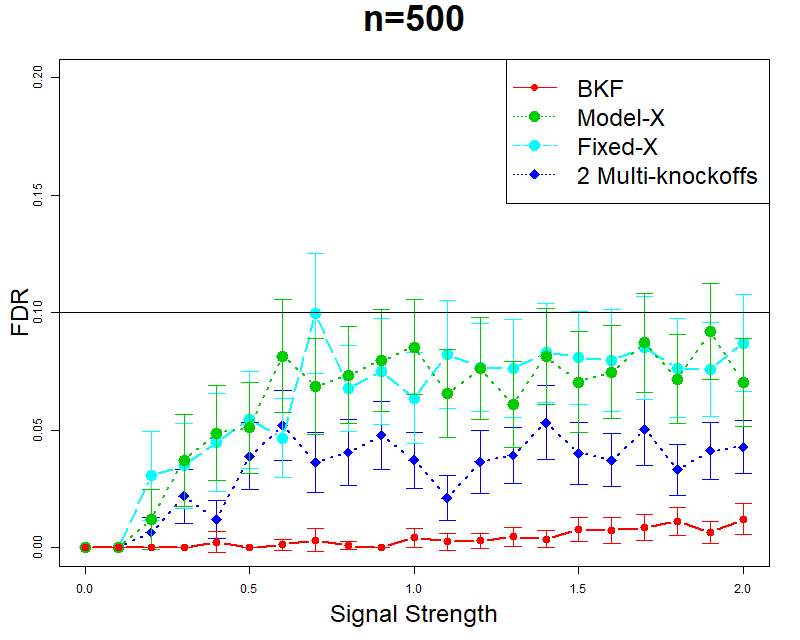}
		\includegraphics[width=0.4\linewidth]{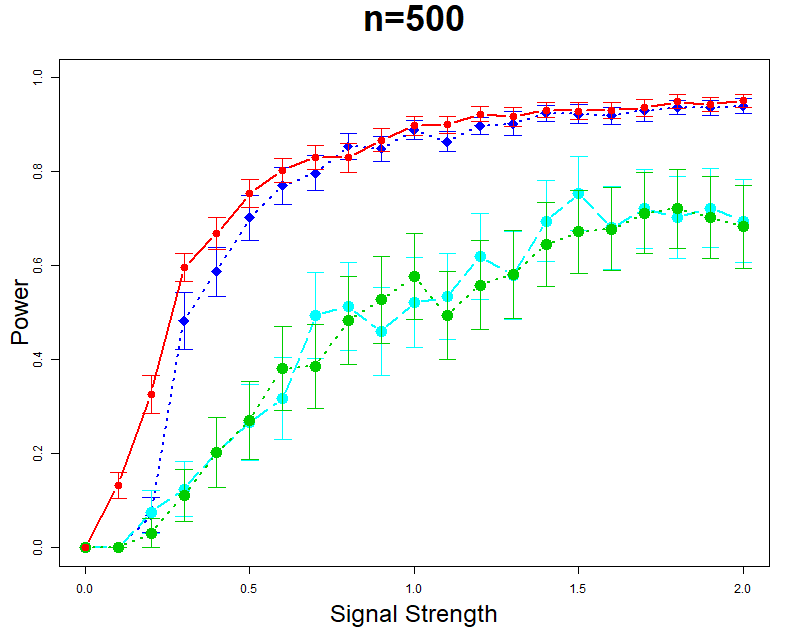}
		\includegraphics[width=0.4\linewidth]{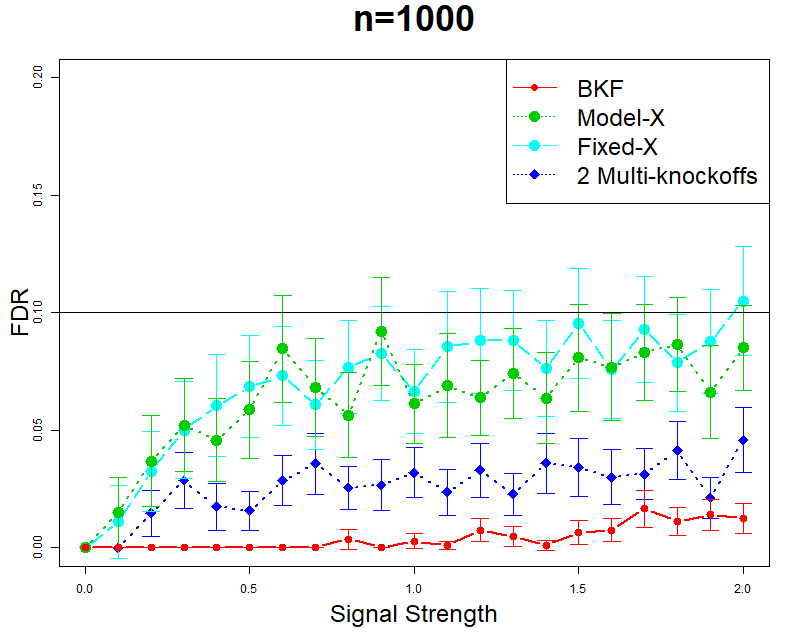}
		\includegraphics[width=0.4\linewidth]{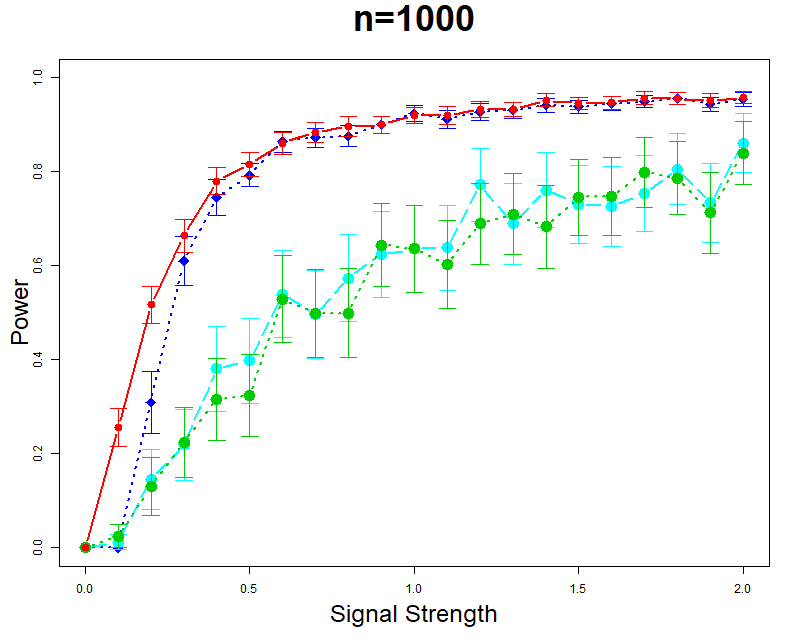}
		\caption{Comparisons of the FDR and power of the BKF and other knockoff methods under different
			sample sizes $n$ and signal strengths $a$ when $f(\textbf{X})$ is misspecified. Each point is averaged over 100 replications.}
		\label{fig:Specification}
	\end{figure}

The power and overall FDR of our BKF as well as
those of existing methods under misspecified $f(\textbf{X})$ are presented in Figure \ref{fig:Specification}.
Not only can our BKF keep the FDR under control but it also maintains the power of distinguishing non-null features
from null features, while existing approaches, especially the 2 multi-knockoffs method, deteriorate in power. Such deterioration is more severe when the sample size is small, where divergence between the Gaussian approximation $\text{MVN}(\hat{\boldsymbol{\mu}},\hat{\boldsymbol{\Sigma}})$ and the true $f(\textbf{X})$ is large. As a result, existing frequentist methods that only generate knockoffs once tend to be conservative in the case of poor-quality knockoffs, while our BKF can sample knockoffs many times in the MCMC and thus reduce the impact of poor-quality knockoffs caused by the misspecification of $f(\textbf{X})$.
	
	\subsection{Comparisons with Bayesian Variable Selection}
	
	As our BKF is a Bayesian approach for feature selection, we also compare it with existing Bayesian variable selection approaches, including the commonly used spike-and-slab regression, Bayesian Lasso, horseshoe estimator and recent IBSS procedure \citep{Wang2020}, to demonstrate the advantage of our method in controlling the proportion of false discoveries. In particular,
	for the spike-and-slab regression, Bayesian Lasso and horseshoe estimator,
	we compute the estimator $\hat{\mathcal{S}}$ using the thresholding procedure with threshold 0.5.
	For IBSS, the estimator $\hat{\mathcal{S}}$ is obtained via the greedy algorithm in Section \ref{bfdr} 
by substituting posterior probabilities $P(H_{0j}|\textbf{D})$ with $1-\text{PIP}_j$, where $\text{PIP}_j$ follows Definition 3.12 in \citet{Wang2020}.
	
	\begin{figure}[b]
		\centering
		\includegraphics[width=0.325\linewidth]{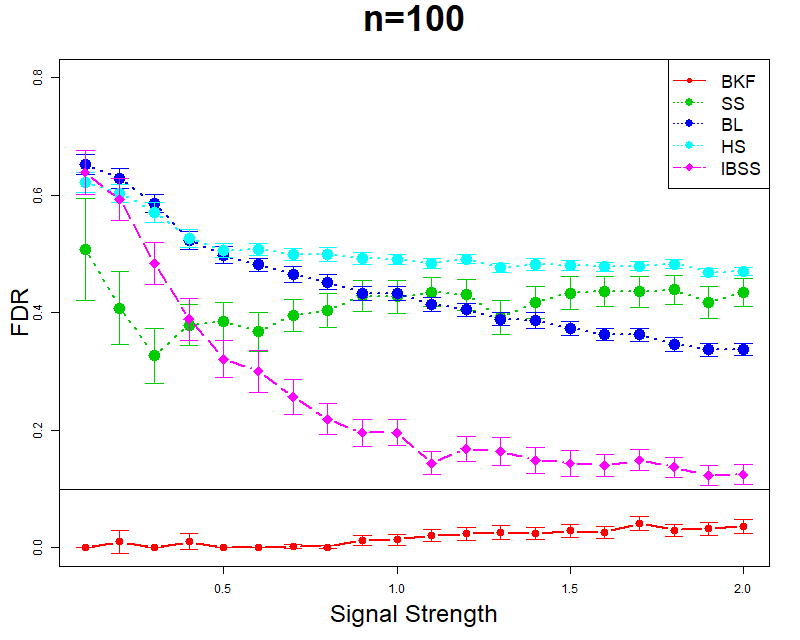}
		\includegraphics[width=0.325\linewidth]{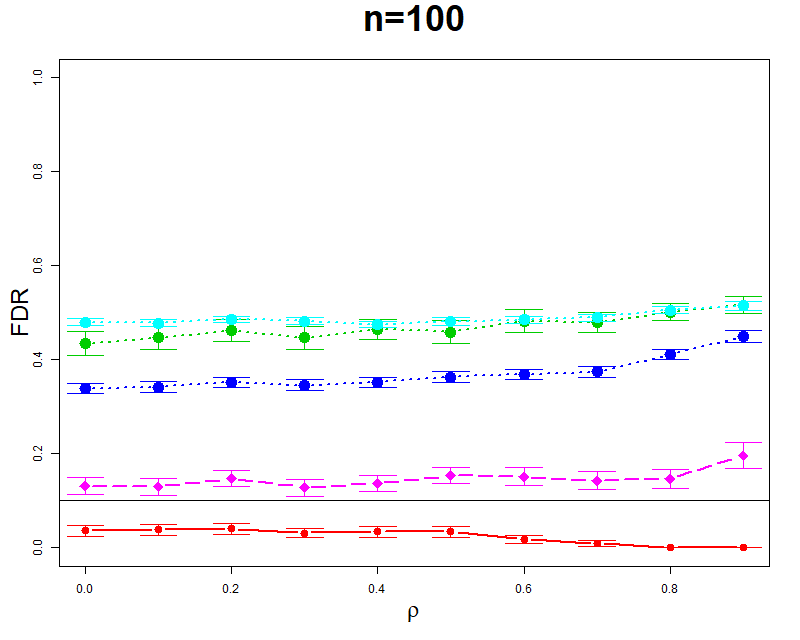}
		\includegraphics[width=0.325\linewidth]{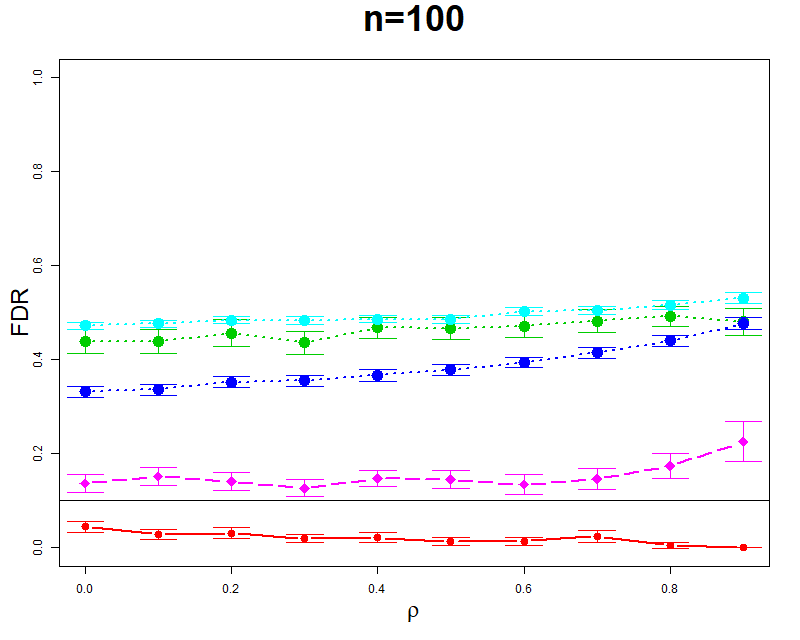}
		\includegraphics[width=0.325\linewidth]{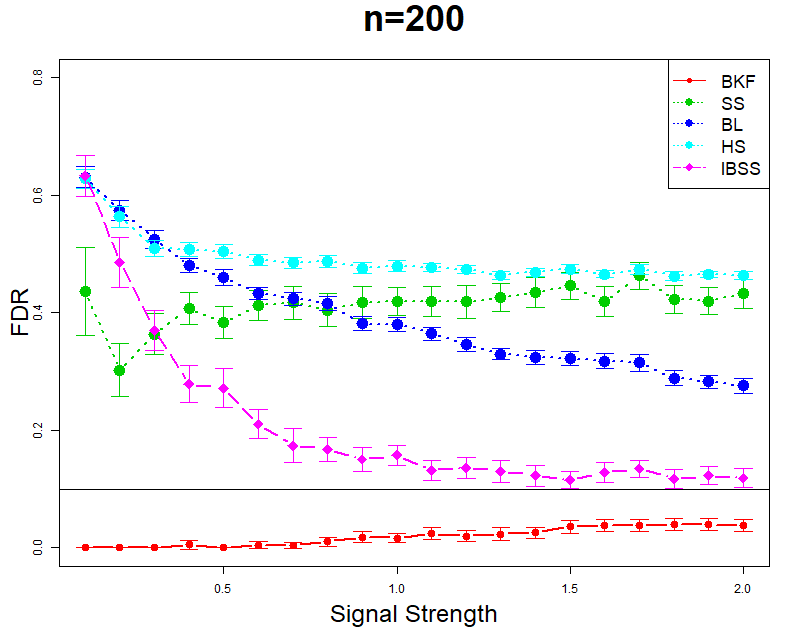}
		\includegraphics[width=0.325\linewidth]{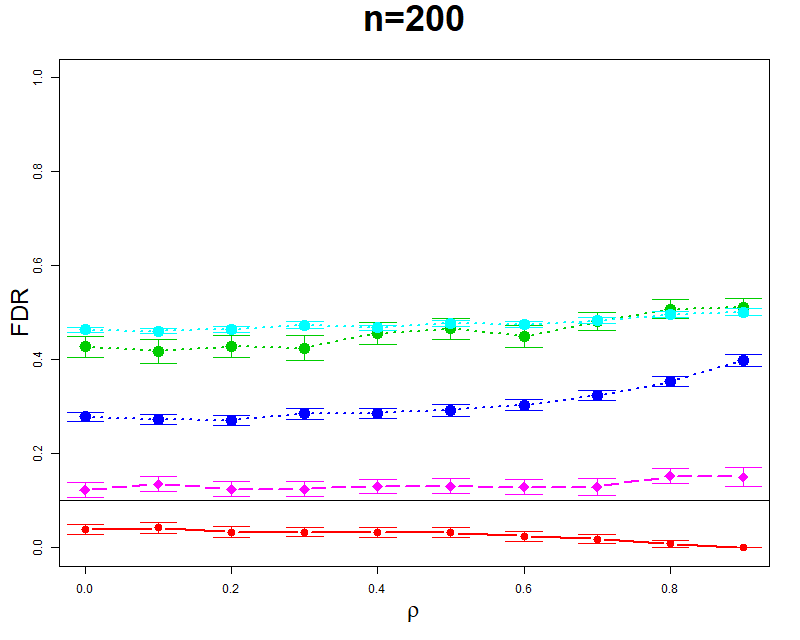}
		\includegraphics[width=0.325\linewidth]{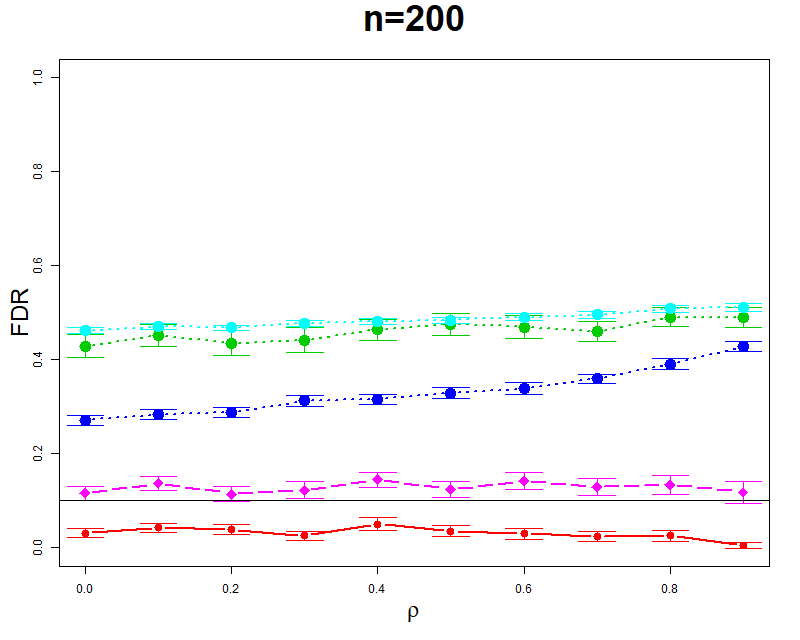}
		\includegraphics[width=0.325\linewidth]{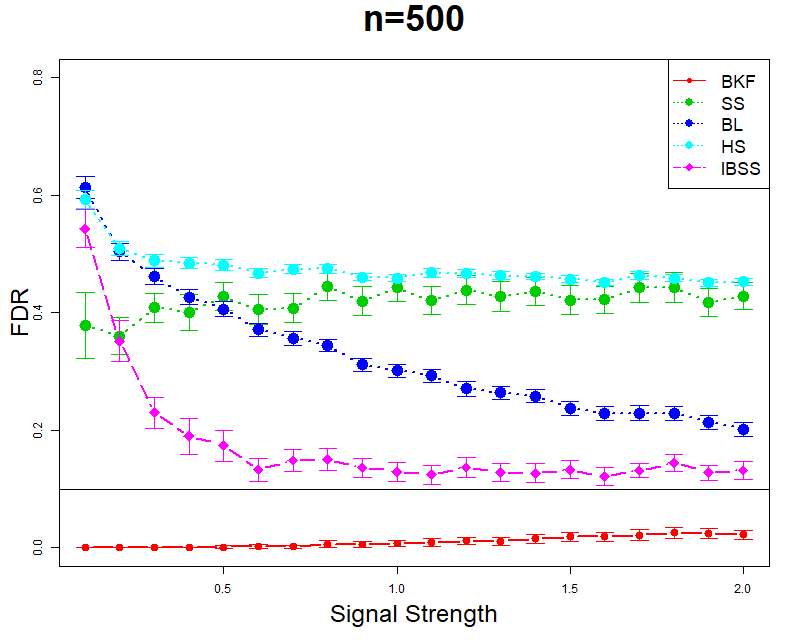}
		\includegraphics[width=0.325\linewidth]{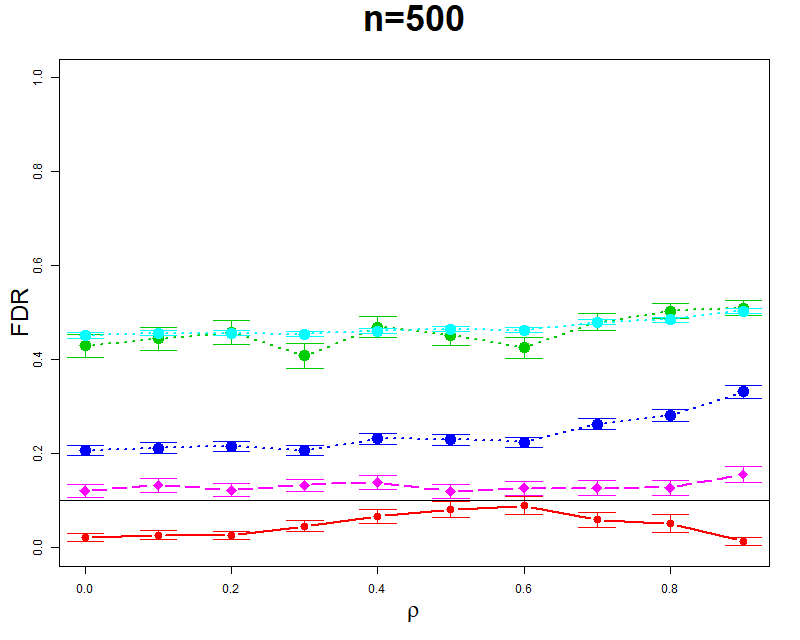}
		\includegraphics[width=0.325\linewidth]{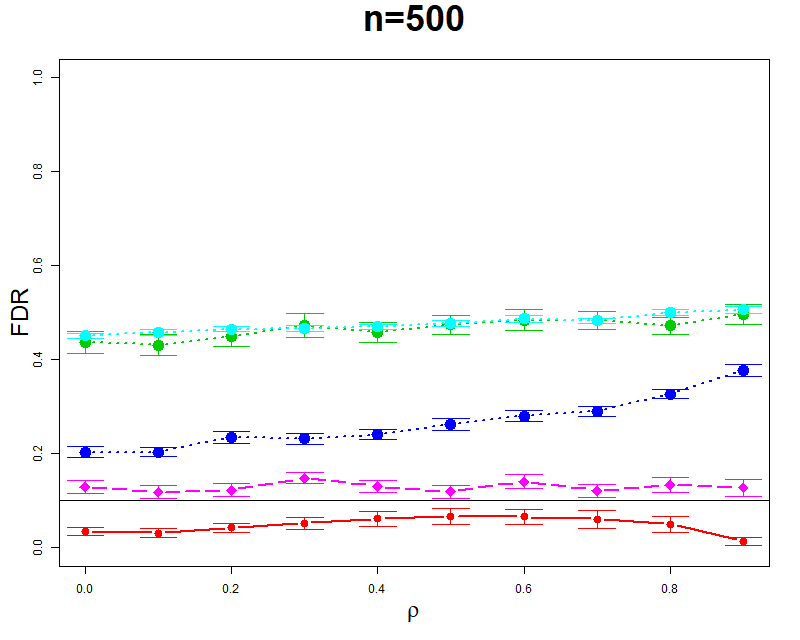}
		\includegraphics[width=0.325\linewidth]{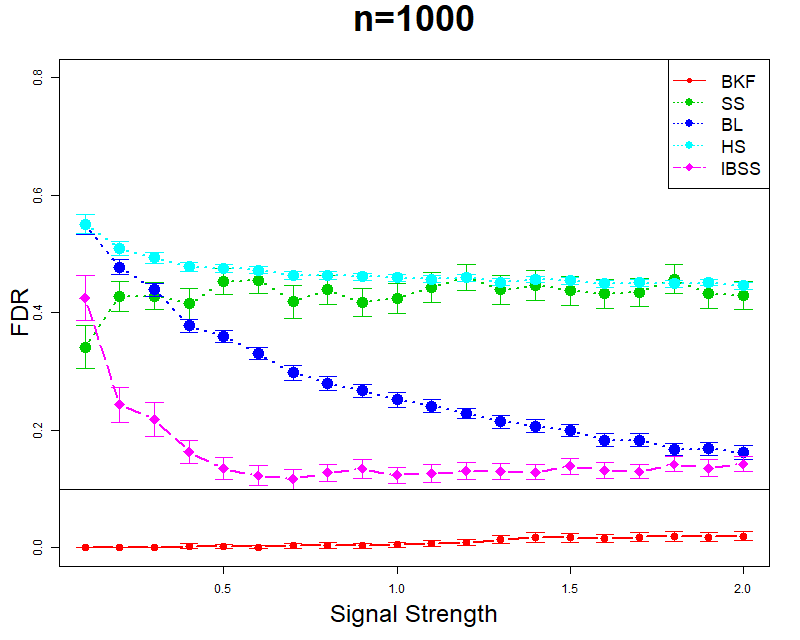}
		\includegraphics[width=0.325\linewidth]{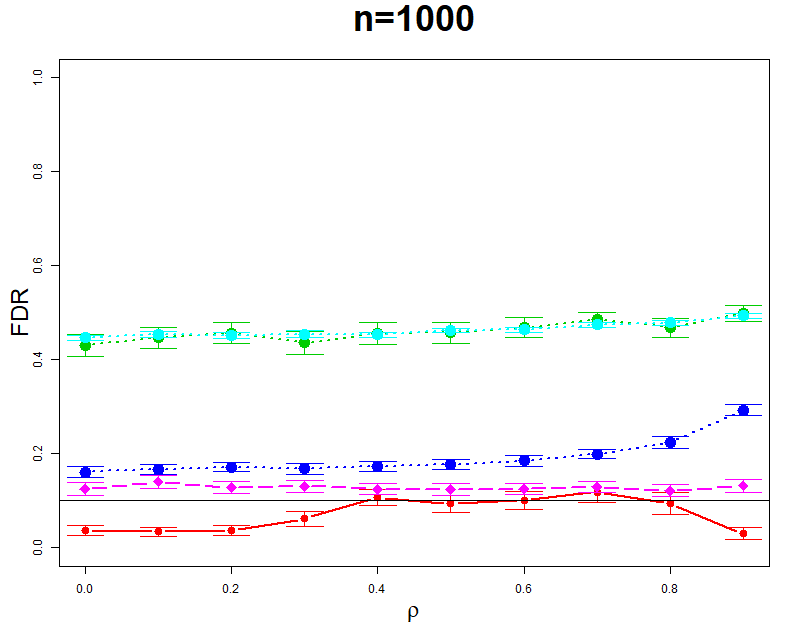}
		\includegraphics[width=0.325\linewidth]{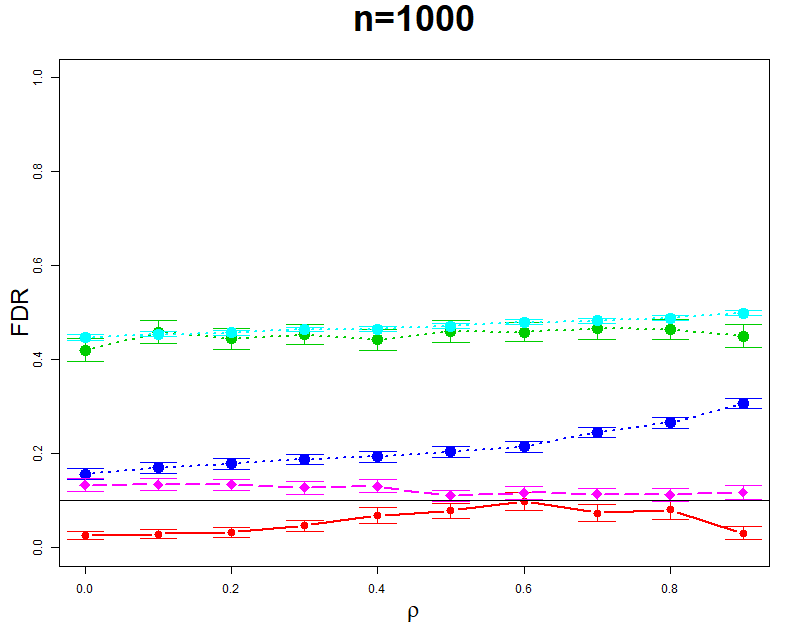}
		\caption{Comparisons of FDRs of the BKF and existing Bayesian
 variable selection methods in low-dimensional settings. The left, middle and right panels correspond to cases of independent, auto-correlated and equal-correlated features. Each point is averaged over 100 replications (SS: spike-and-slab regression; BL: Bayesian Lasso; HS: horseshoe estimator; IBSS: the iterative Bayesian stepwise selection procedure).}
		\label{fig:n100fdrb}
	\end{figure}
	
	\begin{figure}[b]
		\centering
		\includegraphics[width=0.325\linewidth]{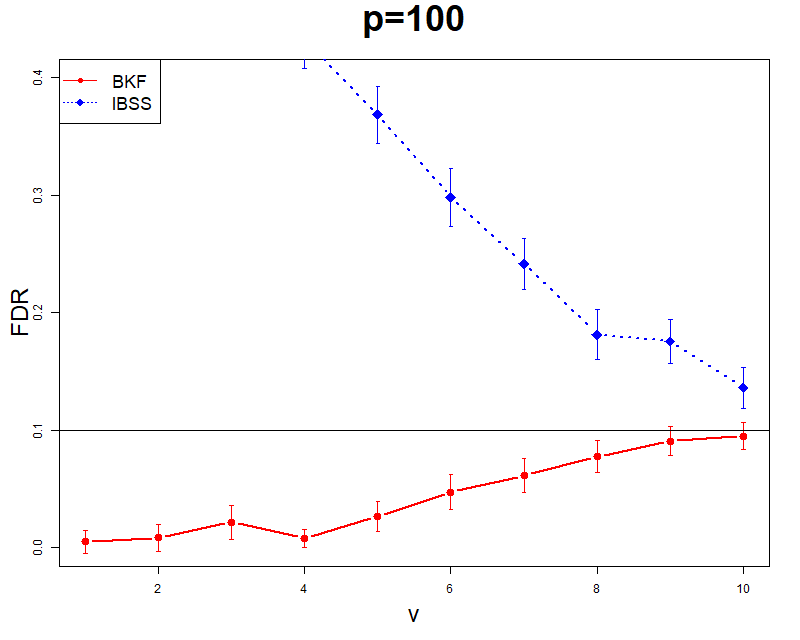}
		\includegraphics[width=0.325\linewidth]{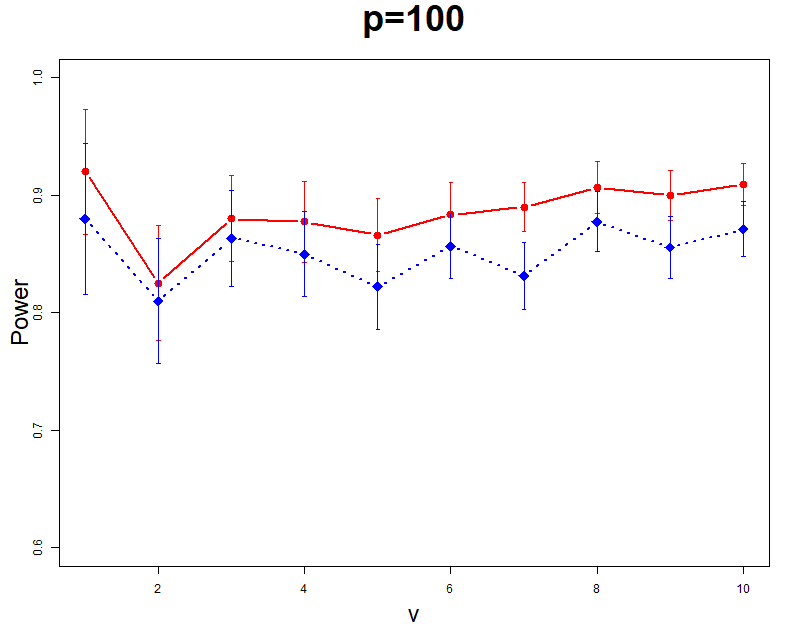}
		\includegraphics[width=0.325\linewidth]{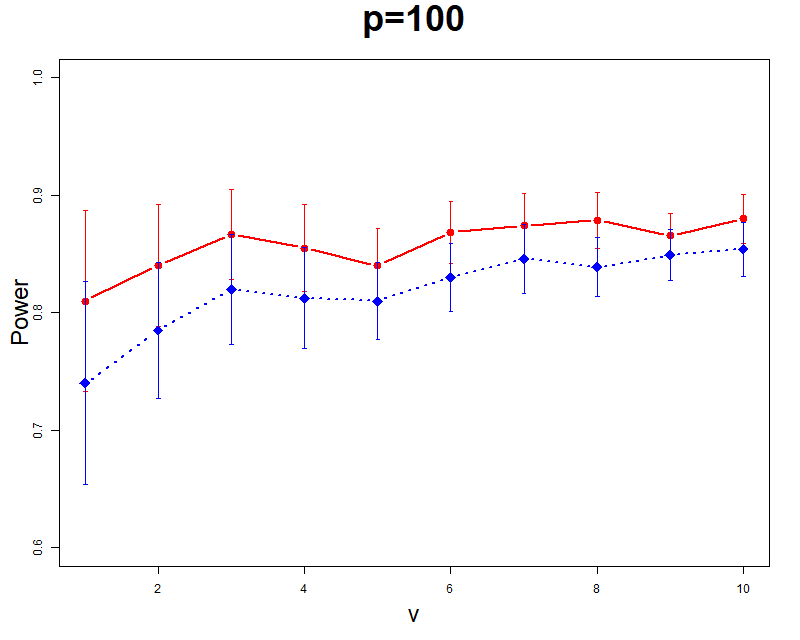}
		\includegraphics[width=0.325\linewidth]{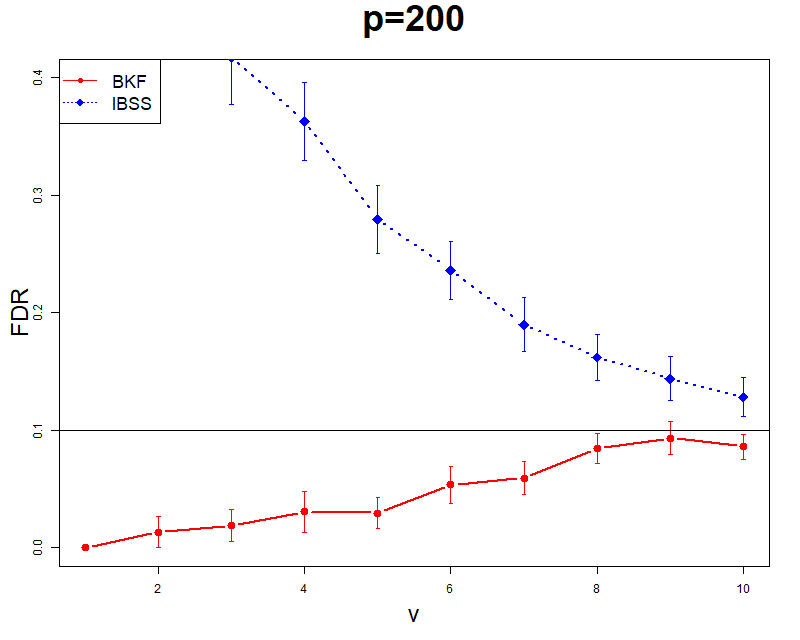}
		\includegraphics[width=0.325\linewidth]{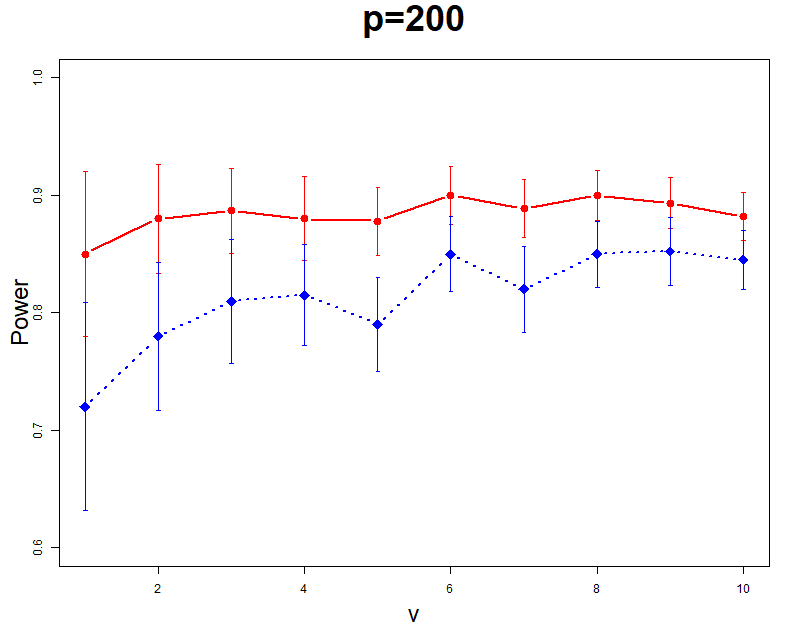}
		\includegraphics[width=0.325\linewidth]{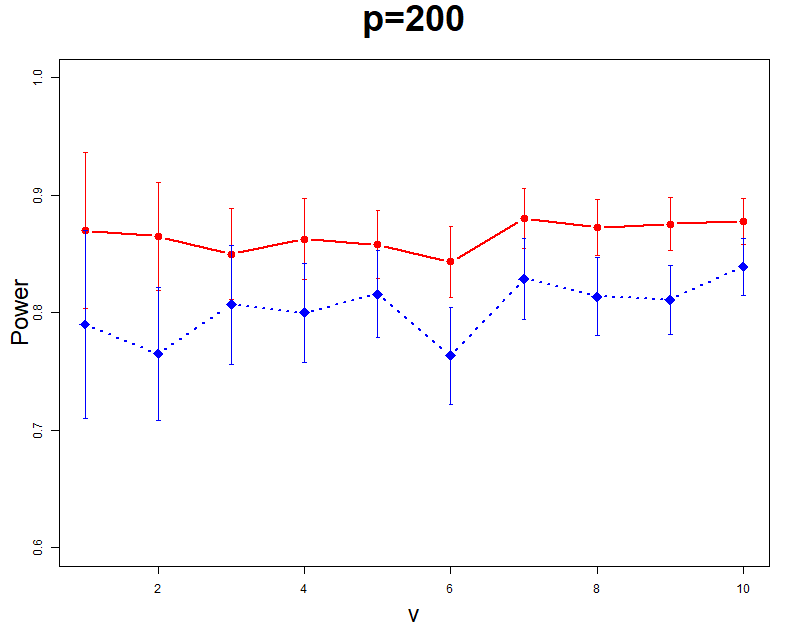}
		\includegraphics[width=0.325\linewidth]{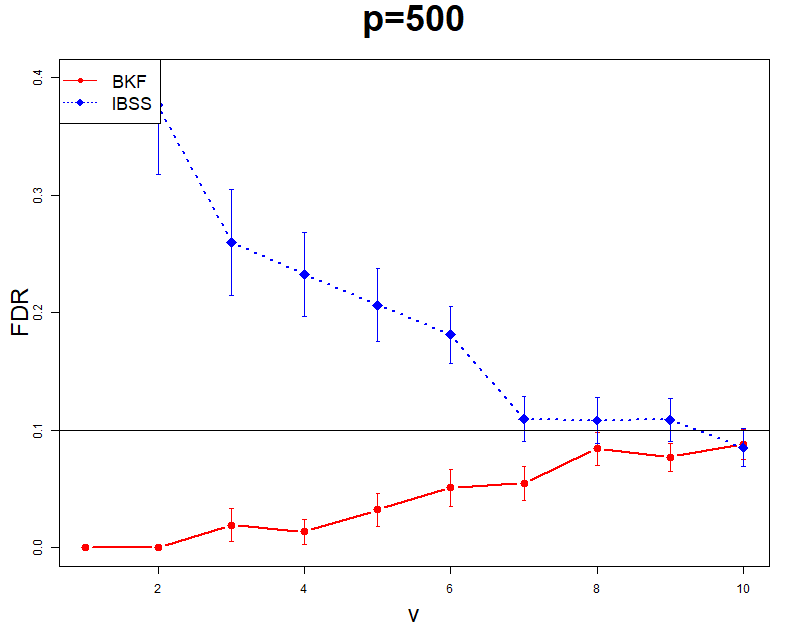}
		\includegraphics[width=0.325\linewidth]{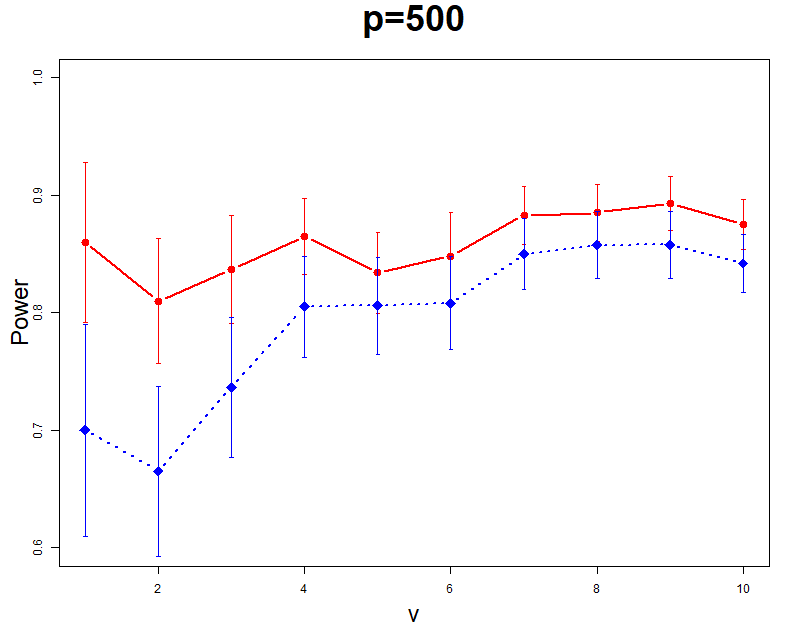}
		\includegraphics[width=0.325\linewidth]{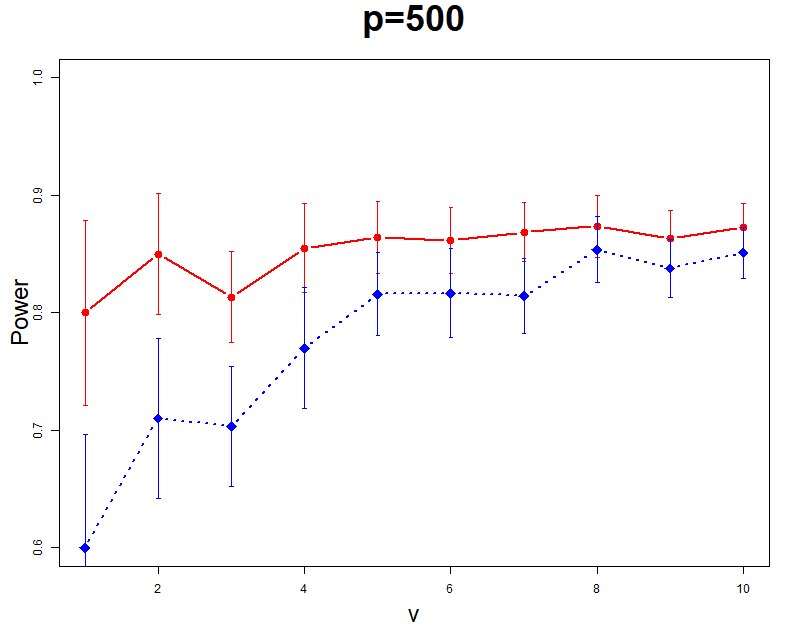}
		\includegraphics[width=0.325\linewidth]{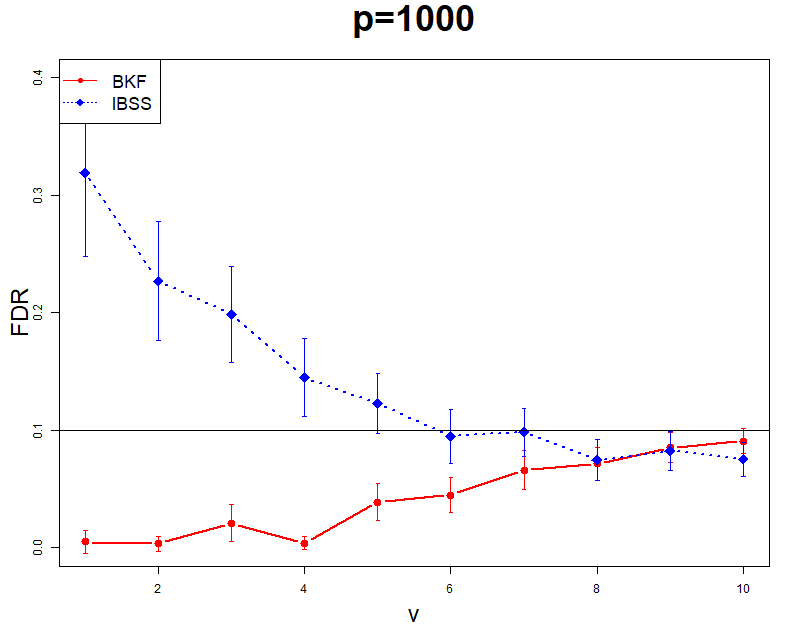}
		\includegraphics[width=0.325\linewidth]{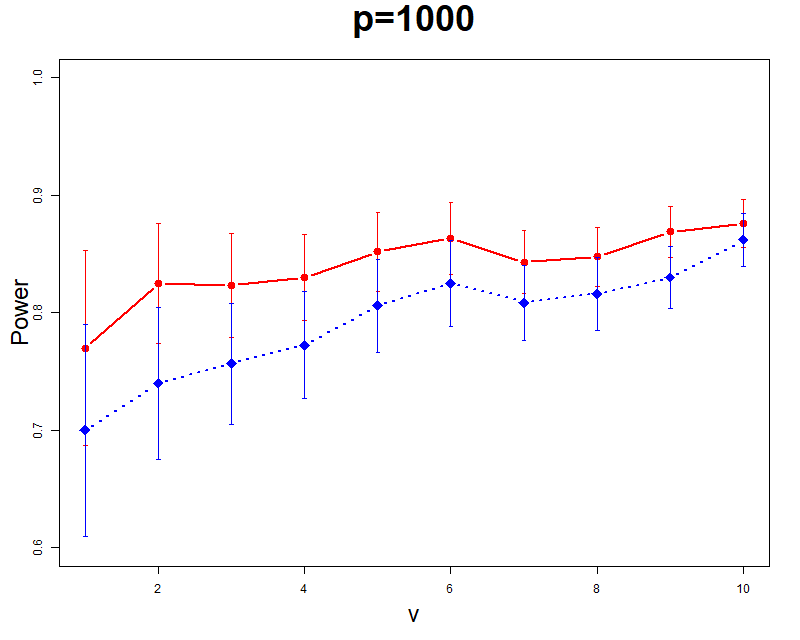}
		\includegraphics[width=0.325\linewidth]{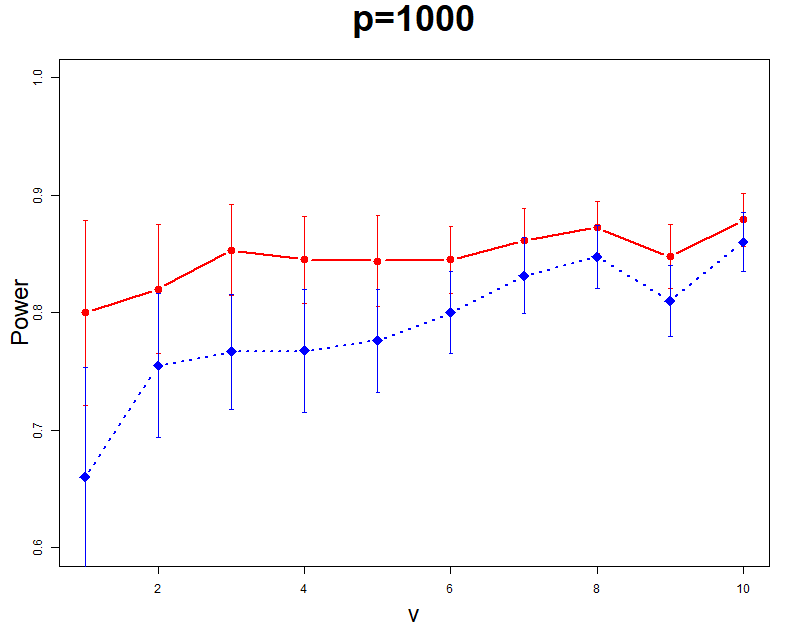}
		\caption{Performances of the BKF and IBSS procedures under different numbers of features ($p$) and
			different sizes of $\mathcal{H}_1$ ($v$). Each point is averaged over 100 replications. The left two columns correspond to the FDR and power under independent features while the right column corresponds to the power under auto-correlated features.}
		\label{fig:p100fdr_b}
	\end{figure}
	
	Figures \ref{fig:n100fdrb}--\ref{fig:p100fdr_b} display the performances of all Bayesian approaches
	in low-dimensional settings, where
	only BKF can control the FDR under the desired level $\alpha=0.1$.
	The IBSS procedure would reduce the proportion of false discoveries as the sample size increases,
	the signal strength amplifies or the feature correlation decreases, while
	FDRs of the Bayesian Lasso and horseshoe estimator are always out of control.
	Although the performance of IBSS is acceptable in low-dimensional settings, its estimator deteriorates for high-dimensional data, as shown in Figure \ref{fig:p100fdr_b}. The IBSS procedure possesses lower power and larger FDR, suggesting its weakness
	in controlling the FDR for high-dimensional settings. On the other hand, our BKF is robust in the FDR control
	and true signal discovery regardless of the dimensionality and the number of non-null features.
	
	\section{Real Data Analysis}\label{Real}
	
	To illustrate its empirical performance, we apply our BKF to the league of legends 2020 esports match data\footnote{https://www.kaggle.com/xmorra/lol2020esports} from Kaggle. The league of legends (LOL)
	is a multiplayer online battle arena video game between two teams.
	In each game, 10 players are assigned to different positions (``top", ``jungle", ``mid", ``adc" and ``support")
	in two teams labeled as ``blue" and ``red".
	Players are asked to select champions with unique abilities and different attributes to battle against the other team.
	This dataset records the selection of champions and results of all games of LOL matches in the year of 2020 as well as
	18 attributes of each selected champion.
	As a result, each game record consists of $180$ attributes in different positions as features $\textbf{X}$.
	We extract all of $726$ records of the LOL pro league in China to investigate which attributes in different positions would substantially affect the results of games.
	
\begin{figure}[b]
		\centering	\includegraphics[width=1\linewidth]{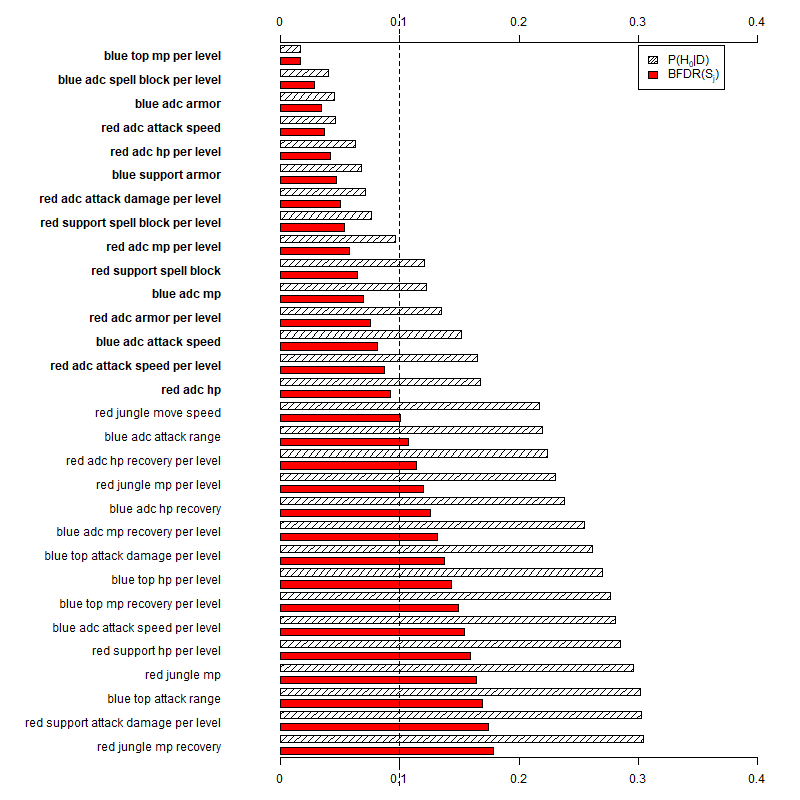}
		\caption{Values of $\widehat{{\mathbb{P}}}(H_{0j}|\textbf{D})$ and
			${\widehat{{\text{BFDR}}}}(\mathcal{S}_j)$ corresponding to all features sorted by $\widehat{{\mathbb{P}}}(H_{0j}|\textbf{D})$.}
		\label{fig:caring}
	\end{figure}

	Given that all the attributes are continuous, we standardize
	all of the $180$ features among $726$ records and
	approximate the joint distribution of original features and knockoff features $f(\textbf{X},\tilde{\textbf{X}})$ with a Gaussian distribution. Taking the binary game result (equal to 1 when the blue team wins and 0 when the red team wins) as the response, we impose a
	probit model to make inference on $f(Y|\textbf{X},\tilde{\textbf{X}})$ and apply our BKF with flat priors as well as other knockoff methods to conduct multiple hypotheses testing (\ref{hypothese}) on
	all the features. Posterior samples are drawn from full conditionals under the probit model as detailed in Appendix \ref{C} and Algorithm \ref{alg3}. We control the FDR at the desired level $\alpha=0.1$ and show
	the results in Table \ref{Betahat} and Figure \ref{fig:caring}, where we only display 30 features with the lowest $\widehat{{\mathbb{P}}}(H_{0j}|\textbf{D})$.
	
	\begin{table}[b]
		\caption{Posterior means and standard deviations (in parentheses)
			of coefficients $\beta_j$ and $\tilde{\beta}_j$ for analysis of the league of legends 2020 esports match data.
		The features selected by BKF are in boldface.}\label{Betahat}
		\centering
		\begin{tabular}{lrrr}
			\toprule
			Features&\multicolumn{1}{c}{$\beta_j$}&\multicolumn{1}{c}{$\tilde{\beta}_j$}
			&$\widehat{{\mathbb{P}}}(H_{0j}|\textbf{D})$\\\midrule
			\textbf{Blue top mp per level}&1.841 (0.481)&-0.114 (0.457)&0.016\\
			\textbf{Blue adc spell block per level}&2.070 (0.612)&-0.007 (0.439)&0.040\\
			\textbf{Blue adc armor}&1.888 (0.615)&-0.054 (0.421)&0.045\\
			\textbf{Red adc attack speed}&-2.546 (0.825)&0.012 (0.458)&0.046\\
			\textbf{Red adc hp per level}&-1.818 (0.548)&0.016 (0.452)&0.063\\
			\textbf{Blue support armor}&1.463 (0.476)&0.016 (0.420)&0.068\\
			\textbf{Red adc attack damage per level}&-2.145 (0.735)&0.165 (0.544)&0.071\\
			\textbf{Red support spell block per level}&-2.623 (1.289)&0.087 (0.596)&0.076\\
			\textbf{Red adc mp per level}&-2.852 (1.242)&-0.002 (0.462)&0.096\\
			\textbf{Red support spell block}&-1.264 (0.554)&0.003 (0.391)&0.121\\
			\textbf{Blue adc mp}&1.326 (0.547)&-0.088 (0.452)&0.122\\
			\textbf{Red adc armor per level}&-2.138 (1.210)&0.142 (0.481)&0.135\\
			\textbf{Blue adc attack speed}&1.353 (0.707)&-0.120 (0.465)&0.152\\
			\textbf{Red adc attack speed per level}&-2.070 (1.431)&-0.038 (0.424)&0.165\\
			\textbf{Red adc hp}&-1.647 (0.901)&0.043 (0.415)&0.168\\
			Red jungle move speed&-0.935 (0.446)&0.086 (0.417)&0.217\\
			Blue adc attack range&1.078 (0.505)&0.060 (0.423)&0.220\\
			Red adc hp recovery per level&-1.353 (0.864)&0.087 (0.415)&0.224\\
			Red jungle mp per level&-1.205 (0.893)&-0.124 (0.457)&0.231\\
			Blue adc hp recovery&1.040 (0.542)&0.048 (0.449)&0.238\\
			Blue adc mp recovery per level&1.176 (0.738)&0.022 (0.467)&0.255\\
			Blue top attack damage per level&0.858 (0.425)&-0.014 (0.395)&0.262\\
			Blue top hp per level&0.945 (0.461)&0.017 (0.442)&0.270\\
			Blue top mp recovery per level&0.938 (0.437)&-0.097 (0.462)&0.277\\
			Blue adc attack speed per level&1.132 (0.800)&0.082 (0.451)&0.281\\
			Red support hp per level&-0.776 (0.450)&-0.055 (0.381)&0.285\\
			Red jungle mp&-1.094 (0.749)&-0.029 (0.411)&0.296\\
			Blue top attack range&0.802 (0.418)&-0.005 (0.398)&0.302\\
			Red support attack damage per level&-0.953 (0.721)&-0.007 (0.405)&0.303\\
			Red jungle mp recovery&-0.822 (0.481)&-0.027 (0.403)&0.305\\
			\bottomrule
		\end{tabular}
	\end{table}
	
	From Table \ref{Betahat}, we can see one obvious pattern that the posterior means of
	$\beta_j$ are all positive (negative) for attributes corresponding to
	the blue (red) team. This is consistent with the common sense that attributes corresponding to
	the blue (red) team would make the blue team more (less) likely to win,
	suggesting the validity of our analysis.
	Fifteen features are chosen in the estimator
	$\hat{\mathcal{S}}$, most of which are attributes in the positions ``adc" and ``support". 
This result is consistent with the consensus among pro players that the strength of
	the bottom duo (``adc" and ``support") is more decisive to match results in 2020. In contrast, when we apply the model-$\textbf{X}$ knockoff procedure to the same data, no features are selected. This
	indicates that our BKF is more powerful in detecting non-null features under the same FDR-controlling level.
	
	\section{Conclusion}\label{Conclusion}
	The knockoff procedure is a powerful tool to select important features with a control
	over the FDR. We incorporate the knockoff method into the Bayesian framework and
	propose a more powerful BKF, where the knockoff variables, instead of
	being generated only once and fixed, are iteratively
	updated in the Gibbs sampling steps. Under the Bayesian model,
	we develop an MCMC data augmentation algorithm to obtain posterior samples of knockoff variables
	and parameters. Based on a probability inequality deduced from the flip-sign
	property of feature statistics and the equivalent definition of the Bayesian FDR,
	we are able to estimate the upper bound of the Bayesian FDR.
	As a result, the estimation of the non-null feature set is translated to
	a knapsack problem and we develop a greedy selection algorithm to obtain the 
Bayesian estimator with control of the Bayesian FDR. Experiments show that
	BKF generally possesses high power in
	identifying non-null features and lower probability of making false discoveries than existing knockoff filters and Bayesian 
variable selection approaches, especially in the cases where the sample size is not large, the true number of non-null features is small and the correlations among original features are strong. Our BKF is also robust against 
misspecification of the distribution of covariates. 
As illustrated in real data analysis, the BKF yields reasonable results which can be
	clearly visualized, allowing easier interpretations
	and decision making than existing knockoff filters.
	
%
	
	\begin{appendices}
		\section{Proof of Theorem \ref{thm1}}\label{proof1}
		Without loss of generality, we assume that $\mathcal{S}=\{1,\ldots,m\}$, $\mathcal{H}_0=\{1,\ldots,v\}$ and $1\leq m\leq v\leq p$ such that $\mathcal{S}\subset\mathcal{H}_0$.
		Let $(\boldsymbol{\beta}_{{\rm Swap}(\mathcal{S})},\tilde{\boldsymbol{\beta}}_{{\rm Swap}(\mathcal{S})})$ denote $(\boldsymbol{\beta},\tilde{\boldsymbol{\beta}})$ by swapping elements $\beta_j$ and $\tilde{\beta}_j$
		for all $j\in\mathcal{S}$. Similarly, $(\textbf{X}_{{\rm Swap}(\mathcal{S})},\tilde{\textbf{X}}_{{\rm Swap}(\mathcal{S})})$
		are also obtained by swapping elements $X_{j}$ and $\tilde{X}_{j}$ of $(\textbf{X},\tilde{\textbf{X}})$ for all $j\in\mathcal{S}$. Since conditional on $\mathcal{H}_0$, for all $j\in\mathcal{S}$,  $X_j\perp Y|\textbf{X}_{-j}$, we have $\textbf{X}_{\mathcal{S}}\perp Y|\textbf{X}_{-\mathcal{S}}$ where $\textbf{X}_{\mathcal{S}}$ and $\textbf{X}_{-\mathcal{S}}$ are the subvectors of $\textbf{X}$ corresponding to indices in and not in the set $\mathcal{S}$ respectively. Thus, we have $\textbf{X}_{\mathcal{S}}\perp Y|\textbf{X}_{-\mathcal{S}},\tilde{\textbf{X}}$. Under Definition \ref{DF:Knockoff}, it is also true that  $\tilde{\textbf{X}}_{\mathcal{S}}\perp Y|\tilde{\textbf{X}}_{-\mathcal{S}},{\textbf{X}}$.
		As a result, it is clear that conditional on $\mathcal{H}_0$, \begin{equation}\label{eq1}
			f(Y|\textbf{X},\tilde{\textbf{X}})=f(Y|\textbf{X}_{{\rm Swap}(\mathcal{S})},\tilde{\textbf{X}}_{{\rm Swap}(\mathcal{S})}).
		\end{equation}
		
		Suppose that the conditional distribution $f(Y|\textbf{X},\tilde{\textbf{X}})$ is parameterized as 
a GLM with density $h(Y|\textbf{X},\tilde{\textbf{X}};\boldsymbol{\beta},\tilde{\boldsymbol{\beta}},\boldsymbol{\phi})$
		in (\ref{EGLM}),	
\begin{equation*} E[Y|\textbf{X},\tilde{\textbf{X}};\boldsymbol{\beta},\tilde{\boldsymbol{\beta}},\boldsymbol{\phi}]=g^{-1}(\eta),\quad\eta=\sum_{j=1}^{p}(X_j\beta_j+\tilde{X}_j\tilde{\beta}_j).
		\end{equation*}
		By (\ref{Exchange}) and (\ref{eq1}), we can deduce that conditional on $\mathcal{H}_0$, 
		\begin{equation}\label{eq2}
			h(Y|\textbf{X},\tilde{\textbf{X}};\boldsymbol{\beta},\tilde{\boldsymbol{\beta}},\boldsymbol{\phi})
			=
			h(Y|\textbf{X}_{{\rm Swap}(\mathcal{S})},\tilde{\textbf{X}}_{{\rm Swap}(\mathcal{S})};\boldsymbol{\beta},\tilde{\boldsymbol{\beta}},\boldsymbol{\phi}).
		\end{equation}
	
Under $h(Y|\textbf{X},\tilde{\textbf{X}};\boldsymbol{\beta},\tilde{\boldsymbol{\beta}},\boldsymbol{\phi})$,
		the response is related to original and knockoff features only via the conditional mean $g^{-1}(\eta)$, 
and thus we have 
\begin{equation}\label{eq3}
			h(Y|\textbf{X}_{{\rm Swap}(\mathcal{S})},\tilde{\textbf{X}}_{{\rm Swap}(\mathcal{S})};\boldsymbol{\beta},\tilde{\boldsymbol{\beta}},\boldsymbol{\phi})=h(Y|\textbf{X},\tilde{\textbf{X}};\boldsymbol{\beta}_{{\rm Swap}(\mathcal{S})},\tilde{\boldsymbol{\beta}}_{{\rm Swap}(\mathcal{S})},\boldsymbol{\phi}).
		\end{equation}
		If $f(\boldsymbol{\beta},\tilde{\boldsymbol{\beta}})$ is invariant
		to swapping for $\mathcal{S}$, i.e.,
		$f(\boldsymbol{\beta},\tilde{\boldsymbol{\beta}})=f(\boldsymbol{\beta}_{{\rm Swap}(\mathcal{S})},\tilde{\boldsymbol{\beta}}_{{\rm Swap}(\mathcal{S})}),$
		it is clear that by (\ref{eq2})--(\ref{eq3}),
		$$
		\begin{aligned}
			&f(\boldsymbol{\beta},\tilde{\boldsymbol{\beta}}|\tilde{\textbf{x}}_1,\ldots,\tilde{\textbf{x}}_n,\sigma^2,
			\textbf{D},\mathcal{H}_0)=f(\boldsymbol{\beta}_{{\rm Swap}(\mathcal{S})},\tilde{\boldsymbol{\beta}}_{{\rm Swap}(\mathcal{S})}|\tilde{\textbf{x}}_1,\ldots,\tilde{\textbf{x}}_n,\sigma^2,
			\textbf{D},\mathcal{H}_0),
		\end{aligned}
		$$ and $$
		\begin{aligned}f(\boldsymbol{\beta},\tilde{\boldsymbol{\beta}}|
			\textbf{D},\mathcal{H}_0)&
			=\int f(\boldsymbol{\beta},\tilde{\boldsymbol{\beta}}|\tilde{\textbf{x}}_1,\ldots,\tilde{\textbf{x}}_n,\sigma^2,
			\textbf{D},\mathcal{H}_0)f(\tilde{\textbf{x}}_1,\ldots,\tilde{\textbf{x}}_n,\sigma^2|
			\textbf{D})d\tilde{\textbf{x}}_1\cdots d\tilde{\textbf{x}}_nd\sigma^2\\&= \int f(\boldsymbol{\beta}_{{\rm Swap}(\mathcal{S})},\tilde{\boldsymbol{\beta}}_{{\rm Swap}(\mathcal{S})}|\tilde{\textbf{x}}_1,\ldots,\tilde{\textbf{x}}_n,\sigma^2,
			\textbf{D},\mathcal{H}_0) f(\tilde{\textbf{x}}_1,\ldots,\tilde{\textbf{x}}_n,\sigma^2|
			\textbf{D})d\tilde{\textbf{x}}_1\cdots d\tilde{\textbf{x}}_nd\sigma^2\\&=f(\boldsymbol{\beta}_{{\rm Swap}(\mathcal{S})},\tilde{\boldsymbol{\beta}}_{{\rm Swap}(\mathcal{S})}|
			\textbf{D},\mathcal{H}_0).
		\end{aligned}
		$$
		
		As long as $W_j$ is antisymmetric with respect to $\beta_j$ and $\tilde{\beta}_j$,
		$j=1,\ldots,p$,  the feature statistics $\textbf{W}=(W_1,\ldots,W_p)^{\dT}$ obey the flip-sign property because	
		$$
		\begin{aligned}
			f(\textbf{W}_{\mathcal{S}}|\textbf{D},\mathcal{H}_0)=&\int_{\boldsymbol{\Theta}_{\textbf{W}_{\mathcal{S}}}} f(\boldsymbol{\beta}_{{\rm Swap}(\mathcal{S})},\tilde{\boldsymbol{\beta}}_{{\rm Swap}(\mathcal{S})}|
			\textbf{D},\mathcal{H}_0)d\boldsymbol{\beta}_{{\rm Swap}(\mathcal{S})}d\tilde{\boldsymbol{\beta}}_{{\rm Swap}(\mathcal{S})}\\=&\int_{\boldsymbol{\Theta}_{\textbf{W}}} f(\boldsymbol{\beta},\tilde{\boldsymbol{\beta}}|
			\textbf{D},\mathcal{H}_0)d\boldsymbol{\beta}d\tilde{\boldsymbol{\beta}}\\=& f(\textbf{W}|\textbf{D},\mathcal{H}_0),\\
		\end{aligned}
		$$ where $\textbf{W}_{\mathcal{S}}$ is defined in Definition \ref{anti-symmetry} and $\boldsymbol{\Theta}_{\textbf{W}}$ is the subspace of all possible values of $(\boldsymbol{\beta},\tilde{\boldsymbol{\beta}})$ corresponding to the value of $\textbf{W}$.
		
		\section{Full Conditionals under Probit Model}\label{C}
		
		In our real data analysis, the joint distribution $f(\textbf{X},\tilde{\textbf{X}})$ is approximated by a Gaussian model (\ref{Joint}) and the probit model $$h(Y|\textbf{X},\tilde{\textbf{X}};\boldsymbol{\beta},\tilde{\boldsymbol{\beta}})=\begin{cases}
			\Phi(\textbf{X}^{\dT}\boldsymbol{\beta}
			+\tilde{\textbf{X}}^{\dT}\tilde{\boldsymbol{\beta}}),&\text{for } Y=1,\\
			1-\Phi(\textbf{X}^{\dT}\boldsymbol{\beta}
			+\tilde{\textbf{X}}^{\dT}\tilde{\boldsymbol{\beta}}),&\text{for } Y=0,
		\end{cases}$$ is imposed. 
		To make inference, we augment variables $u_i$ ($i=1,\ldots,n$), leading to the augmented probit model as follows,
		\begin{equation*}
			\begin{aligned}
				y_i&=\begin{cases}
					1,&\text{if }u_i>0,\\0,&\text{if }u_i\leq0,\\
				\end{cases}\\
				u_i&\sim N(\textbf{x}_i^{\dT}\boldsymbol{\beta}
				+\tilde{\textbf{x}}_i^{\dT}\tilde{\boldsymbol{\beta}},1).\\
			\end{aligned}
		\end{equation*}
		Thus, the joint posterior distribution under the augmented probit model is \begin{equation}\label{C2}
			\begin{aligned}
				&f(\tilde{\textbf{x}}_1,\ldots,\tilde{\textbf{x}}_n,u_1,\ldots,u_n,
				\boldsymbol{\beta},\tilde{\boldsymbol{\beta}}|\textbf{D})\\
\propto&f(\boldsymbol{\beta},\tilde{\boldsymbol{\beta}})\prod_{i=1}^{n}
p(y_i|u_i)p(u_i|\textbf{x}_i,\tilde{\textbf{x}}_i;\boldsymbol{\beta},\tilde{\boldsymbol{\beta}})f(\tilde{\textbf{x}}_i|\textbf{x}_i)\\
				\propto&f(\boldsymbol{\beta},\tilde{\boldsymbol{\beta}})
				\exp\Bigg\{-\frac{1}{2}\sum_{i=1}^{n}(u_i-\textbf{x}_i^{\dT}\boldsymbol{\beta}
				-\tilde{\textbf{x}}_i^{\dT}\tilde{\boldsymbol{\beta}})^2-\frac{1}{2}\sum_{i=1}^{n}(\textbf{x}_i^{\dT},\tilde{\textbf{x}}_i^{\dT})
				\textbf{G}^{-1}\begin{pmatrix}
					\textbf{x}_i\\\tilde{\textbf{x}}_i
				\end{pmatrix}\Bigg\}\\
				&\times \prod_{i=1}^{n}
				\big\{I(y_i=1)I(u_i>0)+I(y_i=0)I(u_i\leq0)\big\}.
			\end{aligned}
		\end{equation}
	
		Under the flat prior, posterior samples are drawn from full conditionals detailed in Algorithm \ref{alg3}.

	\begin{algorithm}
		\caption{Gibbs sampler under the probit model.}\label{alg3}
		\begin{algorithmic}[1]
			\STATE {\bfseries Input:} Observed data $\textbf{D}$.
			\STATE Initialize $$\tilde{\textbf{x}}_i\sim\text{MVN}({\textbf{x}}_i-\boldsymbol{\Sigma}^{-1}\text{diag}\{\textbf{s}\}{\textbf{x}}_i,2\text{diag}\{\textbf{s}\}-\text{diag}\{\textbf{s}\}\boldsymbol{\Sigma}^{-1}\text{diag}\{\textbf{s}\}) $$
			and under the truncated normal (TN) distribution,
$$u_i\sim \begin{cases}
				\text{TN}_{(0,\infty)}(0,1),&\text{if }y_i=1,\\\text{TN}_{(-\infty,0]}(0,1),&\text{if }y_i=0,\\
			\end{cases}$$
			 for $i=1,\ldots,n$.
			\REPEAT
			\STATE Sample $(\boldsymbol{\beta}^\dT,\tilde{\boldsymbol{\beta}}^\dT)\sim\text{MVN}(\boldsymbol{\mu}_{\boldsymbol{\beta},\tilde{\boldsymbol{\beta}}},\boldsymbol{\Sigma}_{\boldsymbol{\beta},\tilde{\boldsymbol{\beta}}})$, where
			$$
			\begin{aligned}
				\boldsymbol{\Sigma}_{\boldsymbol{\beta},\tilde{\boldsymbol{\beta}}}&=\Bigg\{\sum_{i=1}^n\begin{pmatrix}
					 {\textbf{x}}_i{\textbf{x}}_i^\dT&{\textbf{x}}_i\tilde{\textbf{x}}_i^\dT\\\tilde{\textbf{x}}_i{\textbf{x}}_i^{\dT}&\tilde{\textbf{x}}_i\tilde{\textbf{x}}_i^\dT
				\end{pmatrix}\Bigg\}^{-1},\quad
				 \boldsymbol{\mu}_{\boldsymbol{\beta},\tilde{\boldsymbol{\beta}}}=\boldsymbol{\Sigma}_{\boldsymbol{\beta},\tilde{\boldsymbol{\beta}}}\Bigg\{\sum_{i=1}^nu_i\begin{pmatrix}
					{\textbf{x}}_i\\\tilde{\textbf{x}}_i
				\end{pmatrix}\Bigg\}.
			\end{aligned}
			$$
			\FOR{$i=1,\ldots,n$} \STATE {Sample $\tilde{\textbf{x}}_i\sim \text{MVN}(\tilde{\boldsymbol{\mu}}_i,\tilde{\boldsymbol{\Sigma}})$ where $$
				\begin{aligned}
					\tilde{\boldsymbol{\Sigma}}&=\Big(
					\textbf{A}+\frac{1}{\sigma^2}\tilde{\boldsymbol{\beta}}\tilde{\boldsymbol{\beta}}^{\dT}\Big)^{-1},\\
					 \tilde{\boldsymbol{\mu}}_i&=\tilde{\boldsymbol{\Sigma}}\Big[\big(\text{diag}\{\textbf{s}\}^{-1}-\textbf{A}-\frac{1}{\sigma^2}\tilde{\boldsymbol{\beta}}\boldsymbol{\beta}^\dT\big)\textbf{x}_i+\frac{1}{\sigma^2}\tilde{\boldsymbol{\beta}}u_i\Big],\\
					\textbf{A}&=\big(2\text{diag}\{\textbf{s}\}-\text{diag}\{\textbf{s}\}\boldsymbol{\Sigma}^{-1}\text{diag}\{\textbf{s}\}\big)^{-1}.
				\end{aligned}
				$$} \ENDFOR
			\STATE Sample $$u_i\sim \begin{cases}
				\text{TN}_{(0,\infty)}(\textbf{x}_i^{\dT}\boldsymbol{\beta}
				+\tilde{\textbf{x}}_i^{\dT}\tilde{\boldsymbol{\beta}},1),&\text{if }y_i=1,\\\text{TN}_{(-\infty,0]}(\textbf{x}_i^{\dT}\boldsymbol{\beta}
				+\tilde{\textbf{x}}_i^{\dT}\tilde{\boldsymbol{\beta}},1),&\text{if }y_i=0,\\
			\end{cases}$$
			for $i=1,\ldots,n$.
			\UNTIL{convergence}
		\end{algorithmic}
	\end{algorithm}
\newpage

	\end{appendices}
	\bibliographystyle{apalike}
	\bibliography{sample}

\begin{thebibliography}{}

\bibitem[Barber and Cand{\`{e}}s, 2015]{Barber2015}
Barber, R.~F. and Cand{\`{e}}s, E.~J. (2015).
\newblock Controlling the false discovery rate via knockoffs.
\newblock {\em The Annals of Statistics}, 43(5):2055--2085.

\bibitem[Barber and Cand{\`{e}}s, 2019]{Barber2019}
Barber, R.~F. and Cand{\`{e}}s, E.~J. (2019).
\newblock A knockoff filter for high-dimensional selective inference.
\newblock {\em The Annals of Statistics}, 47(5):2504--2537.

\bibitem[Bates et~al., 2020]{Bates2020}
Bates, S., Cand{\`{e}}s, E., Janson, L., and Wang, W. (2020).
\newblock Metropolized knockoff sampling.
\newblock {\em Journal of the American Statistical Association}, pages 1--15.

\bibitem[Benjamini and Hochberg, 1995]{Benjamini1995}
Benjamini, Y. and Hochberg, Y. (1995).
\newblock Controlling the false discovery rate: a practical and powerful
  approach to multiple testing.
\newblock {\em Journal of the Royal Statistical Society: Series B
  (Methodological)}, 57(1):289--300.

\bibitem[Blanchard and Roquain, 2009]{Blanchard2009}
Blanchard, G. and Roquain, E. (2009).
\newblock Adaptive false discovery rate control under independence and
  dependence.
\newblock {\em J. Mach. Learn. Res.}, 10:2837–2871.

\bibitem[Cand{\`{e}}s et~al., 2018]{Candes2018}
Cand{\`{e}}s, E., Fan, Y., Janson, L., and Lv, J. (2018).
\newblock Panning for gold: `model-{X}' knockoffs for high dimensional
  controlled variable selection.
\newblock {\em Journal of the Royal Statistical Society: Series B (Statistical
  Methodology)}, 80(3):551--577.

\bibitem[Carvalho et~al., 2010]{Carvalho2010}
Carvalho, C.~M., Polson, N.~G., and Scott, J.~G. (2010).
\newblock The horseshoe estimator for sparse signals.
\newblock {\em Biometrika}, 97(2):465--480.

\bibitem[Dai and Barber, 2016]{Dai2016}
Dai, R. and Barber, R. (2016).
\newblock The knockoff filter for {FDR} control in group-sparse and multitask
  regression.
\newblock In Balcan, M.~F. and Weinberger, K.~Q., editors, {\em Proceedings of
  The 33rd International Conference on Machine Learning}, volume~48 of {\em
  Proceedings of Machine Learning Research}, pages 1851--1859, New York, New
  York, USA. PMLR.

\bibitem[Efron, 2008]{Efron2008}
Efron, B. (2008).
\newblock Microarrays, empirical bayes and the two-groups model.
\newblock {\em Statistical Science}, 23(1):1--22.

\bibitem[Efroymson, 1960]{Efroymson1960}
Efroymson, M.~A. (1960).
\newblock Multiple regression analysis.
\newblock {\em Mathematical Methods for Digital Computers}, pages 191--203.

\bibitem[Fan and Li, 2001]{Fan2001}
Fan, J. and Li, R. (2001).
\newblock Variable selection via nonconcave penalized likelihood and its oracle
  properties.
\newblock {\em Journal of the American Statistical Association},
  96(456):1348--1360.

\bibitem[Gimenez et~al., 2019]{Gimenez2019}
Gimenez, J.~R., Ghorbani, A., and Zou, J. (2019).
\newblock Knockoffs for the mass: new feature importance statistics with false
  discovery guarantees.
\newblock In Chaudhuri, K. and Sugiyama, M., editors, {\em Proceedings of
  Machine Learning Research}, volume~89 of {\em Proceedings of Machine Learning
  Research}, pages 2125--2133. PMLR.

\bibitem[Gimenez and Zou, 2019]{Gimenez2019b}
Gimenez, J.~R. and Zou, J. (2019).
\newblock Improving the stability of the knockoff procedure: multiple
  simultaneous knockoffs and entropy maximization.
\newblock In Chaudhuri, K. and Sugiyama, M., editors, {\em Proceedings of
  Machine Learning Research}, volume~89 of {\em Proceedings of Machine Learning
  Research}, pages 2184--2192. PMLR.

\bibitem[Katsevich and Sabatti, 2019]{Katsevich2019}
Katsevich, E. and Sabatti, C. (2019).
\newblock Multilayer knockoff filter: controlled variable selection at multiple
  resolutions.
\newblock {\em The Annals of Applied Statistics}, 13(1):1--33.

\bibitem[Leek and Storey, 2008]{Leek2008}
Leek, J.~T. and Storey, J.~D. (2008).
\newblock A general framework for multiple testing dependence.
\newblock {\em Proceedings of the National Academy of Sciences},
  105(48):18718--18723.

\bibitem[Miranda-Moreno et~al., 2007]{Miranda-Moreno2007}
Miranda-Moreno, L.~F., Labbe, A., and Fu, L. (2007).
\newblock Bayesian multiple testing procedures for hotspot identification.
\newblock {\em Accident Analysis {$\&$} Prevention}, 39(6):1192--1201.

\bibitem[Mitchell and Beauchamp, 1988]{Mitchell1988}
Mitchell, T.~J. and Beauchamp, J.~J. (1988).
\newblock Bayesian variable selection in linear regression.
\newblock {\em Journal of the American Statistical Association},
  83(404):1023--1032.

\bibitem[Müller et~al., 2004]{Mueller2004}
Müller, P., Parmigiani, G., Robert, C., and Rousseau, J. (2004).
\newblock Optimal sample size for multiple testing.
\newblock {\em Journal of the American Statistical Association},
  99(468):990--1001.

\bibitem[Nelder and Wedderburn, 1972]{Nelder1972}
Nelder, J.~A. and Wedderburn, R. W.~M. (1972).
\newblock Generalized linear models.
\newblock {\em Journal of the Royal Statistical Society. Series A (General)},
  135(3):370.

\bibitem[Park and Casella, 2008]{Park2008}
Park, T. and Casella, G. (2008).
\newblock The {B}ayesian lasso.
\newblock {\em Journal of the American Statistical Association},
  103(482):681--686.

\bibitem[Sarkar and Chang, 1997]{Sarkar1997}
Sarkar, S.~K. and Chang, C.-K. (1997).
\newblock The {S}imes method for multiple hypothesis testing with positively
  dependent test statistics.
\newblock {\em Journal of the American Statistical Association},
  92(440):1601--1608.

\bibitem[Scott and Berger, 2006]{Scott2006}
Scott, J.~G. and Berger, J.~O. (2006).
\newblock An exploration of aspects of bayesian multiple testing.
\newblock {\em Journal of Statistical Planning and Inference},
  136(7):2144--2162.

\bibitem[Sesia et~al., 2019]{Sesia2019}
Sesia, M., Sabatti, C., and Cand{\`{e}}s, E.~J. (2019).
\newblock Gene hunting with hidden markov model knockoffs.
\newblock {\em Biometrika}, 106(1):1--18.

\bibitem[Storey, 2002]{Storey2002}
Storey, J.~D. (2002).
\newblock A direct approach to false discovery rates.
\newblock {\em Journal of the Royal Statistical Society: Series B (Statistical
  Methodology)}, 64(3):479--498.

\bibitem[Tibshirani, 1996]{Tibshirani1996}
Tibshirani, R. (1996).
\newblock Regression shrinkage and selection via the lasso.
\newblock {\em Journal of the Royal Statistical Society: Series B
  (Methodological)}, 58(1):267--288.

\bibitem[Wang et~al., 2020]{Wang2020}
Wang, G., Sarkar, A., Carbonetto, P., and Stephens, M. (2020).
\newblock A simple new approach to variable selection in regression, with
  application to genetic fine mapping.
\newblock {\em Journal of the Royal Statistical Society: Series B (Statistical
  Methodology)}.

\bibitem[Whittemore, 2007]{Whittemore2007}
Whittemore, A.~S. (2007).
\newblock A {B}ayesian false discovery rate for multiple testing.
\newblock {\em Journal of Applied Statistics}, 34(1):1--9.

\bibitem[Yekutieli and Benjamini, 2001]{Yekutieli2001}
Yekutieli, D. and Benjamini, Y. (2001).
\newblock The control of the false discovery rate in multiple testing under
  dependency.
\newblock {\em The Annals of Statistics}, 29(4):1165--1188.

\end{thebibliography}
\end{document}